\documentclass[reprint,aps,prx,groupedaddress,amsmath,amssymb]{revtex4-2}
\usepackage{graphicx}
\usepackage{dcolumn}
\usepackage{bm}
\usepackage{xcolor}
\usepackage[normalem]{ulem}
\usepackage[USenglish]{babel}
\usepackage{float}
\usepackage{times}
\usepackage{physics}
\usepackage[colorlinks]{hyperref}

\newcommand{\bk}[1]{ \hat{b}_{#1}^{\ } }
\newcommand{\bkd}[1]{ \hat{b}_{#1}^{\dagger} }

\newcommand{\rhoc}{ \hat{\rho}^c }
\newcommand{\rhou}{ \hat{\rho} }
\newcommand{\avg}[1]{ \langle #1 \rangle }
\newcommand{\avgc}[1]{ \langle #1 \rangle^c }
\newcommand{\Cc}[1]{ C^c_{#1} }
\newcommand{\WO}{ \mathbf{W}_{\rm O} }

\newcommand{\varg}{ \text{\it{g}} }

\newcommand{\QTR}{ Q_{\rm Train} }
\newcommand{\QTE}{ Q_{\rm Test} }

\newcommand{\Ltot}{\mathcal{L}}
\newcommand{\Lqrc}{\mathcal{L}_{\rm QRC}}
\newcommand{\Lsys}{\mathcal{L}_{\rm S}}
\newcommand{\Smeas}{\mathcal{S}_{\rm meas}(dW)}
\newcommand{\Lc}{\mathcal{L}_{c}}
\newcommand{\cmax}{\mathcal{C}_{\rm max}}
\newcommand{\nsmax}{N_{\rm S}^{\rm max}}
\newcommand{\tmax}{t_{\rm max}}
\newcommand{\CNL}{\mathcal{N}}
\newcommand{\CNLE}{\mathcal{N}_{\rm eff}}
\newcommand{\BD}{\mathcal{B}^{12}}

\newcommand{\todo}[1]{{\textcolor{red}{[To-do: #1]}}}

\makeatletter
\let\ORIbbl@fixname\bbl@fixname
\def\bbl@fixname#1{%
  \@ifundefined{languagealias@\expandafter\string#1}
    {\ORIbbl@fixname#1}
    {\edef\languagename{\@nameuse{languagealias@#1}}}%
}
\newcommand{\definelanguagealias}[2]{%
  \@namedef{languagealias@#1}{#2}%
}
\makeatother

\definelanguagealias{en}{english}
\definelanguagealias{EN}{english}

\begin{document}



\title[]{Physical reservoir computing using finitely-sampled quantum systems}

\makeatletter

\date{\today}

\author{Saeed A. Khan}
\author{Fangjun Hu}
\author{Gerasimos Angelatos}
\author{Hakan E. T\"{u}reci}
\affiliation{Department of Electrical and Computer Engineering, Princeton University, Princeton, NJ 08544}
\begin{abstract}
The paradigm of reservoir computing exploits the nonlinear dynamics of a physical reservoir to perform complex time-series processing tasks such as speech recognition and forecasting.  Unlike other machine-learning approaches, reservoir computing relaxes the need for optimization of intra-network parameters, and is thus particularly attractive for near-term hardware-efficient quantum implementations. However, the complete description of practical quantum reservoir computers requires accounting for their placement in a quantum measurement chain, and its conditional evolution under measurement. Consequently, training and inference has to be performed using finite samples from obtained measurement records. Here we describe a framework for reservoir computing with nonlinear quantum reservoirs under continuous heterodyne measurement. Using an efficient truncated-cumulants representation of the complete measurement chain enables us to sample stochastic measurement trajectories from reservoirs of several coupled nonlinear bosonic modes under strong excitation. This description also offers a mathematical basis to directly compare the computational capabilities of a given physical reservoir operated across classical and quantum regimes. Applying this framework to the classification of quantum states of systems that are part of the same measurement chain as the quantum reservoir computer, we assess and explain measurement-contingent advantages and disadvantages of reservoir processing in quantum regimes. Our results also identify the vicinity of bifurcation points as presenting optimal nonlinear processing regimes of an oscillator-based quantum reservoir. The considered models are directly realizable in modern circuit QED experiments, while the framework is applicable to more general quantum nonlinear reservoirs.

\end{abstract}

\maketitle


\section{Introduction}

Reservoir computing is a machine-learning paradigm that emphasizes learning efficiency~\cite{Jaeger2004, Jaeger2009, Tanaka2019, gauthier_next_2021}:
a linear combination of the accessible degrees of freedom of a physical `reservoir' is the only quantity trained for a given task. Crucially, foregoing optimization of internal parameters does not necessarily hinder the computational capacity of reservoir computers, which in recent years have enabled efficient approaches to forecasting \cite{pathak_model-free_2018, griffith_forecasting_2019, gauthier_predicting_2021}, inference~\cite{nakai_machine-learning_2018, rohm_model-free_2021}, control~\cite{canaday_model-free_2021}, and similar resource-intensive signal processing tasks~\cite{brunton_discovering_2016, pathak_using_2017, larger_high-speed_2017, rafayelyan_large-scale_2020, barbosa_symmetry-aware_2021}. By relaxing control over microscopic parameters of the learning machine, reservoir computing embraces the fundamental notion of computation with very general dynamical systems: physical or artificial dynamical systems stimulated by incoming time-dependent signals $\mathbf{u}(t)$ perform computation by transforming an input stream into an output stream, effectively computing a function $\mathcal{F}\{\mathbf{u}(t)\}$~[see Fig.~\ref{fig:schematic}(a)]. If the dynamical system is sufficiently complex and high-dimensional, the expressive power of even a linear classifier acting on the output state can enable approximation of a large set of functions $\mathcal{F}\{\cdot\}$ \cite{grigoryeva_echo_2018, gonon_reservoir_2020, nokkala_gaussian_2021}. In this vein, diverse dynamical systems ranging from early mathematical black-box reservoirs~\cite{Jaeger2004} to physical systems across optical~\cite{larger_photonic_2012, brunner_parallel_2013, duport_all-optical_2012, mesaritakis_high-speed_2015, coarer_all-optical_2018, Dong_Rafayelyan_Krzakala_Gigan_2020}, electrical~\cite{soriano_delay-based_2015, haynes_reservoir_2015,canaday_rapid_2018, apostel_reservoir_2021} and recently superconducting~\cite{rowlands_reservoir_2021, angelatos_reservoir_2021} platforms have all been explored as reservoir computers~\cite{nakajima_physical_2020, wright_deep_2021, nakajima_reservoir_2021}. 



%


This flexibility has naturally led to the examination of \textit{quantum} systems as potential physical substrates for reservoir computing~\cite{Fujii2017, ghosh_quantum_2019, Wright2019, Nokkala2020, govia_quantum_2021, martinez-pena_dynamical_2021}, with the promise of exponentially-scaling reservoir dimensionality. Early implementations of quantum reservoirs~[see Fig.~\ref{fig:schematic}(a)] consider computation with explicitly quantum components (e.g. interacting spins, bosonic or fermionic degrees of freedom) and encode the input $\mathbf{u}(t)$ in the Liouvillian governing state dynamics, or even in the quantum reservoir state directly as a `quantum' input. Reservoir output is often constructed assuming access to expectation values of reservoir dynamical variables that correspond loosely to measurable observables (e.g. occupation numbers via photodetectors, quadratures via homodyne measurement).

While such schemes have been used to theoretically show certain performance advantages compared to classical systems~\cite{kalfus_neuromorphic_2021}, critical foundational questions have yet to be addressed. The \textit{experimentally-accessible} degrees of freedom of a quantum system are constrained given a specific input and output scheme, which must be taken into account. Importantly, the role of the quantum theory of measurement in sampling outputs from such a system for machine-learning remains unexplored. Furthermore, a quantum reservoir has yet to be treated from the perspective of a physically-motivated classical limit, the need for which is twofold. First, this provides the basis to unequivocally identify any possible advantages afforded by quantum reservoir processing. Secondly, the energy expended during a given computational task is a fundamental concern in physical computing; when considering computation with decreasing signal powers, the transition from classical to quantum regimes of reservoir operation must naturally be addressed.


This paper introduces a framework for quantum reservoir computing that allows the same physical reservoir to be controllably operated across quantum or classical regimes. To satisfy this requirement, we present an analysis built upon an intrinsically quantum-mechanical description of the complete measurement chain, including not only the physical reservoir but also inputs and measured outputs necessary to define a complete computational framework. The classical limit is then identified where the state of the measurement chain during computation is well-approximated as a product of coherent states. This unified approach allows us to define performance metrics (such as the {\it fidelity} of executing a particular task, or the {\it time-to-solution}) that account for limitations due to a prescribed input-output scheme and measurement noise, and can be compared across classical or quantum regimes of operation of the same physical reservoir. Such a comparative metric provides the theoretical basis to identify task-specific and measurement-contingent advantages (or disadvantages, as we shall see) of operating physical reservoirs in the quantum regime.

The quantum-mechanical treatment of the complete measurement chain we require demands the specification of details often simplified in earlier studies. A representative layout of the quantum measurement chain we wish to describe is shown in Fig.~\ref{fig:schematic}(b), depicted here in the circuit QED (cQED) architecture~\cite{wallraff_strong_2004, blais_circuit_2021} (many components are generalizable to other platforms as well). The quantum reservoir computer (QRC) is modelled as a $K$-node quantum nonlinear device, whose driven-dissipative dynamics are governed by a Liouvillian superoperator $\Lqrc$. Inputs to the QRC must be transmitted via classical control fields used to interact with the quantum measurement chain; while classical inputs can be directly encoded into these control fields (shown here at a single frequency $\omega_d$), the question of quantum inputs (also referred to as 'quantum data' in earlier work~\cite{Wright2019}) is more involved. We define quantum inputs as states of quantum systems (described by a Liouvillian superoperator $\Lsys$) generated in the same cryogenic environment as the QRC. The flow of signals from this system to the reservoir is accurately modeled via a general coupling superoperator $\Lc$, which typically includes directional elements. This description of the composite measurement chain accurately accounts for correlations between the quantum system and the QRC, including the extreme case where the system and the quantum reservoir are \textit{entangled}, a scenario with no analog in classical reservoir computing. Evolution of the QRC under classical or quantum inputs realizes a high-dimensional nonlinear mapping of the input to the reservoir's state space.

Information encoded in the QRC state is extracted from the measurement chain via continuous heterodyne monitoring of the QRC nodes, which requires us to address limitations on measured outputs imposed by quantum theory. These measurements do not directly yield ensemble-averaged expectation values of reservoir state variables, but noisy output field measurement records that include quantum fluctuations non-linearly transformed by the quantum reservoir itself, and are dependent upon the conditional quantum state of the measurement chain. All this is accounted for by formulating and solving for the conditional evolution of all dynamical elements in the quantum measurement chain under the prescribed measurement scheme (often called quantum trajectories~\cite{carmichael_quantum_1993} subject to measurement backaction). The final step in our quantum reservoir computing framework is an output layer applied to these measured outputs, performing \textit{linear} operations that are trained on \textit{finite} measurement records and not directly on expectation values. The latter in principle assumes access to an infinite number of measurement records, ignoring a key resource cost of quantum reservoir computing, and for higher-order expectation values can introduce \textit{classical} nonlinear operations, whose computational influence should be distinguished from that of the nonlinear quantum reservoir alone. This output layer can be implemented in software or, for real-time processing, via a customized closed-loop signal processing hardware such as FPGAs at room temperature.

Starting from the formal mathematical description of the complete measurement chain via a \textit{stochastic master equation} (SME)~\cite{wiseman_quantum_1993, wiseman_interpretation_1993}, we develop an efficient computational framework to simulate its use for quantum reservoir computing. By considering a representative machine-learning task of quantum state classification with small-scale QRCs, we demonstrate computation with quantum reservoirs on quantum inputs across classical and quantum regimes of reservoir operation. Our principal result is that advantages of operating reservoirs in the quantum regime are task-dependent: once limitations due to measurement are taken into account, not all state classification tasks immediately benefit from quantum reservoir processing. Specific tasks, which we identify, can be carried out with reduced measurement resources in quantum regimes of reservoir operation.

Before presenting the complete analysis of this quantum reservoir computing framework that constitutes the main body of this paper, in the remaining subsections of this introduction we preview the technical aspects that make this analysis possible, and a summary of its main results.

\subsection{Technical approach}

\begin{figure}[t]
    \centering
    \includegraphics[scale=1.0]{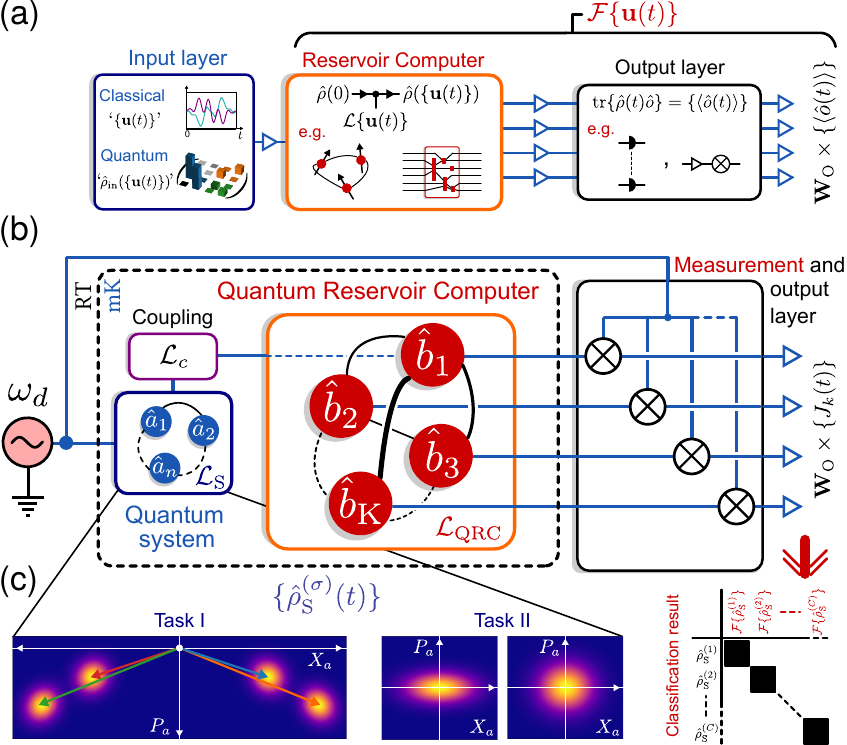}
    \caption{(a) General quantum reservoir computing frameworks. (b) Schematic representation of quantum reservoir computing framework under continuous measurement, describing a $K$-node quantum reservoir computer (QRC) as part of a measurement chain for processing signals encoded in classical control fields such as the single coherent drive (frequency $\omega_d$), or generated by quantum systems in the same quantum environment. The dashed boundary separates classical controls at room temperature (RT) from quantum devices at cryogenic temperatures (mK). The quantum system is governed by Liouvillian $\Lsys$, and is coupled to the QRC via $\Lc$. The QRC nodes undergo heterodyne measurement at the completion of the measurement chain, leading to its conditional evolution under measurement superoperators $\Smeas$. (c) The QRC is used to classify qubit pointer states generated in a cavity (task I), and states with equal phase and mean amplitude, but distinct variances (task II).} 
    \label{fig:schematic}
\end{figure}

The evolution of the conditional quantum state $\rhoc$ of the composite measurement chain under continuous measurement shown in Fig.~\ref{fig:schematic}(b) is formally described by the SME
\begin{align}
d\rhoc =  \Lsys\rhoc~dt + \Lc\rhoc~dt + \Lqrc\rhoc~dt +\mathcal{S}_{\rm meas}(dW)\rhoc
\label{eq:sme}
\end{align}
where $\Smeas$ is the stochastic measurement superoperator. For extended multi-mode and generally nonlinear networks of interest, such as the one presented by the QRC setup of interest in Fig.~\ref{fig:schematic}(b), Eq.~(\ref{eq:sme}) is typically unfeasible to simulate, limited by growing Hilbert space constraints. To accurately yet efficiently account for dynamics of the full measurement chain, a truncated-cumulants representation of its conditional quantum state is developed; the choice of representation is strongly dependent on the measurement scheme and Liouvillian dynamics in Eq.~(\ref{eq:sme}). This description extends earlier work on continuously-monitored linear modes~\cite{breslin_conditional_1997, doherty_feedback_1999} to nonlinear multimode quantum systems, yielding a set of stochastic equations for measured observables that scales much more favourably with increasing system complexity. The accuracy of this representation requires specific conditions on the nonlinearity of the quantum reservoir and its operating conditions which, as discussed below, can be readily satisfied within the cQED architecture. Additionally, this framework admits a well-defined classical limit, where the joint quantum state of the measurement chain is well-approximated as a product of coherent states. Defined in this way, the classical limit can be analyzed without having to redefine the quantum-classical boundary (colloquially, the \textit{Heisenberg-von-Neumann cut}) separating the measured system from the classical observer, which we always place at the termination of the measurement chain.



\subsection{Machine-learning tasks using quantum inputs}


As a quantum information processing task, we consider the problem of classifying $C$ states of a quantum system (governed by $\Lsys$) embedded in the same measurement chain as the QRC. These states $\hat{\rho}^{(\sigma)}_{\rm S}(t)$ (indexed by $\sigma=1,\ldots,C$) are dynamically generated by inputs to the measurement chain, and also evolve under the backaction of quantum measurement of the QRC nodes. As such, they are defined via the \textit{conditional} quantum state of the measurement chain by tracing out the QRC, $\hat{\rho}^{(\sigma)}_{\rm S}(t) = {\rm tr}_{\rm QRC}\{\hat{\rho}^{c(\sigma)}(t)\}$. Note that this is simply a formal definition: our computational approach does not trace out any degrees of freedom of the measurement chain. For concreteness, we consider the classification of system states that (I) differ in their mean amplitudes and phases, and (II) differ only in their variances~[see Fig.~\ref{fig:schematic}(c)]. Task I is closely related to the classification of pointer states for dispersive qubit readout in cQED~\cite{boissonneault_dispersive_2009}, while Task II can be applied to distinguishing entangled and non-entangled quantum states~\cite{sandbo_chang_generating_2018}. In both cases, suitable physical QRCs can process these quantum inputs to produce distinct measured outputs, whose distributions include non-linearly transformed quantum fluctuations due to the action of the nonlinear QRC itself. To successfully implement the function $\mathcal{F}\{\hat{\rho}_{\rm S}^{(\sigma)}(t)\} = \sigma$ for quantum state classification, as our framework does, the linear output layer must be trained to take these non-classical fluctuations into account.


\subsection{Performance across classical and quantum operating regimes}

The quantum reservoir computing framework presented here is constructed to assess computation fidelity as a function of quantum reservoir size $K$, and more importantly to explore the role of quantum resources such as entanglement~\cite{bennett_mixed-state_1996} amongst the reservoir nodes. However, the generation and preservation of entanglement itself requires additional resources and operating constraints~\cite{arnesen_natural_2001} on the QRC. As a first study, we find that comparing QRC performance across the transition from a classical regime of QRC operation to a quantum regime without necessarily demanding entanglement within the quantum reservoir already yields important insights into the role of quantum reservoir processing. Furthermore, this analysis provides the basis for future work to precisely identify any further advantages of reservoir entanglement.



We show that the input power to the measurement chain provides a natural, physically-motivated parameter to explore this classical-to-quantum transition: with decreasing input power, maintaining computational accuracy for quantum state classification requires increasing the QRC nonlinearity, which we show corresponds to the quantum regime of QRC operation. Importantly, across this transition, QRC performance on Tasks I and II shows important qualitative differences. Task I, which requires the QRC to distinguish the amplitudes and phases of quantum states, fares better in the classical regime. Task II, on the other hand, requires the QRC to distinguish information encoded in quantum fluctuations. We find that this task can be performed with reduced measurement resources in increasingly quantum regimes (for details see Sec.~\ref{subsec:cavquantum},~\ref{subsec:ampquantum} respectively).

\subsection{Nonlinear processing near classical bifurcation points}

While the two quantum state classification tasks probe distinct nonlinear processing capabilities of the QRC, we find that useful regimes of reservoir operation are unified by their proximity to classical bifurcation points. Near these bifurcation phases, we find an enhancement of the nonlinear response of the QRC to its inputs; this suggests a general signature to identify parameter regimes of optimal reservoir performance for a broader class of computational tasks.

\subsection{Organization of the paper}

The remainder of this paper is organized as follows. In Sec.~\ref{sec:QRC}, we introduce the specific cQED realization of the $K$-node QRC and the measurement chain. In Sec.~\ref{sec:singleNodeQRC}, we use a minimal version of this measurement chain, a QRC comprised of a single nonlinear mode, to develop our theoretical approach to the simulation of its conditional dynamics, and discuss its regimes of validity. Sec.~\ref{sec:cav} then considers the use of the QRC for classifying pointer states, and Sec.~\ref{sec:amp} for the classification of general Gaussian states with distinct variances (Tasks I and II respectively). In both cases, we provide a characterization of the distinct quantum measurement chains and their measured outputs, a detailed study of QRC parameters that enable successful classification, and a comparison of QRC performance across classical and quantum operating regimes.

\section{Physical Model for a Quantum Reservoir Computer}
\label{sec:QRC}

To describe the QRC, we adopt a model that is realizable within the cQED architecture, based on $K$ coupled Kerr nonlinear modes furnished by Josephson junctions~\cite{squidHandBook}. The nonlinear modes have frequencies $\{\omega_k\}$, nonlinearity strengths $\{\Lambda_k\}$, with linear coupling $\{\varg_{ij}\}$ between modes $i$ and $j$, yielding the Liouvillian governing QRC dynamics,
\begin{align}
\Lqrc\rhoc = -i\Big[ &\sum_k \omega_k \bkd{k}\bk{k} - \sum_k \frac{\Lambda_k}{2}\bkd{k}\bkd{k}\bk{k}\bk{k} ,\rhoc \Big]  \nonumber \\
-i\Big[&\sum_{ij}\varg_{ij}(\bk{i}\bkd{j} + \bkd{i}\bk{j}),\rhoc \Big] + \sum_k \gamma_k \mathcal{D}[\bk{k}]\rhoc.
\label{eq:Lqrc}
\end{align}
The linear damping is described by the standard dissipative superoperator $\mathcal{D}[\hat{o}] = \hat{o}\rhou\hat{o}^{\dagger} - \frac{1}{2}\{ \hat{o}^{\dagger}\hat{o},\rhou\}$ , with rates $\{\gamma_k\}$. The decay channels for each node are continuously monitored with unit efficiency via a heterodyne measurement scheme, described by the stochastic measurement superoperator
\begin{align}
&\mathcal{S}_{\rm meas}(dW)\rhoc = \nonumber \\
&\sum_{k=1}^K \sqrt{\frac{\gamma_k}{2}} \left( \bk{k} \rhoc + \rhoc \bkd{k} - \avgc{\bk{k}+\bkd{k}} \right) dW_{k}^{X}(t) + \nonumber \\
&\sum_{k=1}^K \sqrt{\frac{\gamma_k}{2}} \left( -i\bk{k} \rhoc + i\rhoc \bkd{k} - \avgc{-i\bk{k}+i\bkd{k}} \right) dW_{k}^{P}(t), 
\label{eq:Smeas}
\end{align}
where $dW_k^{X,P}$ are independent Wiener increments, and $\avgc{\hat{o}}$ indicates the expectation value of an arbitrary operator $\hat{o}$ with respect to the conditional quantum state, $\avgc{\hat{o}} = {\rm tr}\{\rhoc\hat{o}\}$. The results of this continuous measurement are heterodyne currents $J_k^{X,P}(t)$ read out from each reservoir node,
\begin{align}
    J_{k}^{X}(t)~dt &= \sqrt{\frac{\gamma_k}{2}}\avgc{\bk{k}  + \bkd{k} }~dt + dW_k^{X}(t), \label{eq:JX} \\
    J_{k}^{P}(t)~dt &= \sqrt{\frac{\gamma_k}{2}}\avgc{-i\bk{k}  + i\bkd{k} }~dt + dW_k^{P}(t). \label{eq:JP} 
\end{align}

As described by Eq.~(\ref{eq:sme}), the QRC is just one part of the measurement chain. The signal-generating quantum system is described by the superoperator $\Lsys$, and will be task-dependent; we will consider specific realizations throughout this paper. The coupling superoperator $\Lc$ completes the measurement chain by allowing the measured quantum system and the QRC to interact. It can once again take an arbitrary form, allowing for both coherent and dissipative, reciprocal and non-reciprocal, and even nonlinear couplings. Depending on the type of coupling interaction, $\Lc$ may also introduce local terms affecting the measured quantum system and the QRC; we will make these clear when they arise.

Having defined every element of the measurement chain in Fig.~\ref{fig:schematic}(b), we will now analyze the nature of its quantum dynamics, introducing an approach to studying this dynamics that is scalable to extended, multimode quantum systems.

\section{Quantum dynamics of the single-node Kerr QRC}
\label{sec:singleNodeQRC}

\begin{figure}[t]
    \centering
    \includegraphics[scale=1.0]{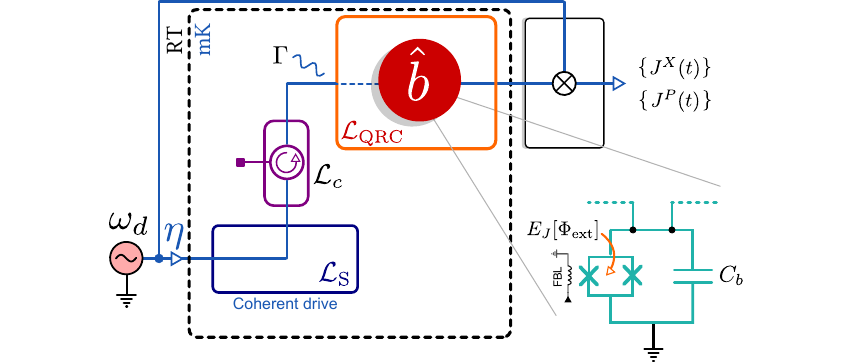}
    \caption{Schematic showing the fundamental node of the proposed QRC as part of a simplified measurement chain: a `system' described by a single frequency coherent drive incident on the QRC node via an input channel that is not measured, and continuous weak measurement of the QRC node output. The QRC node can be realized in cQED as a tunable-frequency Kerr oscillator using a capacitively shunted SQUID loop~\cite{koch_charge-insensitive_2007}.}
    \label{fig:singleQRCClassical}
\end{figure}

To analyze the nonlinear processing capabilities of this QRC, it will prove useful to begin with a simple minimal realization - a single Kerr node - and a single coherent drive incident on the QRC directly, via a separate input port with loss rate $\Gamma$. This situation describes QRC processing of classical signals encoded in the generally time-dependent coherent drive amplitude and phase. The resulting system schematic is shown in Fig.~\ref{fig:singleQRCClassical}. We begin by analyzing the \textit{unconditional} dynamics of the resulting system, described by the quantum master equation
\begin{align}
    d\hat{\rho} \equiv \mathcal{L}\rhou~dt = \Lqrc\hat{\rho}~dt  - i[\hat{\mathcal{H}}_d,\hat{\rho}]~dt
\end{align}
where the drive Hamiltonian in the frame rotating at the drive frequency $\omega_d$ is then
\begin{align}
\hat{\mathcal{H}}_d = \eta(\hat{b} + \hat{b}^{\dagger}).
\end{align}
The Liouvillian $\mathcal{L}_{\rm QRC}$ describing the single-node QRC takes the form
\begin{align}
    \Lqrc\hat{\rho} = -i\left[-\Delta\hat{b}^{\dagger}\hat{b} - \frac{\Lambda}{2}\hat{b}^{\dagger}\hat{b}^{\dagger}\hat{b}\hat{b},\hat{\rho} \right] + (\gamma+\Gamma) \mathcal{D}[\hat{b}]\hat{\rho}
    \label{eq:singleQRCME}
\end{align}
where $\Delta = \omega_d - \omega$ is the detuning of the drive frequency from the bare QRC node frequency $\omega$. For simplicity of notation, in what follows we introduce for convenience the total linear damping $\widetilde{\gamma} = \gamma + \Gamma$ of the nonlinear mode.


\subsection{Equations of motion approach based on truncation of higher-order cumulants}

The quantum state described by an arbitrary density matrix $\rhou$ can always be written in terms of normal-ordered cumulants, defined for a single quantum mode $\hat{b}$ as
\begin{equation}
    C_{b^{\dagger p} b^q} \equiv \left. \frac{\partial^{p+q}  }{\partial (i z^{\ast})^p \partial (i z)^q} \ln \mathrm{tr} \{\hat{\rho} e^{i z^{\ast} \hat{b}^{\dagger}} e^{i z \hat{b}}\} \right|_{z=z^{\ast}=0}.
\end{equation}
The term in curly braces may immediately be recognized as the characteristic function of $\rhou$~\cite{carmichael_statistical_2002}, associated with normal-ordered phase-space distributions such as the $P$-representation; this connection is elaborated upon in Appendix~\ref{app:cumulants}. The above definition generalizes straightforwardly to states $\rhou$ describing multiple bosonic modes. 

Cumulants are defined by their order $p+q$, and provide an equivalent set of variables to describe the dynamics of a quantum state. Specifically, we can write down the equation of motion for the first-order cumulant of the single-node QRC ($C_b \equiv \avg{\hat{b}} = {\rm tr}\{\mathcal{L}\rhou \hat{b}\}$),
\begin{align}
    \avg{\dot{\hat{b}}} &= \left( i\Delta -\frac{\widetilde{\gamma}}{2} \right)\avg{\hat{b}} + i\Lambda \avg{\hat{b}^{\dagger}}\avg{\hat{b}}\avg{\hat{b}} -i \eta  \nonumber \\
    &~~~~+ i\Lambda \left( C_{bb}\avg{\hat{b}^{\dagger}} + 2C_{b^{\dagger}b}\avg{\hat{b}} + C_{b^{\dagger}bb} \right)
    \label{eq:b}
\end{align}
where $C_{bb} = \avg{\hat{b}\hat{b}}-\avg{\hat{b}}^2, C_{b^{\dagger}b} = \avg{\hat{b}^{\dagger}\hat{b}} - \avg{\hat{b}^{\dagger}}\avg{\hat{b}}$, and $C_{b^{\dagger}bb} = \avg{\hat{b}^{\dagger}\hat{b}\hat{b}} - 2\avg{\hat{b}^{\dagger}\hat{b}}\avg{\hat{b}} - \avg{\hat{b}\hat{b}}\avg{\hat{b}^{\dagger}} + 2\avg{\hat{b}^{\dagger}}\avg{\hat{b}}^2$. Expressions for higher-order cumulants become increasingly more unwieldy, but can be systematically obtained, as discussed in Appendix~\ref{app:mToC}. We see immediately that the first-order cumulant $\avg{\hat{b}}$ couples to cumulants of second and higher order. We can similarly obtain equations of motion for unique second-order cumulants $C_{b^{\dagger}b}, C_{bb}$,
\begin{subequations}
\begin{align}
    \dot{C}_{b^{\dagger}b} &= -\widetilde{\gamma}C_{b^{\dagger}b} -i\Lambda (C_{bb}\avg{\hat{b}^{\dagger}}^2-C_{bb}^*\avg{\hat{b}}^2) \nonumber \\
    &~~~~- i\Lambda (C_{b^{\dagger}bb}\avg{\hat{b}}-C_{b^{\dagger}bb}^*\avg{\hat{b}^{\dagger}} ), \label{eq:CbdbFull} \\
    \dot{C}_{bb} &= \left( i2\Delta - \widetilde{\gamma} + i\Lambda(1+4|\avg{\hat{b}}|^2\!\! +6C_{b^{\dagger}b}) \right)C_{bb}     \nonumber \\  
    + i\Lambda &\avg{\hat{b}}^2 (1+2C_{b^{\dagger}b}) +i2\Lambda ( C_{b^{\dagger}bbb} +2\avg{\hat{b}}C_{b^{\dagger}bb} + C_{bbb}\avg{\hat{b}^{\dagger}}  ). \label{eq:CbbFull} 
\end{align}
\end{subequations}
Clearly, second-order cumulants couple to cumulants of third and fourth order. This is a general feature of nonlinear quantum systems: moments and thus cumulants of a certain order can couple to those of higher-order, leading to an infinite hierarchy of equations that do not form a closed set.




However, an infinite set of normal-ordered cumulants is not necessarily required to describe all multimode quantum states. In particular, a multimode quantum system in a product of coherent states is described entirely by its nonzero first-order cumulants; all cumulants of order $p+q > 1$ vanish~(see Appendix~\ref{app:cumulantsQS} for derivations). Multimode quantum states that are defined entirely by their first and second-order cumulants admit Gaussian phase-space representations, and are labelled Gaussian states. States with nonzero cumulants of third or higher-order are thus by definition non-Gaussian states~\cite{boutin_effect_2017, ra_non-gaussian_2020}.



Our numerical approach leverages this efficient representation of specific multimode quantum states in terms of cumulants. In particular, we consider an ansatz wherein the quantum state of the complete measurement is described entirely by cumulants up to a finite order $p+q \leq n_{\rm trunc}$; all cumulants of order $p+q > n_{\rm trunc}$ are thus set to zero, truncating the hierarchy and yielding a closed set of equations for the retained nonzero cumulants. In this paper, we choose $n_{\rm trunc} = 2$ for a quantum measurement chain defined entirely by its first and second-order cumulants, although the truncation can similarly be carried out at higher order. The resulting Truncated Equations Of Motion (TEOMs), and their generalization to \textit{conditional} dynamics of multimode quantum systems introduced in Sec.~\ref{subsec:condQRCdyn}, form the basis of our computational approach in this paper. 

Since we are specifically interested in \textit{nonlinear} quantum systems for reservoir computing, which can generate higher-order cumulants in dynamics, one must ask when such an ansatz may hold. In the two following subsections, we identify system-specific drive and parameter regimes where this ansatz is valid, and verify our conclusions via numerical comparisons with exact results and (S)ME simulations. 




\subsection{Application to unconditional QRC node dynamics}

We will use the single coherently-driven QRC node modeled as a Kerr oscillator to benchmark the TEOMs approach. In doing so, we make extensive use of the exact quantum steady-state solution of the coherently-driven Kerr oscillator, obtained via the complex-$P$ representation~\cite{drummond_quantum_1980}. This provides access to steady-state cumulants of arbitrary order~(see Appendix~\ref{app:complexP} for details), which can be used to identify regions of parameter space where a truncation of cumulants beyond second-order can be justified.

One such regime should be furnished by the \textit{classical limit} of the single Kerr QRC node, where the QRC is always in a coherent state and thus described entirely by its first-order cumulants. To identify this classical limit, and thus explore deviations from it due to quantum effects, it proves convenient to work with specific scaled quantities: after introducing dimensionless time $t' = \widetilde{\gamma}t$ and energy scales $(\Delta',\Lambda',\eta') = (\Delta,\Lambda,\eta)/\widetilde{\gamma}$, we scale $\avg{\hat{b}}'\to \sqrt{\Lambda'}\avg{\hat{b}}$, and impose our truncated ansatz, following which Eq.~(\ref{eq:b}) becomes
\begin{align}
    \frac{d}{dt'}\avg{\dot{\hat{b}}}' &= \left( i\Delta' -\frac{1}{2} \right)\avg{\hat{b}}' + i \avg{\hat{b}^{\dagger}}'\avg{\hat{b}}'\avg{\hat{b}}' -i \CNL  \nonumber \\
    & + i\Lambda' \left( C_{bb}\avg{\hat{b}^{\dagger}}' + 2 C_{b^{\dagger}b}\avg{\hat{b}}' \right). 
    \label{eq:transfb}
\end{align}
The first line in Eq.~(\ref{eq:transfb}) makes no reference to second-order cumulants of the QRC node state; it thus describes dynamics when the QRC is in a coherent state (for which such cumulants vanish), corresponding to the classical limit. Dynamics in this limit depend on the drive and the nonlinearity via a single dimensionless parameter~\cite{dykman_theory_1979, dykman_fluctuating_2012},
\begin{align}
    \CNL = \eta'\sqrt{\Lambda'} = \frac{\eta}{\widetilde{\gamma}} \sqrt{\frac{ \Lambda}{\widetilde{\gamma}}}.
    \label{eq:CNL}
\end{align}

The second line of Eq.~(\ref{eq:transfb}) does however depend on cumulants; these terms describe quantum dynamics that lead to a deviation from purely coherent state evolution. In these scaled units, the TEOMs for the second-order cumulants obtained from Eq.~(\ref{eq:CbdbFull}),~(\ref{eq:CbbFull}) take the form
\begin{subequations}
\begin{align}
    \dot{C}_{b^{\dagger}b} &= -C_{b^{\dagger}b} -i (C_{bb}(\avg{\hat{b}^{\dagger}}')^2-C_{bb}^*(\avg{\hat{b}}')^2), \label{eq:transfCbdb} \\
    \dot{C}_{bb} &= \left( i2\Delta' - 1 + i(\Lambda'+4|\avg{\hat{b}}'|^2\!\! +6\Lambda'C_{b^{\dagger}b}) \right)C_{bb}  \nonumber \\
    &~~~~+ i(\avg{\hat{b}}')^2(1+2C_{b^{\dagger}b}).  \label{eq:transfCbb}
\end{align}
\end{subequations}
Consider now the transformation where $\Lambda' \to 0$, while $\CNL$ is held fixed (by increasing the drive strength $\eta'$ simultaneously). Terms in the first line of Eq.~(\ref{eq:transfb}) describing classical dynamics remain unchanged, while those in the second line describing quantum deviations become smaller, \textit{provided} second-order cumulants do not grow with $\Lambda'$. From Eqs.~(\ref{eq:transfCbdb}),~(\ref{eq:transfCbb}), it is clear that $C_{b^{\dagger}b}$ is independent of $\Lambda'$, while $C_{bb}$ has a dependence that is negligible as $\Lambda' \to 0$ for $|\avg{\hat{b}}'| \neq 0$. Hence by keeping $\CNL$ fixed while decreasing the nonlinearity strength, the influence of second-order cumulants on dynamics of the QRC node amplitude is suppressed, describing the classical limit of QRC operation. Physically, this transformation leads to an increase in the unscaled drive $\eta'$ and QRC node amplitude $|\avg{\hat{b}}|$, and hence occupation number, decreasing the relative impact of quantum fluctuations in agreement with conventional notions of classicality.
Conversely, increasing nonlinearity strength for fixed $\CNL$ should enhance the impact of quantum fluctuations, marked by a systematic increase in second-order cumulants, and taking the Kerr QRC gradually towards a quantum regime of operation.



\begin{figure}[t]
    \centering
    \includegraphics[scale=1.0]{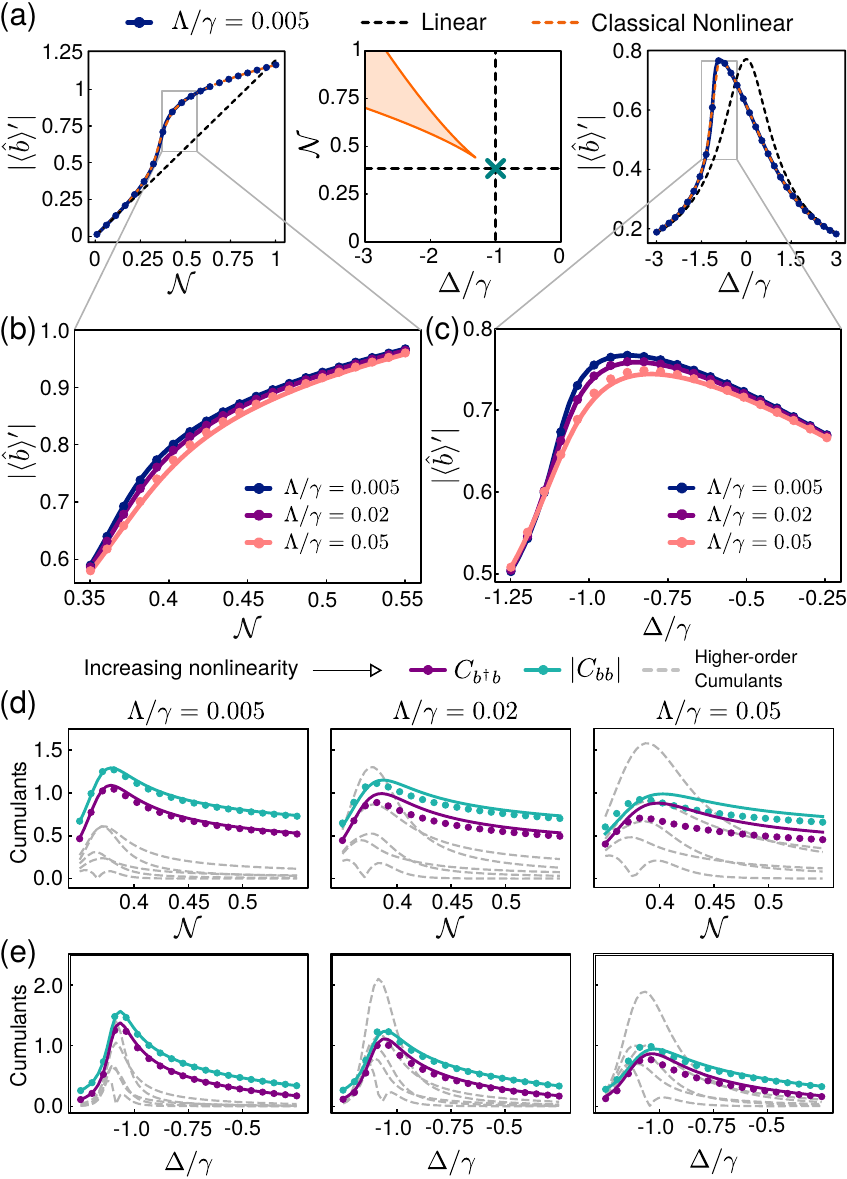}
    \caption{Classical and quantum dynamics of a single Kerr QRC node. In this plot, colored solid curves are exact complex-$P$ results, colored dots are TEOMs simulations in the long-time limit, and dashed orange curves are classical results. Other curves are specified when relevant. (a) Center panel: \textit{Classical} phase diagram of the single coherently-driven Kerr QRC node. Left panel: QRC amplitude as a function of $\CNL$ for fixed detuning $\Delta/\gamma = -1.0$ (across indicated vertical cross-section of the phase diagram). Right panel: QRC amplitude as a function of detuning $\Delta$ for fixed $\CNL = 0.385$ (across indicated horizontal cross-section of the phase diagram). Dashed black lines in both plots indicate the response for a linear oscillator. (b) Steady-state scaled amplitude $|\avg{\hat{b}'}|$ for nonlinearity strengths $\Lambda/\gamma \in [0.005,0.02,0.05]$, as a function of $\CNL$ for fixed $\Delta/\gamma = -1.0$ (region within the gray box in (a), left panel). (c) Same as (b) but as a function of detuning $\Delta$ for fixed $\CNL = 0.385$ (region within the gray box in (a), right panel). (d) Steady-state second-order cumulants ($C_{b^{\dagger}b}$, purple, and $|C_{bb}|$, green) and higher (third, fourth) order cumulants (dashed gray) across the cross-section in (b); nonlinearity increases from left to right. (e) Same as (d) but now for the cross-section in (c).
    }
    \label{fig:singleQRCCumulants}
\end{figure}

To explore this transition, we start with the classical limit of the QRC as determined by terms in the first line of Eq.~(\ref{eq:transfb}), for which the steady-state phase diagram can be easily computed (see Appendix~\ref{app:phaseDiagram}). This phase diagram, shown in $(\Delta,\CNL)$-space in the center panel of Fig.~\ref{fig:singleQRCCumulants}(a), is the well-known result for a coherently-driven classical Kerr oscillator~\cite{dykman_theory_1979, dykman_fluctuating_2012}: the orange region depicts the classical bistability, which emerges for sufficiently negative detuning given our choice of the sign of the Kerr nonlinearity.


This bistability is a feature specific to nonlinear systems; in its vicinity, we can expect an enhanced nonlinear response of the QRC observables to QRC inputs. This is explicitly observed in the scaled field amplitude $|\avg{\hat{b}}'|$, plotted in Fig.~\ref{fig:singleQRCCumulants}(b) as a function of $\CNL$ for fixed $\Delta/\gamma = -1.0$ (that is, across the indicated vertical dashed line of the classical phase diagram). The solid blue curve is the exact complex-$P$ result for $\Lambda/\gamma = 0.005$, while blue dots are obtained by evolving the TEOMs, Eqs.~(\ref{eq:b}),~(\ref{eq:CbdbFull}),~(\ref{eq:CbbFull}), to their steady state numerically. For small $\CNL$, the response follows that of a linear oscillator (dashed black line), but becomes nonlinear with increasing $\CNL$. Also shown in dashed orange is the response of the classical nonlinear oscillator (explicitly dropping second-order cumulants), which does not depend on the actual nonlinearity strength but only on $\CNL$. 

The influence of nonlinearity is similarly evident when plotting $|\avg{\hat{b}'}|$ as a function of $\Delta$, for fixed $\CNL=0.385$ in Fig.~\ref{fig:singleQRCCumulants}(c) (across the horizontal dashed line of the classical phase diagram). The black dashed curve indicates the standard Lorentzian response of a linear oscillator, centered around the bare frequency $\omega_b$. The response of the nonlinear QRC node is clearly that of a Lorentzian deformed towards lower frequencies due to the Kerr-induced frequency shift. 

In both results, we find excellent agreement between the exact quantum results, the TEOMs, \textit{and} classical nonlinear dynamics. This is indicative of operating parameters where not only is a truncated description using first and second-order cumulants sufficient, but the influence of second-order cumulants on first-order cumulants is negligible as well. Consequently, we expect the QRC node here to function as a classical reservoir node, still allowing for nonlinear processing of information encoded in the drive amplitude, phase, and/or frequency.

If we now keep $\CNL$ fixed but increase the nonlinearity strength $\Lambda$, the QRC can be moved to a quantum regime of operation. We plot the QRC response as a function of $\CNL$ for $\Delta/\gamma = -1.0$ in Fig.~\ref{fig:singleQRCCumulants}(b), and as a function of $\Delta$ for $\CNL = 0.385$ in Fig.~\ref{fig:singleQRCCumulants}(c), for three different nonlinearity strengths $\Lambda/\gamma \in [0.005,0.02,0.05]$. In both cases, we see that with increasing nonlinearity, the response deviates from that of the classical nonlinear oscillator (which recall is essentially equivalent to the quantum result for weak $\Lambda/\gamma = 0.005$). From Eq.~(\ref{eq:transfb}), it is clear that this deviation is due to the coupling of first-order moments to second-order cumulants; this is precisely the effect accounted for by the TEOMs, which thus agree with the full quantum result.

Also shown in Figs.~\ref{fig:singleQRCCumulants}(d) and (e) are steady-state cumulants plotted across the same cross-sections as Figs.~\ref{fig:singleQRCCumulants}(b) and (c) respectively, with panels from left to right indicating stronger nonlinearities. Second-order cumulants are plotted in green and purple (lines are exact complex-$P$ results, dots are TEOMs); they clearly display non-monotonic behaviour as a function of $\CNL$ and $\Delta$, acquiring large magnitudes for specific operating parameters in the vicinity of the classical bistability. Physically, second-order cumulants are related to the maximum and minimum (dimensionless) quadrature variances of the QRC nodes: $\frac{1}{2} + C_{b^{\dagger}b} \pm |C_{bb}|$ respectively, and their magnitudes are thus measures of amplification and squeezing of quantum fluctuations due to the underlying Kerr nonlinearity. 

Third and fourth order cumulants are shown in the same plots in dashed gray (exact complex-$P$ only).  There are two features of note. Firstly, the overall magnitude of these higher-order cumulants relative to second-order cumulants appears to increase for stronger nonlinearities. This is an indicator of the emergence of non-Gaussian features with increasing $\Lambda$, as expected, and generally leads to greater deviation between TEOMs and exact quantum results for first and second-order cumulants. Secondly, even for a fixed nonlinearity strength, the magnitude of cumulants depends on operating parameters ($\CNL$ and $\Delta$) that lead to operation in the vicinity of the classical bistability. Hence, while results obtained from TEOMs for such operating parameters can deviate from the exact quantum solution, away from these regions the agreement improves, even for strong nonlinearities.

\subsection{Conditional QRC node dynamics}
\label{subsec:condQRCdyn}
We have so far used steady-state, unconditional quantities to analyze the nonlinear processing capabilities of a single Kerr-based QRC node, and identified parameter regimes where an approach based on truncated EOMs is valid. However, to allow for processing of time-dependent information by the QRC, and to analyze the output from the QRC obtained as \textit{individual} measurement records, an approach is needed that is able to accurately capture the \textit{conditional, dynamical} evolution of the QRC node. This is described by the SME
\begin{align}
    d\rhoc = \mathcal{L}\rhoc~dt + \Smeas\rhoc
    \label{eq:singleQRCSME}
\end{align}
where $\Smeas$ is defined in Eq.~(\ref{eq:Smeas}), and our current system assumes the $K=1$ case. Expectation values with respect to the conditional quantum state under measurement $\rhoc$ are thus also conditional, which we indicate by the superscript $c$. For example, the equation of motion for $\avgc{\hat{b}}$ is now given by ${\rm tr }\{(\mathcal{L}~dt+\Smeas)\rhoc\hat{b} \}$, which takes the form (see Appendix~\ref{app:stochcumulants} for details)
\begin{align}
    &d\avgc{\hat{b}} = {\rm tr }\{(\mathcal{L}\rhoc\hat{b} \}~dt + \nonumber \\
    &\sqrt{\frac{\gamma}{2}} \left(\Cc{b^{\dagger}b}+\Cc{bb} \right)dW^X(t) + i\sqrt{\frac{\gamma}{2}}\left(\Cc{b^{\dagger}b}-\Cc{bb} \right)dW^P(t). 
    \label{eq:bstoch}
\end{align}
The first line $\propto \mathcal{L}$ simply includes terms from Eq.~(\ref{eq:b}), with expectation values replaced by their conditional counterparts. Terms in the second line are due to the measurement superoperator, and render the equation of motion for $\avgc{\hat{b}}$ \textit{stochastic}, conditioned on the trajectory-specific realizations of $dW^{X,P}(t)$. Dropping cumulants higher than second order in Eq.~(\ref{eq:bstoch}), arising in the first line, yields the \textit{stochastic} truncated EOMs (STEOMs) for $\avgc{\hat{b}}$. 


We can similarly obtain truncated equations of motion for the conditional cumulants $\Cc{bb}$, $\Cc{b^{\dagger}b}$ under heterodyne measurement:
\begin{subequations}
\begin{align}
    d\Cc{b^{\dagger}b} &=  \left[{\rm tr }\{\mathcal{L}\rhoc\hat{b}^{\dagger}\hat{b} \}\! - \!\avgc{\hat{b}^{\dagger}}{\rm tr }\{\mathcal{L}\rhoc\hat{b} \}\! - \!\avgc{\hat{b}}{\rm tr }\{\mathcal{L}\rhoc\hat{b}^{\dagger} \}\right]\! dt \nonumber \\
    &~~~~-{\gamma}\left[ (\Cc{b^{\dagger}b})^2 + \Cc{bb}(\Cc{bb})^* \right]~dt, \label{eq:Cbbstoch} \\
    d\Cc{bb} &=  \left[ {\rm tr }\{\mathcal{L}\rhoc\hat{b}\hat{b} \} - 2\avgc{\hat{b}}{\rm tr }\{\mathcal{L}\rhoc\hat{b} \}\right]dt \nonumber \\
    &~~~~-2{\gamma}\Cc{bb}\Cc{b^{\dagger}b}~dt. \label{eq:Cbdbstoch}
\end{align}
\end{subequations}
Again, terms $\propto \mathcal{L}$ are as found in Eqs.~(\ref{eq:CbdbFull}),~(\ref{eq:CbbFull}) post-truncation, with expectation values replaced by their conditional counterparts. The second line of each equation describes the evolution due to measurement, which at first glance appears deterministic: note that Wiener increments $dW^{X,P}(t)$ make no appearance. For \textit{linear} quantum systems under continuous weak measurement, this is in fact the case: second-order cumulants form a closed set described by the above equations, and no stochastic terms arise~\cite{breslin_conditional_1997, doherty_feedback_1999, cernotik_adiabatic_2015, zhang_prediction_2017}. However, for \textit{nonlinear} quantum systems of interest here, second-order cumulants can couple to the \textit{stochastic} first-order moments (here, via terms $\propto \mathcal{L}$), rendering the conditional evolution of second-order cumulants generally stochastic as well.



\begin{figure}
    \centering
    \includegraphics[scale=1.0]{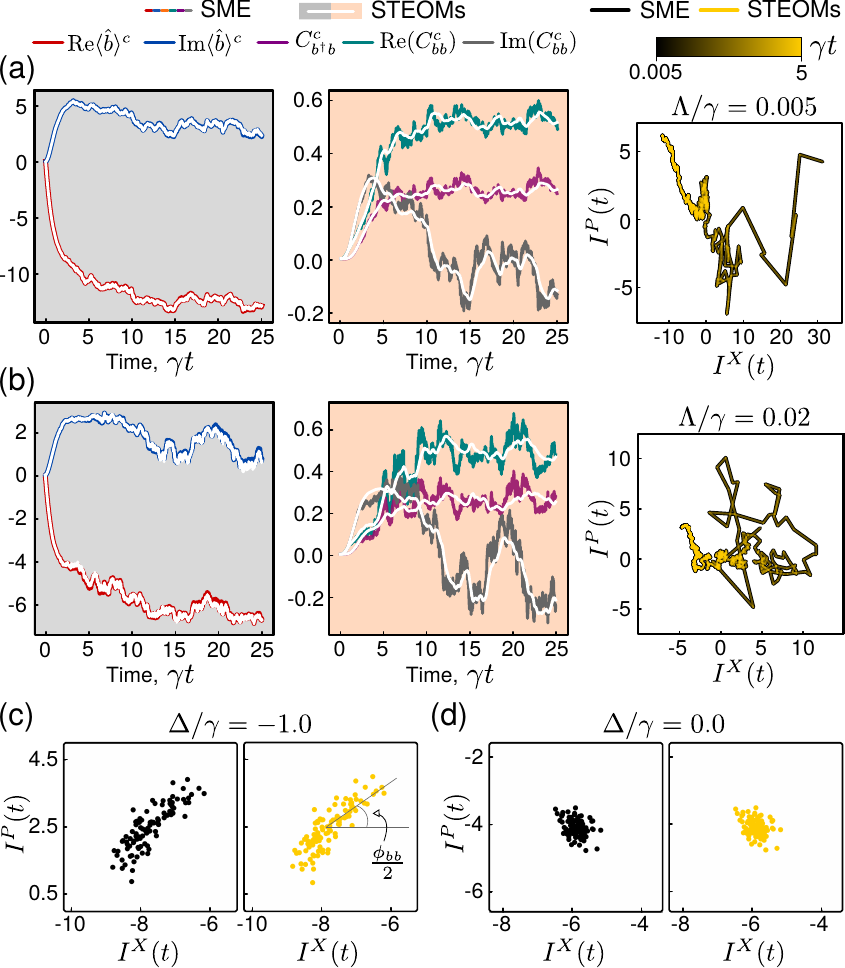}
    \caption{Benchmarking STEOMs against full SME simulations. Simulated parameters are as marked by the teal `X' in the phase diagram of Fig.~\ref{fig:singleQRCCumulants}(a): $\CNL = 0.385$, $\Delta/\gamma =-1.0$, for (a) $\Lambda/\gamma = 0.005$ and (b) $\Lambda/\gamma = 0.02$. Left panel shows real and imaginary parts of the conditional expectation value $\avg{\hat{b}}$, and center panel shows conditional second-order cumulants $C_{bb}, C_{b^{\dagger}b}$; colored (white) curves are SME (STEOMs) simulations. Right panel shows measured quadratures $I^X(t)$, $I^P(t)$: the SME result is plotted in solid black, while the STEOMs result is plotted in yellow, with a lighter shade indicating earlier times. (c) Measured quadratures $\{I^X(t),I^P(t)\}$ at $\gamma t= 20.0$ obtained using SME (left panel) and STEOMs (right panel), for $\CNL = 0.385$, $\Lambda/\gamma = 0.02$, and $\Delta/\gamma = -1.0$. (d) Same as (c) but for $\Delta/\gamma = 0.0$.}
    \label{fig:singleQRCSim}
\end{figure}


Eqs.~(\ref{eq:bstoch}),~(\ref{eq:Cbbstoch}),~(\ref{eq:Cbdbstoch}) define the STEOMs for the single Kerr QRC node. To assess their validity, we compare their simulation results against integration of the SME, Eq.~(\ref{eq:singleQRCSME}). For concreteness, we consider the operating point marked by the teal `X' in the phase diagram of Fig.~\ref{fig:singleQRCCumulants}(a): $\CNL = 0.385$, $\Delta/\gamma =-1.0$. Note that the same noise realizations $dW^{X,P}(t)$ are used for both simulations to ensure the compared trajectories are conditioned on the same measurement record. Results are included in Fig.~\ref{fig:singleQRCSim}(a),~(b) for $\Lambda/\gamma \in [0.005,0.02]$ respectively, showing first-order moments (left panel) and second-order cumulants (center panel), with STEOMs results in white, and SME results obtained using QuTiP~\cite{johansson_qutip_2013} in color. Both methods agree very well, especially for first-order cumulants. The increasing fluctuations in the second-order cumulants are a direct signature of the nonlinearity of the quantum system under study.


However, in a real experiment, such information about the QRC state must be extracted from measurement records. For the case of heterodyne measurement, the obtained records $J^{X,P}(t)$ are defined in Eqs.~(\ref{eq:JX}),~(\ref{eq:JP}). These records are typically processed to reduce noise, most commonly via temporal filtering; the processed records define \textit{measured quadratures}
\begin{align}
    I^{X,P}(t) &= \frac{1}{t}\int_0^t d\tau~J^{X,P}(\tau)
    \label{eq:IXIP} 
\end{align}
which here describe processing of single-shot readout records. Measured quadratures obtained using a single set of measurement records are shown in the \textit{measured phase space} given by $(I^X(t),I^P(t))$ in the right panel of Fig.~\ref{fig:singleQRCSim}(a),~(b) $\Lambda/\gamma \in [0.005,0.02]$ as before; SME results are shown in solid black, with the STEOMs results plotted on top, using a colorscale going from black to yellow indicating increasing time. Both show excellent agreement; this is unsurprising since $J^{X,P}(t)$ yield information about the underlying conditional dynamics of $\avgc{\hat{b}}$, which also agree very well between the two methods.

Note that the measured quadratures are themselves stochastic quantities, whose statistics are correlated with the underlying measured system state. This can be seen via the measured quadrature distributions in Fig.~\ref{fig:singleQRCSim}(c), obtained by simulating several measurement records (here $100$ in total) and plotting $\{I^X(t),I^P(t)\}$ at $\gamma t = 20.0$, for $\Delta/\gamma = -1.0$ as before. At this operating point near the classical bistable region, the Kerr nonlinearity amplifies the magnitude of cumulants $C_{bb} \equiv |C_{bb}|e^{i\phi_{bb}}$~[see Fig.~\ref{fig:singleQRCCumulants}~(d)] leading to squeezing of the internal QRC field along the axis determined by $\phi_{bb}/2$ (and amplification along the orthogonal quadrature). This internal QRC squeezing manifests as squeezing of the measured quadrature distribution; the squeezing axis is unchanged since the temporal filter defining measured quadratures via Eq.~(\ref{eq:IXIP}) is quadrature-agnostic, and thus only reduces the overall noise power while preserving its relative strength amongst measured quadratures. On the other hand, for $\Delta/\gamma = 0.0$, further away from the bistable region, $C_{bb}$ is much smaller in magnitude and the measured quadrature distribution in Fig.~\ref{fig:singleQRCSim}(d) exhibits no squeezing. Excellent agreement between SME (left) and STEOMs (right) is observed for both cases.



\subsection{Validity and scalability of (S)TEOMs}

The key observations from our analysis of Kerr QRC operating regimes and the generation of higher-order cumulants (summarized in Fig.~\ref{fig:singleQRCCumulants}) are twofold. Primarily, the magnitude of nonzero higher-order cumulants increases with the strength of the Kerr nonlinearity relative to the QRC node damping rate, $\Lambda/\widetilde{\gamma}$, which coincides with the QRC under coherent driving transitioning from classical to quantum regimes. However, operating regimes in the vicinity of the classical bistability also lead to an enhancement in higher-order cumulants, regardless of nonlinearity strength. 

To be confident of the validity of the truncated cumulants approach, we therefore analyze QRCs with nonlinearity strengths $\Lambda/\widetilde{\gamma} \leq 0.02$, and operate \textit{near} but not within the classically-bistable region. This truncated cumulants approach provides a highly efficient mathematical description when applied to multimode quantum systems. For a measurement chain comprising an $N$-mode quantum system coupled to a $K$-node QRC, the composite system of $N_{\rm T}=N+K$ total quantum modes is described by $2N_{\rm T}^2 + 3N_{\rm T}$ unknowns using (S)TEOMs, scaling quadratically with $N_{\rm T}$ instead of the exponential growth in required Hilbert space size for full (S)ME simulations. This description also places no constraints on modal occupation numbers, enabling our exploration of a well-defined classical limit of the coherently-driven Kerr QRC. Our (S)TEOMs approach is built to be a scalable theoretical framework via an efficient computer-algebra implementation, which allows the calculation of equations of motion of retained cumulants up to order $n_{\rm trunc}$ (here, second-order) for nonlinear quantum systems comprising arbitrary numbers of bosonic modes under continuous measurement. Supplementary benchmarking simulations for systems with more than one mode are included in Appendix~\ref{app:verify}. 

In the remainder of this paper, we employ (S)TEOMs to simulate dynamics of measurement chains comprising measured quantum systems coupled to multimode nonlinear QRCs. We find that such QRCs are able to successfully perform the quantum state classification tasks we have set out as objectives in the introduction, while operating within the aforementioned constraints.



\section{Cavity measurement: classifying qubit pointer states}
\label{sec:cav}

We are now in a position to apply our quantum reservoir computing framework to the task of quantum state classification, beginning with Task I as identified in Fig.~\ref{fig:schematic}(c): classifying Gaussian states with mean values that differ in amplitude or phase. A useful example of this task is the classification of pointer states generated during cavity-mediated dispersive-qubit measurements~\cite{boissonneault_dispersive_2009, filipp_two-qubit_2009}, when qubit evolution during measurement is negligible. Classification for nontrivial qubit evolution during measurement has been considered using classical reservoirs in Ref.~\cite{angelatos_reservoir_2021}. We begin with a description of all elements of the measurement chain included in Eq.~(\ref{eq:sme}) for this task, depicted schematically in Fig.~\ref{fig:classifyCavStates}(a).





\subsection{Hardware measurement chain}


\subsubsection{Single cavity mode and nonreciprocal coupling to QRC}

Our measured quantum system is described by the Liouvillian superoperator $\mathcal{L}_{\rm S}$ defined as
\begin{align}
    \mathcal{L}_{\rm S}\rhoc = -i[-\Delta^{(\sigma)}_{a} \hat{a}_1^{\dagger}\hat{a}_1 + \eta(\hat{a}_1+\hat{a}_1^{\dagger}),\rhoc] + \kappa\mathcal{D}[\hat{a}_1]\rhoc 
\end{align}
which defines a single cavity mode $\hat{a}_1$, with frequency $\omega_{a_1}$, damping rate $\kappa$, and coherently driven at frequency $\omega_d$ with drive strength $\eta$. The superoperator is written in the frame rotating at the drive frequency, rendering the drive term time-independent and introducing the cavity mode detunings
\begin{align}
    \Delta^{(\sigma)}_{a} = \omega_d - (\omega_{a_1} + \delta^{(\sigma)}).
\end{align}
Here frequency shifts $\delta^{(\sigma)}$ indexed by $\sigma$ model the case of dispersive qubit readout: for the measurement of two qubits $\hat{\sigma}_{1,2}^z$ dispersively coupled to the cavity with strengths $\chi_{1,2}$ respectively, the dispersive shifts corresponding to the four joint qubit $\hat{z}$-basis eigenstates are given by $\delta^{(\sigma)} \in \{\chi_1-\chi_2,\chi_1+\chi_2,-\chi_1-\chi_2,-\chi_1+\chi_2\}$. Under coherent driving, the cavity state evolves to a distinct amplitude and phase for each dispersive shift, serving as a pointer to the corresponding qubit state. The computational task set for the QRC is then to distinguish these pointer states. For concreteness, we choose the drive to be resonant with the bare measured cavity frequency, $\omega_d = \omega_a$, and choose cavity detunings that map to effective dispersive shifts $\chi_1 = 2\kappa, \chi_2=0.5\kappa$.


The description of this quantum system alone is insufficient to define its conditional dynamics when embedded within a measurement chain, which includes the measured QRC and a coupling element. The specific form of the coupling is important as well: ideally, we require the information from the measured quantum system to be seen by the QRC, but for the QRC to not strongly couple to, and thus renormalize the properties of, the measured quantum system.


These requirements can be met by ensuring that $\Lc$ describes a non-reciprocal coupling between the measured quantum system and the QRC. Here we couple the single cavity mode $\hat{a}_1$ to a single node $\hat{b}_1$ of the QRC via a directional amplifier interaction, realized by balancing a coherent QND interaction with its appropriately chosen dissipative counterpart~\cite{metelmann_nonreciprocal_2015}. This is defined by the specific coupling superoperator $\mathcal{L}_c$ defined as
\begin{align}
    \Lc\rhou = -i[-\varg_c \hat{P}_1 \hat{X}_{a} ,\hat{\rho}] + \Gamma_c\mathcal{D}[\hat{X}_a + i \hat{P}_1]\hat{\rho}
    \label{eq:couplingDirAmp}
\end{align}
where $\hat{X}_a = \frac{1}{\sqrt{2}}(\hat{a}_1+\hat{a}_1^{\dagger})$, $\hat{P}_a = -\frac{i}{\sqrt{2}}(\hat{a}_1-\hat{a}_1^{\dagger})$ are the canonically conjugate quadratures of the measured cavity mode, and $\hat{X}_1,\hat{P}_1$ are analogously-defined quadratures of the QRC $\hat{b}_1$ mode. The form of the interaction, realizable in cQED using multiple parametric pumps~\cite{sliwa_reconfigurable_2015, chien_multiparametric_2020}, ensures that the quadratures $\hat{P}_a$, $\hat{X}_1$ undergo coupled evolution, as defined by the equations of motion
\begin{subequations}
\begin{align}
    \frac{d}{dt}\avgc{\hat{P}_a}\! &=\! {\rm tr}\{[\Lsys+\Smeas]\rhoc\hat{P}_a\}  - \left[\varg_c-\Gamma_c \right]\avgc{\hat{P}_1},  \\
    \frac{d}{dt}\avgc{\hat{X}_1}\! &=\! {\rm tr}\{[\Lqrc+\Smeas]\rhoc\hat{X}_1\} - \left[\varg_c+\Gamma_c \right]\avgc{\hat{X}_a}\!\! .
    \label{eq:dX1dt}
\end{align}
\end{subequations}
The first term in each equation describes the uncoupled dynamics of the corresponding system, while the second term includes the conditional evolution due to the measurement superoperator. The third term is the coupling contribution: under the specific choice of coupling strength and phase
\begin{align}
    \varg_c = \Gamma_c
    \label{eq:cavnrcond}
\end{align}
it is clear from Eq.~(\ref{eq:dX1dt}) that $\Lc$ realizes a directional amplifier: the cavity $\hat{X}_a$ quadrature is amplified by a factor $\propto \Gamma_c$ and impinges on the QRC, while the cavity itself is not driven by the QRC in the opposite direction.







Even if the cavity-QRC coupling is nonreciprocal, measurement of the QRC nodes downstream leads to measurement-conditioned evolution of the cavity $\propto \Smeas$~\cite{hatridge_quantum_2013}. This is evident from the conditional equation of motion for the cavity $\avgc{\hat{X}_a}$ quadrature (which drives the QRC),
\begin{align}
    \frac{d}{dt}&\avgc{\hat{X}_a} = {\rm tr}\{\Lsys\rhoc\hat{X}_a\}  \nonumber \\
    +&\sum_k \frac{\sqrt{\gamma_k}}{2}\left[ \Cc{ab_k}+(\Cc{ab_k})^*+\Cc{a^{\dagger}b_k}+(\Cc{a^{\dagger}b_k})^*\right] dW^X_k(t) \nonumber \\ -i&\sum_k \frac{\sqrt{\gamma_k}}{2}\left[ \Cc{ab_k}-(\Cc{ab_k})^*+\Cc{a^{\dagger}b_k}-(\Cc{a^{\dagger}b_k})^*\right] dW^P_k(t)
    \label{eq:cohXa}
\end{align}
where the noted stochastic terms are visible in the second and third lines. These terms depend on joint cumulants of the cavity-QRC quantum state and are thus influenced by the QRC evolution; in particular, they can be nonzero even if Eq.~(\ref{eq:cavnrcond}) is satisfied. This principle of measurement-conditioned evolution has been analyzed in cQED experiments, for example in quantum trajectories of dispersively measured qubits nonreciprocally coupled to a quantum-limited phase-sensitive amplifier~\cite{murch_observing_2013}, and is accurately captured by the joint description of the measurement chain we employ here. In the present case, typical conditional cavity dynamics are illustrated for $\avgc{\hat{X}_a}$ in Fig.~\ref{fig:classifyCavStates}(b), color-coded for the four states defined by $\Delta_{a}^{(\sigma)}$.


\subsubsection{QRC}

\begin{figure*}[t]
    \centering
    \includegraphics[scale=1.0]{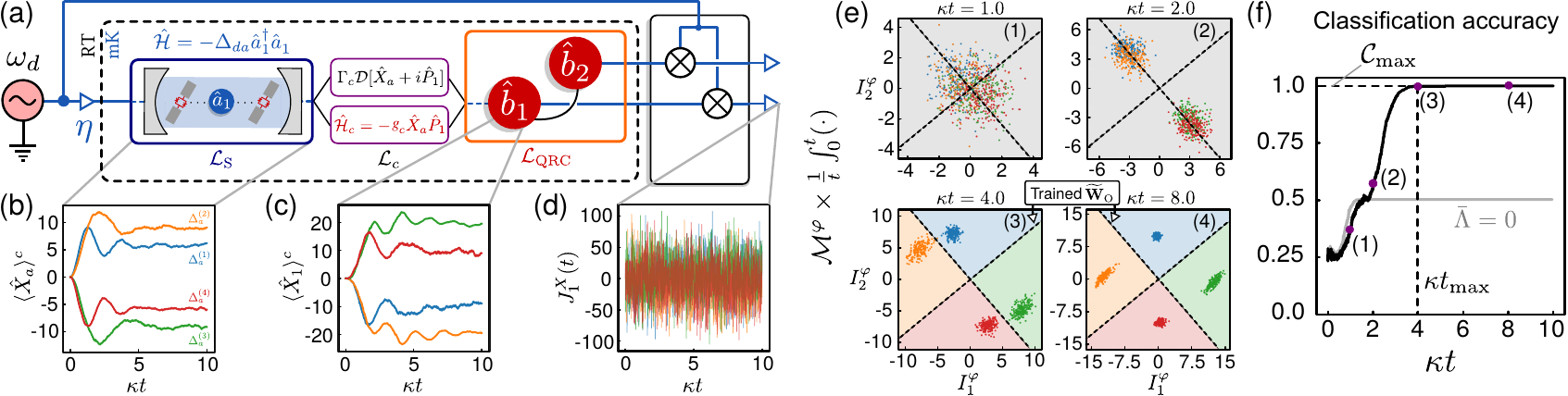}
    \caption{(a) Measurement chain using a 2-node QRC for the classification of states of a single cavity mode. The cavity mode is coupled to the QRC using a quantum-limited directional amplifier. Each node of the QRC is continuously monitored using separate output channels via heterodyne measurement. Conditional dynamics of (b) the cavity quadrature $\avgc{\hat{X}_a}$, (c) the QRC node quadrature $\avgc{\hat{X}_1}$ it drives, and (d) measurement records $J_1^X(t)$, are shown for one initialization of the measurement chain per state, with colors blue, orange, green, and red corresponding to cavity mode detunings $\Delta_{a}^{(\sigma)} \in \{2.5,1.5,-1.5,-2.5\}\kappa$ respectively. (e) Dynamics of QRC outputs projected onto the measured subspace $\{I_1^{\varphi},I_2^{\varphi}\}$ as a function of time, with panels (1) through (4) corresponding to $\kappa t \in [1.0,2.0,4.0,8.0]$ respectively. $100$ trajectories are shown for each state $\sigma$.  (f) Classification accuracy as a function of time, computed for a test set of size $Q_{\rm test}=200$ measured outputs per state. Black and gray curves correspond to a typical nonlinear QRC and a linear QRC respectively.
    }
    \label{fig:classifyCavStates}
\end{figure*}

Our choice of coupling leads to the cavity $\hat{X}_a$ quadrature driving the QRC via its $\hat{X}_1$ quadrature, whose resulting conditional evolution is given by:
\begin{align}
    \frac{d}{dt}\avgc{\hat{X}_1} &= {\rm tr}\{\Lqrc\rhoc\hat{X}_1\} - 2\Gamma_c \avgc{\hat{X}_a} \nonumber \\
    +&\sum_k \frac{\sqrt{\gamma_k}}{2}\left[ \Cc{b_1b_k}+(\Cc{b_1b_k})^*+2\Cc{b_1^{\dagger}b_k}\right] dW^X_k(t) \nonumber \\ 
    -i&\sum_k \frac{\sqrt{\gamma_k}}{2}\left[ \Cc{b_1b_k}-(\Cc{b_1b_k})^*\right] dW^P_k(t)
    \label{eq:cohX1}
\end{align}
The QRC nodes thus evolve under the cavity signal, but also the QRC's intrinsic network dynamics governed by $\Lqrc$, and conditional evolution that depends on cumulants of the joint QRC state. Typical conditional dynamics of the $\avgc{\hat{X}_1}$ quadrature of QRC node $\hat{b}_1$ for each generated cavity state is shown in Fig.~\ref{fig:classifyCavStates}(c). For this illustration, we have chosen two-node QRC parameters: $\gamma_1 = \gamma_2 = \kappa$, $\Lambda_1 = \Lambda_2 = 0.005\gamma_1$, $\Delta_1 = \Delta_2 = 0.0$, and $\varg_{12} = \gamma_1$.

Nonlinear processing of the cavity signal is encoded in this time evolution of the multimode QRC state. However, processed information about this state must be extracted from noisy measurement records $J_k^{X,P}(t)$ defined in Eqs.~(\ref{eq:JX}),~(\ref{eq:JP}) obtained from the measurement chain, which we discuss next.



\subsection{Training and classification using finitely-sampled measurement distributions}
\label{subsec:cavtraining}


 
Measurement records $\{J_k^{X,P}(t)\}$ contain experimentally-accessible information about the multimode QRC state  for a given initialization of the measurement chain. We consider the case of \textit{single-shot reservoir readout}, where the output $\mathbf{x}(t)$ of the QRC is constructed as the vector of all measured quadratures obtained via linear boxcar filtering of single-shot heterodyne measurement records~\cite{walter_single_2017},
\begin{align}
    \mathbf{x}(t) \equiv
    \begin{pmatrix}
     I_1^{X}(t) \\
     I_1^{P}(t) \\
     I_2^{X}(t) \\
     I_2^{P}(t) 
    \end{pmatrix},~I_k^{X,P}(t) = \frac{1}{t}\int_0^t d\tau~J_k^{X,P}(\tau).
    \label{eq:IkXP}
\end{align}
Measured quadratures are themselves stochastic, so that the QRC output provides a distinct stochastic trajectory for repeated runs of the measurement chain; QRC output corresponding to $\Delta_a^{(\sigma)}$ is thus labelled $\mathbf{x}^{(\sigma)}_{(q)}$, indexed by both $\sigma$ and a trajectory label $q$.


Having defined each element of the measurement chain and the measured QRC output, we can finally describe its operation for reservoir computing. We employ an output layer configured to yield a predicted class label $\sigma^p$ using the measured QRC output, $\sigma^p = f_N\{ \mathbf{W}_{\rm O}\mathbf{x}^{(\sigma)}_{(q)} + \mathbf{b} \}$, where $\mathbf{W}_{\rm O}$ defines a matrix of time-independent trainable weights, $\mathbf{b}$ a vector of trainable biases, and where $f_N\{\cdot\}$ is a fixed (that is, untrained) normalizing function. Perfect classification yields $\sigma^p = \sigma$ for all $\sigma = 1,\ldots,C$, irrespective of $q$. 



Formally, training is performed by minimizing a cost function, here the squared error over a training set whose output is known, and importantly is a convex minimization problem that is guaranteed to converge; full details of this training are included in Appendix~\ref{app:training}. Here we instead present a visual representation of training and classification, shedding light into how the QRC dynamics enable classification. Classification using linear weights $\mathbf{W}_{\rm O}$ is equivalent to constructing \textit{linear} decision boundaries to separate reservoir outputs $\mathbf{x}^{(\sigma)}_{(q)}$ for different $\sigma$ in the four-dimensional measured phase space spanned by $\mathbf{x} = (I_1^X,I_1^P,I_2^X,I_2^P)$, as shown in Appendix~\ref{app:training}. While this four-dimensional space is difficult to visualize, we find that for the present task two optimally chosen quadratures prove to be sufficient for successful classification. We thus rewrite the matrix of trainable weights as $\mathbf{W}_{\rm O}\mathbf{x} = \widetilde{\mathbf{W}}_{\rm O}\mathcal{M}^{\varphi}\mathbf{x}$, where $\mathcal{M}^{\varphi}$ acts to project the measured QRC output onto a two-dimensional measured subspace $\{I_1^{\varphi},I_2^{\varphi}\}$,
\begin{align}
\begin{pmatrix}
I_1^{\varphi} \\
I_2^{\varphi} 
\end{pmatrix} \equiv
\mathcal{M}^{\varphi}
\mathbf{x}
\!
= 
\!
\begin{pmatrix}
\cos \varphi_1 & \sin \varphi_1 & 0 & 0 \\
0 & 0 & \cos \varphi_2 & \sin \varphi_2  
\end{pmatrix}
\!\!\!
\begin{pmatrix}
I_1^{X} \\
I_1^{P} \\
I_2^{X} \\
I_2^{P}  
\end{pmatrix} .
\label{eq:Mphi}
\end{align}
We note that all QRC node quadratures are still measured in each measurement run;. The projection phases are simply applied as part of the output layer, and are learned during training.

In Fig.~\ref{fig:classifyCavStates}(d), we project QRC outputs $\{\mathbf{x}^{(\sigma)}_{(q)}\}$ onto the measured subspace to visualize their evolution. Panels (1) through (4) show these outputs as a function of time, $\kappa t \in [1.0,2.0,4.0,8.0]$ respectively. We see clearly that for short times (1), all four states are indistinguishable and the measured quadrature distributions overlap. This is not surprising, as the measurement chain is initialized to vacuum for all states. As the cavity is populated, distributions for states with distinct phases first separate (2) while the QRC response is still effectively linear. By (3), the cavity signal corresponding to higher amplitude states (orange, green) has increased sufficiently to drive the QRC into a nonlinear regime. This is clear from the rotation of these distributions relative to those corresponding to the low amplitude states, as well as their visible distortion; the latter is a signature of amplified and squeezed QRC node fluctuations due to the Kerr nonlinearity, also seen in Fig.~\ref{fig:singleQRCCumulants}(c). The effect of the nonlinearity clearly separates all four distributions into (correspondingly shaded) distinct phase space regions.

It is then clear that linear decision boundaries (dashed black) can be drawn in the measured subspace to separate distributions for all $C$ states. Training using a set of QRC output trajectories $\{\mathbf{x}^{(\sigma)}_{(q)}\}$ for $\QTR=100$ trajectories per state ($q=1,\ldots,\QTR$) simply learns the parameters $\{\widetilde{\mathbf{W}}_{\rm O}, \mathbf{\varphi} \}$ uniquely specifying the optimal decision boundaries. Note that these decision boundaries, and therefore the training protocol to learn them, must take into account both the mean \textit{and} variance of the finitely-sampled QRC output distributions. These quantities are affected by factors such as measurement time and filtering; therefore, training on measured output distributions is tailored to the finite experimental resources available for output processing. This is in sharp contrast to training on unconditional expectation values, which in principle assumes access to infinitely-many measurement records~(for further discussion, see Appendix~\ref{app:trainingTypes}). Finally, we emphasize that the linear output layer employed here serves two important, yet simple purposes: (i) filtering out measurement noise obscuring the underlying QRC dynamics, and (ii) drawing linear decision boundaries that optimally separate regions of the measured phase space into classes. The nontrivial processing step that enables classification - the evolution of measured outputs into linearly-separable regions of the measured subspace - is performed by the \textit{nonlinear} QRC. 

Once the QRC is trained, we can quantify its performance by predicting state labels $\sigma^p$ for a distinct test set of reservoir outputs $\{\mathbf{x}^{(\sigma)}_{(q)}\}$ for $q=1,\ldots,Q_{\rm Test}=200$. Using the trained output layer, we determine $\sigma^p$ for each $\mathbf{x}^{(\sigma)}_{(q)}$ and calculate the classification accuracy as a function of time, defined as the fraction of $C\times Q_{\rm Test}$ total trajectories that were correctly classified. The classification accuracy is plotted in Fig.~\ref{fig:classifyCavStates}(f), and is characterized by two metrics: the maximum classification accuracy $\cmax$ achieved within $\kappa t = 10.0$, and the time $\tmax$ taken to reach this maximum. We note that the specific QRC shown here achieves perfect classification ($\cmax = 1$) before the measurement chain dynamics have settled into steady-state [see Figs.~\ref{fig:classifyCavStates}(b),~(c)], indicating the ability of the QRC to process the information encoded in time-evolving quantum signals. 

Also shown in gray is the classification accuracy for a linear QRC, $\bar{\Lambda} = 0$, which saturates to $0.5$. The typical dynamics in the measured phase space are analogous to that shown in panel (2) of Fig.~\ref{fig:classifyCavStates}(e); the linear QRC is able to separate the pairs of states with different phases (blue-orange from red-green) but not distinguish states within this subset. Having used pecisely the same form of output layer for both linear and nonlinear QRCs, we are able to unequivocally identify the QRC nonlinearity as being critical for the present classification task.

It is important to note, however, that a completely arbitrary nonlinear QRC will not succeed at this classification task; internal QRC parameters introduced in Eq.~(\ref{eq:Lqrc}) must satisfy some task-dependent constraints, which we explore next.

\subsection{Classification performance as a function of hyperparameters}

\textit{Hyperparameters.}$-$ For large classical reservoirs, it is conventional - and practical - to describe reservoir parameters in terms of average hyperparameters instead of individual microscopic values. Here we adopt a similar convention despite the small size of the two-node QRC, parameterizing it as
\begin{align}
    \Lambda_k &= \bar{\Lambda} + [-\epsilon,\epsilon]\cdot\bar{\Lambda},~\gamma_k = \bar{\gamma} = \kappa~\forall~k, \nonumber \\
    \Delta_k &= \bar{\Delta} + [-\epsilon,\epsilon]\cdot\bar{\gamma},~\varg_{12} = \bar{\varg}
    \label{eq:randomParams}
\end{align}
where $[x,y]$ denotes random samples from a uniform distribution with probability density between $x$ and $y$; we set $\epsilon = 0.1$ to observe the effect of random variations in QRC nonlinearities and frequencies that may occur in experiment. For convenience, the QRC damping rates are all set equal to the measured quantum system decay rate $\kappa$. 


We analyze QRC classification performance as a function of the above hyperparameters in terms of the maximum classification accuracy $\cmax$, and the time taken to reach this maximum, $\tmax$. In doing so, we ensure that all other aspects of the measurement chain, training of the output layer, and testing remain unchanged from Sec.~\ref{subsec:cavtraining}, so that the particular role of QRC hyperparameters can be highlighted. Our key findings are recounted in a summary at the end of this section.

\subsubsection{Dependence on nonlinearity $\bar{\Lambda}$}

\begin{figure}[t]
    \centering
    \includegraphics[scale=1.0]{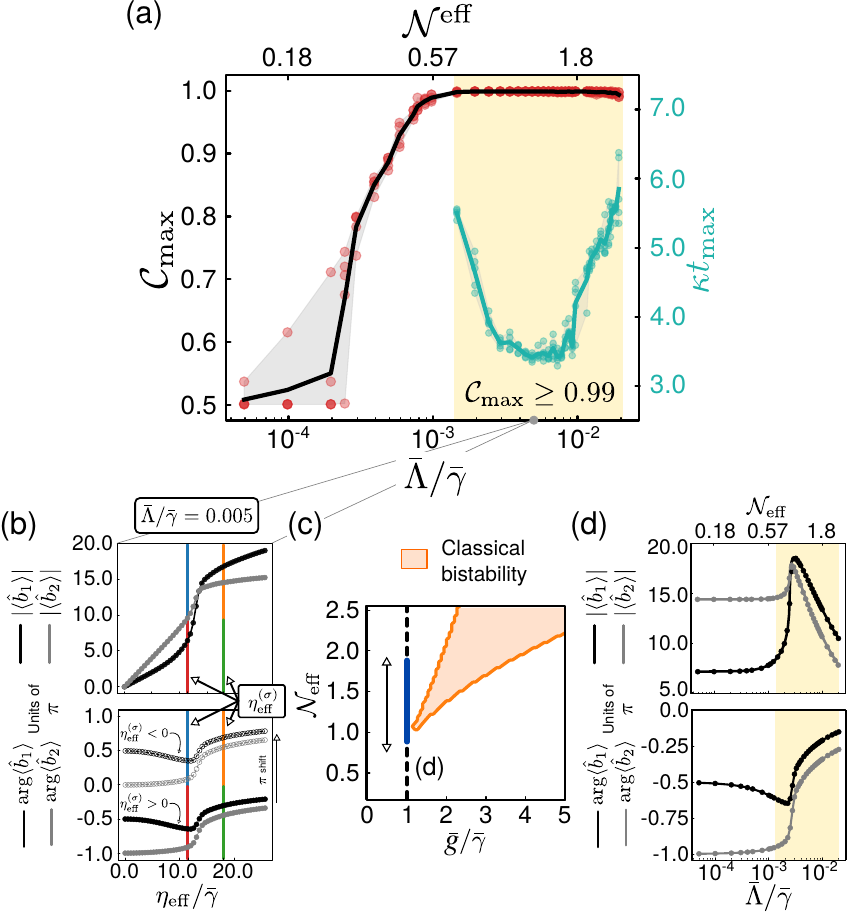}
    \caption{(a) $\cmax$ (left-hand axis) vs. nonlinearity hyperparameter $\bar{\Lambda}$, for fixed coupling $\bar{\varg}/\bar{\gamma}=1.0$ 
    and detuning $\bar{\Delta}/\bar{\gamma} = 0.0$. For $\cmax \geq 0.99$ (yellow shaded region), $\kappa\tmax$ is plotted along the right-hand axis (green). Dots indicate different random QRCs, solid curves are averages over these QRCs, and shaded regions mark the range between worst and best performing QRCs (see text for more details). (b) Steady-state response of $|\avg{\hat{b}_{1,2}}|$, ${\rm arg}\avg{\hat{b}_{1,2}}$  as a function of constant drive strength $\eta_{\rm eff}$. (c) Classical phase diagram of the two-node QRC in $(\bar{\varg},\CNLE)$ space, for fixed detuning $\bar{\Delta}/\bar{\gamma} = 0.0$. (d) Steady-state response of $|\avg{\hat{b}_{1,2}}|$ as a function of increasing nonlinearity $\bar{\Lambda}$ across the vertical dashed line ($\bar{\varg}/\bar{\gamma} = 1.0$) in (c). }
    \label{fig:cohClassifyLambda}
\end{figure}

We begin by simulating QRCs with different values of the nonlinearity hyperparameter $\bar{\Lambda}$, while fixing $\bar{\varg}/\gamma = 1.0$, $\bar{\Delta}/\bar{\gamma} = 0.0$, and calculating the maximum classification accuracy $\cmax$. The results are shown in Fig.~\ref{fig:cohClassifyLambda}(a). Each red dot corresponds to a different random QRC, as defined by Eq.~(\ref{eq:randomParams}). The solid black curve is the mean over these QRCs, while the gray shaded region marks the range between worst and best performing QRCs. 5 random QRCs are simulated for each set of hyperparameters, amounting to over 200 distinct QRCs shown in Fig.~\ref{fig:cohClassifyLambda}(a).

For weak nonlinearities, $\cmax$ approaches $0.5$, the limit of a linear QRC as seen earlier. With increasing nonlinearity however, the QRCs are able to perform the classification task, and $\cmax$ approaches unity. The time $\tmax$ to reach $\cmax$ can further distinguish operating hyperparameters: we plot $\kappa \tmax$ in Fig.~\ref{fig:cohClassifyLambda}(a) (right-hand axis) in green, for parameters where $\cmax \geq 0.99$ (yellow shaded region); green dots again indicate different random QRCs with the same hyperparameter values, while the solid green curve indicates the average, and the (small) green shaded region marks the range between the worst and best performing QRCs. We see that there exists a subset of optimal nonlinearity strengths for which the QRC correctly classifies all states in the shortest amount of time.

This calculation of $\cmax$ and $\tmax$ obtained from simulation of the complete measurement chain captures the full complexity of QRC evolution under time-dependent cavity signals with distinct amplitudes and phases. However, we find that a heuristic understanding of QRC performance characteristics can be developed by focusing on its \textit{steady-state} response to a \textit{time-independent} drive signal. For the pointer state classification task, we can define an effective drive strength $\eta_{\rm eff}^{(\sigma)}$ as the long-time limit of the cavity signal corresponding to $\Delta_a^{(\sigma)}$ incident on the QRC,
\begin{align}
    \eta_{\rm eff}^{(\sigma)} &= -2\Gamma_c\!\lim_{t\to\infty}\!\{ \sqrt{2}\avg{\hat{X}_a^{(\sigma)}(t)} \} = -2\Gamma_c\eta\left[\!\frac{\Delta_{a}^{(\sigma)}}{ (\Delta_{a}^{(\sigma)})^2 + \frac{\kappa^2}{4} }\! \right]\!,
    \label{eq:coheffetaJ}
\end{align}
where the second equality is obtained from the steady-state solution of the unconditional version of Eq.~(\ref{eq:cohXa}). As visible in cavity signals in Fig.~\ref{fig:classifyCavStates}(b), for $|\Delta_{a}^{(\sigma)}| \geq \kappa$, $\sigma=1,4$ (blue, red) correspond to larger $|\Delta_{a}^{(\sigma)}|$ and yield weaker effective drives than $\sigma=2,3$ (orange, green).

We now analyze the steady-state QRC response in a regime enabling perfect classification, $\bar{\Lambda}/\bar{\gamma} = 0.005$ in Fig.~\ref{fig:cohClassifyLambda}(a) (and setting $\epsilon=0$). To do so, we plot the steady-state values of $|\avg{\hat{b}_{1,2}}|$ (top panel) and ${\rm arg}\avg{\hat{b}_{1,2}}$ (bottom panel) under a constant drive strength $\eta_{\rm eff}$ in Fig.~\ref{fig:cohClassifyLambda}(b), with the particular effective drives $\eta_{\rm eff}^{(\sigma)}$ indicated by correspondingly colored lines. Clearly, the QRC amplitude and phase response to higher amplitude cavity states $\sigma = 2,3$ (orange, green) is non-linearly related to its response to lower amplitude states $\sigma = 1,4$ (blue, red). This leads to the nonlinear amplitude response and phase rotation observed in the measured quadrature distributions for higher amplitude states in Fig.~\ref{fig:classifyCavStates}(e), which enables successful classification of the pointer states.

We find that the nonlinear QRC response observed in Fig.~\ref{fig:cohClassifyLambda}(b) is typical when the two-node QRC is successful at the pointer state classification task; it is therefore helpful to locate such regimes in terms of general QRC hyperparameters. The \textit{classical} phase diagram of the two-node QRC ends up providing the pathway to identifying such regimes. Furthermore, this phase diagram depends on the effective QRC drive and nonlinearity strengths only via the effective parameter~(for details see Appendix~\ref{app:phaseDiagram})
\begin{align}
    \CNLE = \frac{\eta_{\rm eff}}{\bar{\gamma}}\sqrt{\frac{\bar{\Lambda}}{\bar{\gamma}}}.
    \label{eq:cnleCoh}
\end{align}
To construct the QRC phase diagram informed by the pointer state classification task, we choose $\eta_{\rm eff} \equiv {\rm max}_{\sigma}\{|\eta_{\rm eff}^{(\sigma)}|\}$, corresponding to higher amplitude cavity states that drive the QRC into nonlinear processing regimes.

We plot the resulting phase diagram in Fig.~\ref{fig:cohClassifyLambda}(c) as a function of $\CNLE$ and coupling $\bar{\varg}$, for fixed $\bar{\Delta}/\bar{\gamma} = 0.0$. The orange region indicates the classical bistability, analogous to that found for the single QRC node. The vertical dashed line indicates $\bar{\varg}/\bar{\gamma} = 1.0$, namely the cross-section along which Fig.~\ref{fig:cohClassifyLambda}(a) is plotted. Also shown in Fig.~\ref{fig:cohClassifyLambda}(d) is the steady-state response of the QRC node amplitudes and phases along this cross-section; the solid blue line in (c) and yellow shaded region in (d) mark $\CNLE$ values for which $\cmax \geq 0.99$. For weak nonlinearities where classification is unsuccessful, the mode amplitudes and phases are effectively constant, indicating the linear regime of operation where QRC states generated by different $\eta_{\rm eff}$ cannot be separated. For increasing $\bar{\Lambda}$, approaching the classical bistability results in a nonlinear QRC response enabling successful classification.  When the nonlinearity is further increased however, $|\avg{\hat{b}_{1,2}}|$ begins to decrease; this is a result of the increasing Kerr shift moving the node resonances further from the steady-state input signals. The reduced QRC amplitude is responsible for the increase in $\tmax$ seen in Fig.~\ref{fig:cohClassifyLambda}(a).


\subsubsection{Dependence on coupling $\bar{\varg}$ and detuning $\bar{\Delta}$}

We can similarly analyze the performance of QRCs as a function of coupling and detuning hyperparameters, for fixed nonlinearity (here, $\bar{\Lambda}/\bar{\gamma} = 0.002$, which fixes $\CNLE = 0.8$). In cQED realizations of the proposed QRCs, these hyperparameters can be rendered \textit{in-situ} tunable (via flux-modulated SQUIDs and parametric couplers respectively), thus providing experimental control over QRC performance.



\begin{figure}[t]
    \centering
    \includegraphics[scale=1.0]{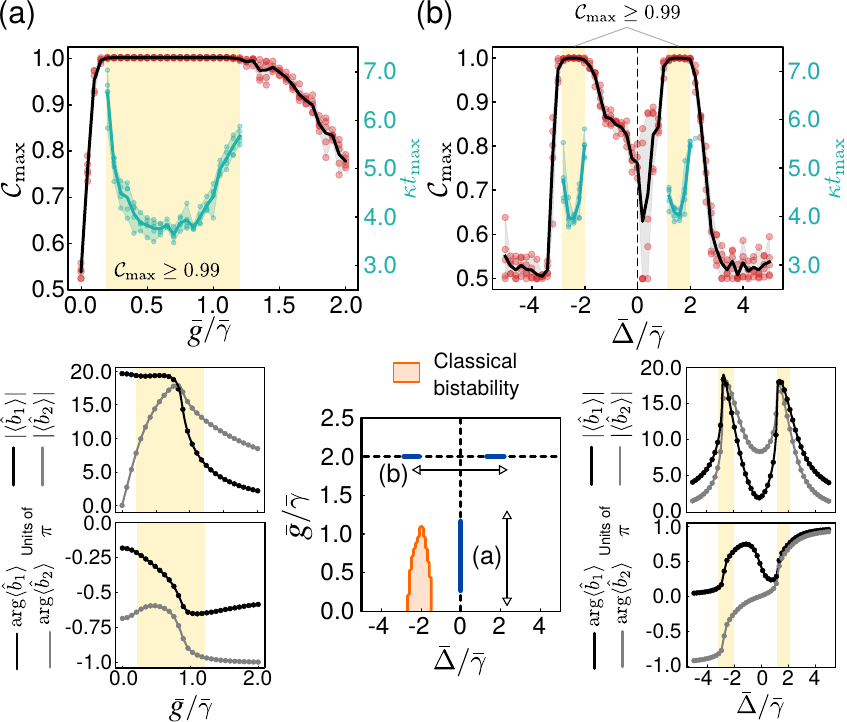}
    \caption{$\cmax$ (left-hand axis) for fixed nonlinearity $\bar{\Lambda}/\bar{\gamma}=0.002$ (and hence fixed $\CNLE=0.8$) vs. (a) coupling hyperparameter $\bar{\varg}$ for fixed detuning $\bar{\Delta}/\bar{\gamma} = 0.0$, and vs. (b) average detuning hyperparameter $\bar{\Delta}$ for fixed coupling $\bar{\varg}/\bar{\gamma} = 2.0$. For both plots, where $\cmax \geq 0.99$ (yellow shaded region), $\tmax$ is plotted along the right-hand axis (green). Dots indicate different random QRCs, solid curves are averages over these QRCs, and shaded regions mark the range between worst and best performing QRCs (see text for more details). (c) Classical phase diagram of the two-node QRC in $(\bar{\Delta},\bar{\varg})$ space for fixed $\CNLE=0.8$. Cross-sections of the steady-state response of QRC nodes are shown in (d) for fixed detuning $\bar{\Delta}/\bar{\gamma} = 0.0$ as a a function of coupling $\bar{\varg}$, and in (e) for fixed coupling $\bar{\varg}/\bar{\gamma} = 2.0$ as a a function of detuning $\bar{\Delta}$.  }
    \label{fig:cohClassifyGDelta}
\end{figure}

We begin by calculating $\cmax$ as a function of the coupling hyperparameter $\bar{\varg}$ in Fig.~\ref{fig:cohClassifyGDelta}(a), fixing the detuning  $\bar{\Delta}/\bar{\gamma} = 0.0$. This cross-section of $(\bar{\Delta},\bar{\varg})$ space is indicated in the center panel classical phase diagram with a vertical dashed line, with the corresponding QRC steady-state amplitude and phase response plotted in the left panel of Fig.~\ref{fig:cohClassifyGDelta}(c). For weak couplings, the $\hat{b}_2$ is correspondingly weakly excited, since the input couples only to $\hat{b}_1$.  Only the $\hat{b}_1$ is effectively available for processing; this is insufficient for the task at hand and the QRC does not attain $\cmax = 1$ within the measurement time.

With increasing coupling, the QRC achieves perfect classification; furthermore, an optimal range of couplings exists where $\tmax$ (green, right-hand axis) is shortest. From the left panel of Fig.~\ref{fig:cohClassifyGDelta}(c), we see that here the QRC response for both nodes is strongest. Curiously, the performance dips again for strong couplings: here the dynamics due to the linear coupling dominate over the nonlinearity, as the QRC polariton modes are further split and respond weakly to the signal emanating from the cavity, resulting in the decreased response visible in the steady state plots.


To demonstrate the advantage of \textit{in-situ} control, we now consider QRCs with sub-optimal coupling
$\bar{\varg}/\bar{\gamma} = 2.0$ (where $\cmax \simeq 0.8$, see Fig.~\ref{fig:cohClassifyGDelta}(a)). Figure~\ref{fig:cohClassifyGDelta}(b) shows the resulting $\cmax$ as a function of the detuning hyperparameter $\bar{\Delta}$, a range  indicated via the horizontal dashed line in the phase of diagram of Fig.~\ref{fig:cohClassifyGDelta}(c). The corresponding steady-state QRC amplitude and phase responses are shown in the right panel. We clearly see that $\cmax$ approaches unity once more for specific positive and negative detunings where the QRC response is enhanced.


\subsubsection{Dependence on damping $\bar{\gamma}$}

Thus far, the QRC node damping rate has appeared as the natural scaling parameter for hyperparameters in our analysis. Hence if $\bar{\gamma}$ is varied but the remaining QRC hyperparameters are suitably adjusted so that $\{\CNLE,\bar{\varg}/\bar{\gamma},\bar{\Delta}/\bar{\gamma}\}$ are unchanged, the QRC will operate in a fixed regime of the classical phase diagram; provided this regime enables nonlinear processing as previously analyzed, successful classification will be possible.

However, two aspects of QRC dynamics controlled by $\bar{\gamma}$ are not captured by its steady-state phase diagram: the response time of QRC nodes to \textit{time-dependent} input signals, as well as the magnitude of QRC response for fixed input signal strengths. To isolate the influence of $\bar{\gamma}$ on QRC performance via these specific factors, we must ensure that the QRC operates in a fixed nonlinear regime of the classical phase diagram even as $\bar{\gamma}$ is varied.

This analysis is carried out in Appendix~\ref{app:damping}, with our main results summarized here. We find that for $\bar{\gamma} \ll \kappa$, classification is still possible in nonlinear QRC regimes; however, as the QRC responds much slower than the evolution timescale of cavity signals (set by $\kappa$), $\tmax$ for classification increases. As $\bar{\gamma}$ increases the QRC response time is reduced, leading to a decrease in $\tmax$ until a minimum is attained around $\bar{\gamma} \simeq 2\kappa$. For larger $\bar{\gamma}$, while the QRC response time is still fast relative to the quantum system evolution timescale, the QRC node amplitude is suppressed such that the amplitude of extracted output signals from the measurement chain is reduced relative to the measurement noise, thereby increasing $\tmax$ once more.

\subsubsection{Summary}
The results of this section indicate that a range of hyperparameter values $\{\CNLE,\bar{\varg}/\bar{\gamma},\bar{\Delta}/\bar{\gamma}\}$ exist for which the QRC is able to perform nonlinear processing of input signals, and consequently succeed at the task of pointer state classification considered here. Importantly, regions of classical bistability - key signatures of classical nonlinearity - are strong (although not exclusive) indicators of such parameter regimes. The mechanism of classification is the nonlinear response of QRC node amplitudes with respect to the amplitude and phase of incident signals from the cavity mode.



\subsection{Dependence on input signal strength: classical vs. quantum regimes of operation}
\label{subsec:cavquantum}


Our analysis of QRC performance as a function of hyperparameters highlights the importance of operating in dynamical regimes that enable nonlinear processing of input signals. Importantly, the effective nonlinearity parameter $\CNLE$ that is central to such regimes depends not only on the QRC nonlinearity $\bar{\Lambda}$, but also on the effective drive $\eta_{\rm eff}$. The latter is ultimately related to the strength of the drive $\eta$ incident on the measurement chain, which in turn can be constrained by various physical considerations. For composite measurement chains considered here, the incident field initially drives the measured quantum system upstream of the QRC. Hence, its amplitude is dictated by the desired response of this quantum system, namely the amplitude required to generate quantum states of interest, as well as to avoid possible non-idealities that may appear under strong driving. Secondly, to decrease the energy of physical computation, one must naturally consider a reduction in the input power to the measurement chain containing the QRC.

With this view, we consider separate instances of pointer state classification tasks for the same values of detunings $\Delta_{a}^{(\sigma)}$ defining the pointer states as considered earlier, but for different values of drive input to the measurement chain, $\eta/\bar{\gamma} \in \{26.0,15.0,9.0\}$. These input drive values translate to effective QRC drives $\eta_{\rm eff}/\bar{\gamma} \in \{31.2,18.0,10.8\}$. We then determine classification metrics $\cmax$ and $\kappa\tmax$ as a function of QRC nonlinearity strength $\bar{\Lambda}$ as before, holding the remaining QRC hyperparameters fixed, $\bar{\varg}/\bar{\gamma} = 1.0$, $\bar{\Delta}/\bar{\gamma} = 0.0$. 




\begin{figure}[t]
    \centering
    \includegraphics[scale=1.0]{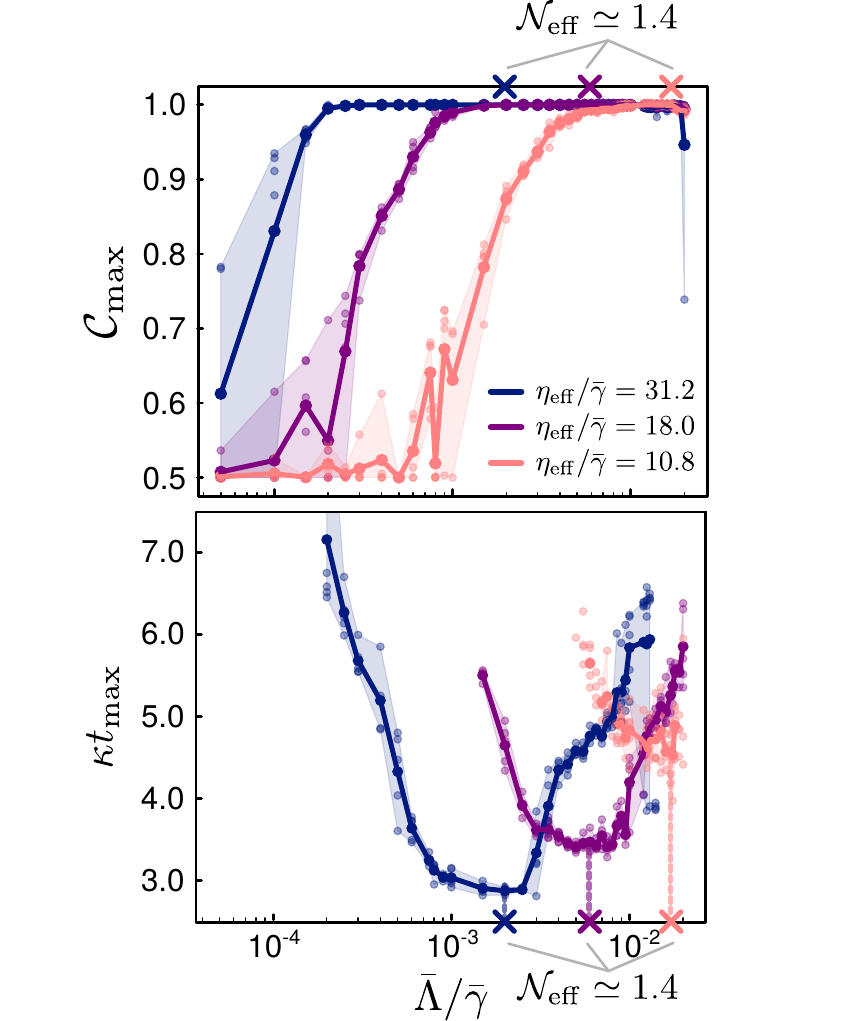}
    \caption{$\cmax$ (top panel) and $\tmax$ (bottom panel) as a function of average nonlinearity hyperparameter $\bar{\Lambda}$, for instances of the pointer state classification task corresponding to three different effective QRC drives $\eta_{\rm eff}/\bar{\gamma}$. Dots indicate different random QRCs, solid curves are averages over these QRCs, and shaded regions mark the range between worst and best performing QRCs. Crosses indicate QRCs with fixed $\CNLE\simeq 1.4$ for the three considered task instances; for weaker $\eta_{\rm eff}/\bar{\gamma}$, corresponding QRCs have stronger nonlinearity and operate in more quantum regimes (see text for more details).}
    \label{fig:cohClassifyDrive}
\end{figure}


The achieved $\cmax$ is plotted in the top panel of Fig.~\ref{fig:cohClassifyDrive} for the aforementioned $\eta_{\rm eff}/\bar{\gamma}$ values (blue, purple, and peach respectively). Immediately, we note that for weaker input drives, a larger QRC nonlinearity $\bar{\Lambda}$ is required for successful pointer state classification. When interpreted in the context of the effective nonlinearity $\CNLE$, this result makes sense: the latter depends on the product of the nonlinearity $\bar{\Lambda}$ of the QRC, and the effective drive $\eta_{\rm eff}$ it sees. To operate QRCs in a nonlinear processing regime given by a fixed value of $\CNLE$ (for example, $\CNLE \simeq 1.4$ marked by colored crosses in Fig.~\ref{fig:cohClassifyDrive}), a weaker $\eta_{\rm eff}$ will require stronger $\bar{\Lambda}$.


This intuitive result encourages a complementary perspective on reservoir processing across varying input drive strengths, enabled by our full quantum treatment of the measurement chain using truncated cumulants. As shown in Sec.~\ref{sec:singleNodeQRC}, increasing $\bar{\Lambda}$ for fixed $\CNLE$ controllably transitions Kerr QRCs from classical to quantum regimes of operation. Therefore, considering instances of the pointer state classification task across drive strengths allows us to analyze and compare QRC performance on the task throughout this transition.

In this context, extrapolating the results of Fig.~\ref{fig:cohClassifyDrive} to weaker nonlinearities leads to an immediate observation: even a QRC operating in the classical nonlinear regime ($\bar{\Lambda}\to 0$ for fixed $\CNLE$) can perform the pointer state classification task we consider here, albeit requiring increasing input strengths. In this limit, the quantum state of the measurement chain is described as a product of coherent states, defined entirely by its first-order cumulants. Fortunately, the information that distinguishes qubit pointer states is also encoded in their first-order cumulants (amplitudes and phases) and not their second-order cumulants (quantum fluctuations). As a result, the nonlinear mapping of this information to measured outputs by the nonlinear QRC is described primarily by dynamics of first-order cumulants, and is therefore possible in the classical limit of QRC operation.

Fig.~\ref{fig:cohClassifyDrive} indicates that for weaker incident drive strengths, classification is enabled by QRCs with stronger nonlinearities, operating in more quantum regimes; however, not all classification metrics are unchanged. In particular, the time $\tmax$ to reach maximum classification accuracy (plotted in in the lower panel of Fig.~\ref{fig:cohClassifyDrive}) \textit{increases} for weaker input drives. This can be understood as follows. Decreasing the input drive strength reduces the amplitude (and hence first-order cumulants) of the pointer states to be classified. This reduced amplitude can be compensated by increasing the QRC nonlinearity to perform the desired nonlinear mapping to measured outputs; however, the amplitude of QRC outputs is still correspondingly reduced. In terms of the measured subspace $\{I_1^{\varphi},I_2^{\varphi}\}$ shown in Fig.~\ref{fig:classifyCavStates}(e), the QRC output distributions to be separated have centroids that are closer to each other. The spread of these distributions, on the other hand, is determined by measurement noise (independent of drive strength) and QRC-mediated amplified or squeezed quantum fluctuations (mostly unchanged for fixed $\CNLE$); it primarily decreases with integration time $t$ due to filtering. Hence a decrease in input strengths reduces the relative separation of the distributions for a given measurement time, making them harder to separate, and demanding longer integration times $\tmax$ to enable classification.

The above results have an important consequence for the analysis of quantum reservoir processing for general computational tasks. Information processing with reduced input signals necessitates strongly-nonlinear QRCs, and thus provides a physically-motivated probe of QRC operation in quantum regimes. However, by using a framework that accounts for the complete measurement chain, we explicitly observe that the decrease in input power can lead to an increase in required measurement resources for computation; in this particular case, the measurement time. Such costs must be accounted for in any analysis of the possible advantages of reservoir computing in the quantum regime.



Finally, we note that processing with low input signals is commonplace in cQED experiments. Constraints on measurement time due to qubit decay and the use of noisy HEMT amplifiers~\cite{bishop_circuit_2010} in cQED experiments means that the reduced output signal is overcome by including quantum-limited amplifiers~\cite{hatridge_dispersive_2011, macklin_nearquantum-limited_2015, roy_introduction_2016} as part of the measurement chain. Such amplifiers can be included as part of our QRC framework for operation with low signal powers, as additional quantum measurement resources. The key considerations in doing so are discussed in Appendix~\ref{app:postamp} (albeit applied to our second computational task, which is considered next).

\section{Amplifier measurement: classifying Gaussian states with equal means and distinct variances}
\label{sec:amp}

We will now apply our quantum reservoir computing framework to Task II: the classification of Gaussian states that differ only in their second-order moments, or variances. To generate such states, we consider a minimal system based on a two-mode amplifier; as a shorthand, we refer to this task as that of amplifier state classification from here on.

\subsection{Hardware measurement chain}
\label{subsec:ampChain}


The quantum measurement chain is depicted schematically in Fig.~\ref{fig:classifyAmpStates}(a). The quantum system is coupled to a single-node QRC, which we will see proves sufficient for the classification task at hand. We again begin with a description of each element of this measurement chain, depicted in Fig.~\ref{fig:classifyAmpStates}(a).

\subsubsection{Two-mode amplifier and nonreciprocal coupling to QRC}

Our measured quantum system is a general two-mode amplifier, described by the Liouvillian superoperator $\Lsys$ defined as
\begin{align}
    \Lsys\hat{\rho} = \! -i[\hat{\mathcal{H}}_0 + \hat{\mathcal{H}}_{\rm S}^{(\sigma)} +  \eta^{(\sigma)}(-i\hat{a}_1+i\hat{a}_1^{\dagger}),\hat{\rho}  ]\! + \!\sum_n\kappa_n \mathcal{D}[\hat{a}_n]\hat{\rho},
    \label{eq:lsysamp}
\end{align}
where the system Hamiltonians take the form:
\begin{align}
    \hat{\mathcal{H}}_0 &= -\sum_{n={1,2}}\Delta_{an}\hat{a}_n^{\dagger}\hat{a}_n, \nonumber \\
    \hat{\mathcal{H}}_{\rm S}^{(\sigma)} &= \frac{1}{2}G_1^{(\sigma)}e^{-i\pi/2}\hat{a}_1^{2} + G_{12}^{(\sigma)}\hat{a}_1\hat{a}_2 + h.c.,
\end{align}
where $\Delta_{an} = \omega_d - \omega_{an}$. The two-mode system is defined by a degenerate parametric amplifier interaction $\propto G_1$ acting on mode $\hat{a}_1$, and a non-degenerate interaction $\propto G_{12}$. Note that these interaction strengths can also be set by tunable parametric drives in cQED implementations. Defining for convenience $\hat{X}_{a_1} = \frac{1}{\sqrt{2}}(\hat{a}_1+\hat{a}_1^{\dagger})$, $\hat{P}_{a_1} = -\frac{i}{\sqrt{2}}(\hat{a}_1-\hat{a}_1^{\dagger})$ as the canonically conjugate quadratures of the $\hat{a}_1$ mode, note that our choice of drive Hamiltonian $\propto \hat{P}_{a_1}$ drives the amplifier $\hat{X}_{a_1}$ quadrature only. Distinct states of this system are thus defined by the drive strength and values of the two interaction parameters, $(\eta^{(\sigma)},G_1^{(\sigma)},G_{12}^{(\sigma)})$. 

As before, understanding the dynamics of this two-mode amplifier requires specifying how it is embedded within the measurement chain including the QRC. We choose to couple the system $\hat{a}_1$ mode to the $\hat{b}_1$ node of the QRC; distinct from Sec.~\ref{sec:cav}, however, we now employ a non-reciprocal hopping interaction (modeling  a standard circulator), defined by the coupling superoperator $\mathcal{L}_c$,
\begin{align}
    \mathcal{L}_c\hat{\rho} = -i\left[\frac{i}{2}\varg_c\hat{a}_1^{\dagger}\hat{b}_1 + h.c. ,\hat{\rho}\right] + \Gamma_c\mathcal{D}[\hat{a}_1 + \hat{b}_1]\hat{\rho}.
    \label{eq:couplingCirc}
\end{align}
To understand the role of this coupling, we can once again analyze the conditional dynamics of the $\hat{X}_{a_1}$ quadrature that couples to the QRC:
\begin{align}
    \frac{d}{dt}&\avgc{\hat{X}_{a_1}} = {\rm tr}\{\Lsys\rhoc\hat{X}_{a_1}\} + \frac{\varg_c-\Gamma_c}{2}\avgc{\hat{X}_1} - \frac{\Gamma_c}{2}\avgc{\hat{X}_{a_1}}  \nonumber \\
    +& \frac{\sqrt{\gamma_1}}{2}\left[ \Cc{a_1b_1}+(\Cc{a_1b_1})^*+\Cc{a_1^{\dagger}b_1}+(\Cc{a_1^{\dagger}b_1})^*\right] dW^X_1(t) \nonumber \\ -i& \frac{\sqrt{\gamma_1}}{2}\left[ \Cc{a_1b_1}-(\Cc{a_1b_1})^*+\Cc{a_1^{\dagger}b_1}-(\Cc{a_1^{\dagger}b_1})^*\right] dW^P_1(t)
    \label{eq:ampXa}
\end{align}
Setting $\varg_c = \Gamma_c$ once again ensures that the amplifier does not see the QRC field, while the QRC is driven by the field from mode $\hat{a}_1$. However, in addition to this coupling term, the form of $\Lc$ introduces local damping $\propto \Gamma_c$ of the $\hat{a}_1$ mode, which modifies the system's dynamics. For convenience, we choose damping parameters such that the total damping rate of system modes $\hat{a}_1$ and $\hat{a}_2$ is equal after taking this additional damping term into account; more precisely, we set $\kappa_1 + \Gamma_c = \kappa_2 \equiv \kappa$. For simplicity, we further set $\kappa_1 = \Gamma_c = \kappa/2$, and $\omega_d = \omega_{a1} = \omega_{a2}$ so that the detunings $\Delta_{a1} = \Delta_{a2} = 0$. 

\begin{figure*}[t]
    \centering
    \includegraphics[scale=1.0]{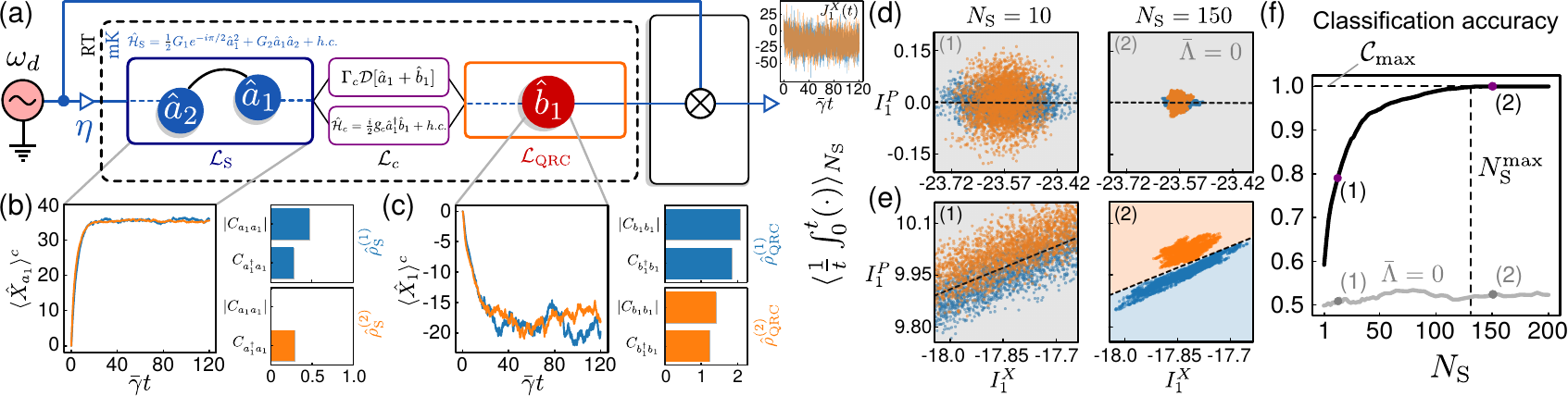}
    \caption{(a) Measurement chain using a single-node QRC for the classification of Gaussian states generated by a two-mode amplifier. A single mode $\hat{a}_1$ of the system is coupled to the QRC using a directional hopping interaction (a circulator). The single QRC node is continuously monitored via heterodyne measurement. (b) Conditional dynamics of the $\hat{a}_1$ mode's $\avgc{\hat{X}_a}$ quadrature, and steady-state unconditional cumulants. (c) Same as (b), for the $\hat{b}_1$ QRC node's $\avgc{\hat{X}_1}$ quadrature and unconditional cumulants. Conditional dynamics are shown for one initialization of the measurement chain per state. Colors blue and orange correspond to parameter sets $(\eta^{(\sigma)},G_1^{(\sigma)},G_{12}^{(\sigma)}) \in  \{(5.0,0.3,0.0), (8.0,0.0,0.3)\}\kappa$ respectively. (d) Measured phase space quadratures $\{I_1^X(t_f),I_1^P(t_f)\}$ for (1) $N_{\rm S} = 10$, and (2) $N_{\rm S} = 150$, for a linear QRC, $\bar{\Lambda}=0.0$. (e) Same as (d) but for a nonlinear QRC (see text for parameters). (f) Classification accuracy as a function of $N_{\rm S}$, computed for a test set of size $\QTE = 3000$; gray and black curves correspond to the linear and nonlinear QRCs of (d) and (e) respectively. }
    \label{fig:classifyAmpStates}
\end{figure*}

We are now in a position to define our classification task: we wish to use a QRC to classify $C=2$ different quantum states of the two-mode amplifier, defined by specific values of the drive and amplifier interaction strengths $(\eta^{(\sigma)},G_1^{(\sigma)},G_{12}^{(\sigma)}) \in  \{(5.0,0.3,0.0), (8.0,0.0,0.3)\}\kappa$. Leaving details for Appendix~\ref{app:amp}, these parameter choices generate amplifier states with $\avg{\hat{X}_{a_1}^{(1)}} =\avg{\hat{X}_{a_1}^{(2)}} \neq 0$, and $\avg{\hat{P}_{a_1}^{(1)}}=\avg{\hat{P}_{a_1}^{(2)}} = 0$, namely with equal first-order moments for the mode $\hat{a}_1$ that the QRC is directly coupled to. This is visible in the conditional dynamics of $\avgc{\hat{X}_{a_1}}$ plotted in Fig.~\ref{fig:classifyAmpStates}(b), with blue and orange colors corresponding to the two states. However the two states are distinct, as is clear from their different second-order cumulants, whose unconditional steady-state values are shown in the bar plots in Fig.~\ref{fig:classifyAmpStates}(b).


\subsubsection{QRC}

Under directional coupling $\varg_c = \Gamma_c$, the amplifier $\hat{a}_1$ mode drives the QRC via the $\hat{b}_1$ mode. For specificity, we can look at the conditional evolution of the $\hat{X}_1$ quadrature of the QRC:
\begin{align}
    \frac{d}{dt}\avgc{\hat{X}_1} &= {\rm tr}\{\Lqrc\rhoc\hat{X}_1\} - \Gamma_c \avgc{\hat{X}_{a1}} - \frac{1}{2}\Gamma_c \avgc{\hat{X}_{1}} \nonumber \\
    +& \frac{\sqrt{\gamma_1}}{2}\left[ \Cc{b_1b_1}+(\Cc{b_1b_1})^*+2\Cc{b_1^{\dagger}b_1}\right] dW^X_1(t) \nonumber \\ 
    -i& \frac{\sqrt{\gamma_1}}{2}\left[ \Cc{b_1b_1}-(\Cc{b_1b_1})^*\right] dW^P_1(t)
    \label{eq:ampX1}
\end{align}
The QRC nodes thus evolve under the amplifier signal and also experiences local damping due to the coupling superoperator, in addition to its internal dynamics governed by $\Lqrc$, and conditional evolution that depends on cumulants of the QRC state. Typical conditional dynamics of $\avgc{\hat{X}_1}$ are shown in Fig.~\ref{fig:classifyCavStates}(c) for the two amplifier states. The specific single-node QRC parameters we have chosen here are $\gamma_1 = \kappa$, $\Lambda_1 = 0.0027\gamma_1$, and $\Delta_1 = -\gamma_1$.

The QRC is driven to distinct quantum states under the two amplifier state inputs, as is clear from the unconditional second-order QRC cumulants shown in the bar plots in Fig.~\ref{fig:classifyCavStates}(c). What is not immediately clear is that the resulting QRC states can \textit{also} differ in their first-order moments, even though the two amplifier states do not; this has implications for state classification, as we analyze next.



\subsection{Training and classification using finitely-sampled measurement distributions}

Measurement records $\{J_1^{X,P}(t)\}$ obtained from the measurement chain contain information about QRC dynamics that will enable state classification. In using these measurement records to construct the measured QRC output, we will consider a slightly different scheme that is also experimentally relevant, to highlight the flexibility afforded by our description.

In contrast to the pointer state classification task, the measured QRC output $\mathbf{x}(t)$ here is constructed as the vector of measured quadratures obtained using both temporal filtering \textit{and} ensemble-averaging of heterodyne measurement records, calculated over $N_{\rm S}$ distinct runs of the measurement chain, or `shots',
\begin{align}
    \mathbf{x}(t) \equiv
    \begin{pmatrix}
     I_1^{X}(t) \\
     I_1^{P}(t) 
    \end{pmatrix},~I_1^{X,P}(t) = \Big< \frac{1}{t}\int_{t_0}^{t} d\tau~J_1^{X,P}(\tau) \Big>_{N_{\rm S}}.
    \label{eq:IkXPNS}
\end{align}
This more general form of readout (which reduces to the single-shot case when $N_{\rm S} = 1$) can be used to reduce noise power in readout using multiple measurement records obtained for the same inputs to the measurement chain. Measured quadratures are themselves stochastic, so that the QRC output provides a distinct stochastic trajectory for repeated runs of the measurement chain. QRC output corresponding to different parameters $\{\eta^{(\sigma)},G_1^{(\sigma)},G_{12}^{(\sigma)}\}$ is thus labelled $\mathbf{x}^{(\sigma)}_{(q)}$, where $q$ is the trajectory label. Note that each trajectory $q$ for a given $\sigma$ is constructed using $N_{\rm S}$ measurement records, each obtained from a distinct run of the measurement chain for the same input parameters $\{\eta^{(\sigma)},G_1^{(\sigma)},G_{12}^{(\sigma)}\}$.
 
 
Secondly, instead of integrating the entire obtained measurement records, we introduce a waiting time $t_0$. By choosing $t_0$ to be much longer than the amplifier response time for the two states (specifically, $t_0 = 35.0/\kappa \gg \frac{2}{\kappa-2G_1^{(1)}} \simeq \frac{2\kappa}{\kappa^2-4(G_{12}^{(2)})^2}$), we ensure that initial transients have settled, so that any information about the states contained in their slightly-distinct transient dynamics is lost and cannot influence classification.

The classification task is then set up as in Sec.~\ref{sec:cav}. We employ an output layer to predict a class label $\sigma^p$ using the measured QRC output, $\sigma^p = f_N\{\mathbf{W}_{\rm O}\mathbf{x}^{(\sigma)}_{(q)}(t_f)+\mathbf{b}\}$. Here, $\mathbf{W}_{\rm O}$ defines a matrix of time-independent trainable weights, $\mathbf{b}$ a vector of trainable biases, while $f_N\{\cdot\}$ is the same fixed (that is, untrained) normalizing function from earlier. Perfect classification yields $\sigma^p = \sigma$ for all $\sigma=1,\ldots,C$. Note that here we consider the measured phase space as being spanned by measured quadratures evaluated at the \textit{end} of the measurement, namely $\mathbf{x}(t_f) = \{I_1^X(t_f),I_1^P(t_f)\}$ with $t_f = 120.0/\kappa$. This corresponds to classification using a fixed measurement time; we will instead analyze performance as a function of the other measurement resource, the number of shots $N_{\rm S}$.






Beginning with the case of a complete linear QRC ($\bar{\Lambda} = 0$), the QRC outputs $\{\mathbf{x}^{(\sigma)}_{(q)}\}$ are shown in measured phase space in Fig.~\ref{fig:classifyAmpStates}(d), for $q=1,\ldots,1500$ per state $\sigma$ (blue and orange). Each panel shows outputs for the indicated value of $N_{\rm S}$. We see (1) that for $N_{\rm S} = 10$, the distributions corresponding to the two states have distinct profiles but are overlapping, and hence not separable by a linear decision boundary (black dashed line). The spread of the distributions can be decreased by averaging over an increasing number of shots; this is seen in (2), where $N_{\rm S} = 150$ (the phase space area shown is the same as in (1), allowing direct comparison of the distribution area). However, the centers of the measured distributions are unmoved, so that they still overlap and their linear separation is not possible.

The observed distributions are very different for an appropriately-chosen nonlinear QRC. The measured phase space is shown in Fig.~\ref{fig:classifyAmpStates}(e), again for increasing values of $N_{\rm S}$. For (1) $N_{\rm S} = 10$, we see that the distributions are significantly distorted in comparison to the linear QRC, showing amplified and squeezed axes due to the Kerr nonlinearity. However the distributions still overlap significantly and are not linearly separable. With increasing (2) $N_{\rm S} = 150$, the width of the distributions decreases, and a relative displacement of their centers (means) ensures that they can be separated by the dashed black linear decision boundary, allowing for perfect classification.

Note that both the linear and nonlinear QRC's are driven to distinct quantum states under input from the two states to be classified; this is clear from the distinct distributions observed in the measured phase space. However, that this difference manifests in a change to first order moments that are linearly related to the measured quadratures is a signature of nonlinear quantum systems, and not specific to the Kerr model we have chosen here: first-order moments can couple directly to higher-order moments, and can thus be modified by differences in these higher-order moments. Precisely this effect was observed in the steady-state response of a single driven Kerr QRC node in Figs.~\ref{fig:singleQRCCumulants}(b),~(c). 

As before, an arbitrary nonlinear QRC will not generally carry out this classification task optimally. To understand the required operating regimes, it proves useful to analyze the classification performance more quantitatively. Using a training set $\{\mathbf{x}^{(\sigma)}_{(q)}\}$ for $q=1,\ldots,Q_{\rm Train}=1500$ QRC outputs per state, we learn the optimal weights ${\mathbf{W}}_{\rm O}$ and biases $\mathbf{b}$ defining the linear decision boundary separating measured output distributions. Then, the trained QRC is used to predict state labels $\sigma^p$ for a distinct test set $\{\mathbf{x}^{(\sigma)}_{(q)}\}$ for $q=1,\ldots,Q_{\rm Test}=3000$, and the classification accuracy as a function of $N_{\rm S}$ is calculated. The result is plotted in Fig.~\ref{fig:classifyAmpStates}(f), in gray and black for the linear and the specific nonlinear QRC respectively, and is characterized by two metrics: the maximum classification accuracy $\cmax$ as before, and the number of shots used in constructing measured quadratures to reach this maximum, $\nsmax$. In the following section, we will analyze these performance metrics as a function of QRC hyperparameters to determine constraints that enable the amplifier state classification task.


\begin{figure*}[t]
    \centering
    \includegraphics[scale=1.0]{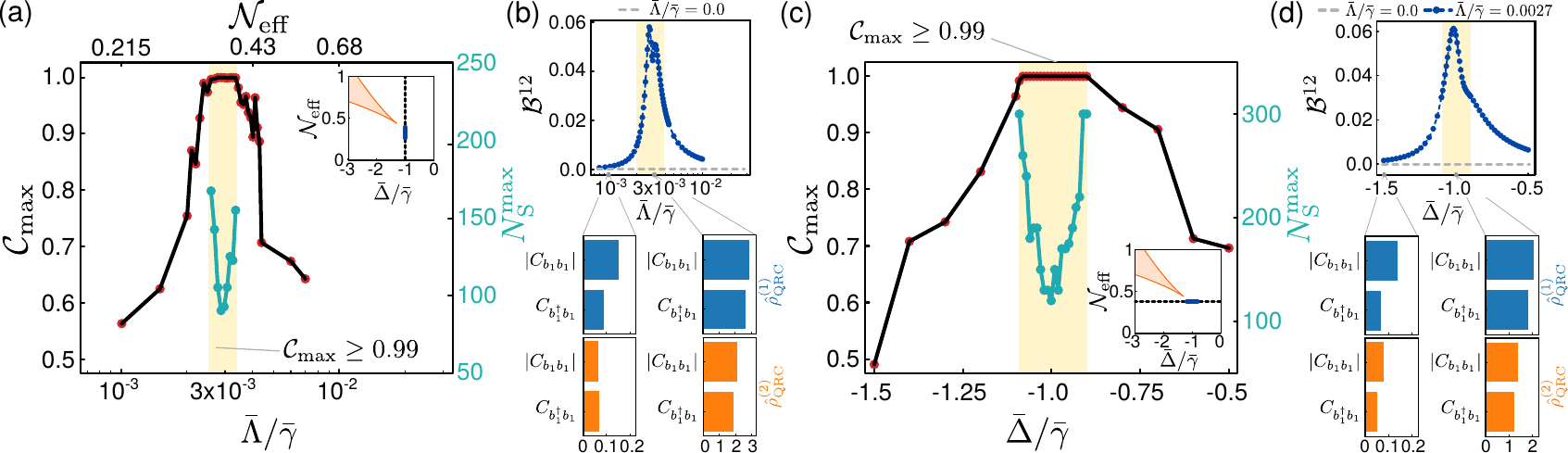}
    \caption{(a) $\cmax$ (left-hand axis) as a function of nonlinearity $\bar{\Lambda}$, for fixed detuning $\bar{\Delta}/\bar{\gamma} = -1.0$. Where $\cmax \geq 0.99$ (yellow shaded region), $\nsmax$ is plotted along the right-hand axis (green). Inset indicates the classical phase diagram. (b) Unconditional QRC node amplitude difference $\BD =|\avg{\hat{b}_1^{(1)}}-\avg{\hat{b}_1^{(2)}}|$ calculated using TEOMs as a function of nonlinearity $\bar{\Lambda}$. Lower panel shows unconditional QRC cumulants corresponding to the states $\hat{\rho}_{\rm QRC}^{(1),(2)}$ (blue and orange respectively) in a region of small $\BD$ (left) and large $\BD$ (right). Note the difference in magnitude of cumulants. (c), (d) Same as (a), (b) respectively, but now as a function of detuning $\bar{\Delta}$, for fixed nonlinearity $\bar{\Lambda}/\bar{\gamma} = 0.0027$, $\CNLE\simeq 0.354$. }
    \label{fig:ampClassifyJoint}
\end{figure*}


\subsection{Classification performance as a function of hyperparameters}

\textit{Hyperparameters.}$-$ For a single-node QRC, the parameter space to vary is reduced relative to the task considered in Sec.~\ref{sec:cav}, which employed a two-node QRC. Regardless, for consistency with our prior analysis we employ a similar notation, setting $\Lambda_1 \equiv \bar{\Lambda}$, $\Delta_1 \equiv \bar{\Delta}$, and setting the damping rate equal to the (common) decay rate $\kappa$ of amplifier modes, $\gamma_1 \equiv \bar{\gamma} = \kappa$. Once again, all remaining aspects of the measurement chain and training of the output layer remain unchanged as hyperparameters are varied and QRC performance observed. Our key findings are summarized at the end of this section.

\subsubsection{Dependence on nonlinearity $\bar{\Lambda}$}

We begin by fixing the QRC detuning $\bar{\Delta}/\bar{\gamma} = -1.0$, and calculating $\cmax$ for varying QRC nonlinearity $\bar{\Lambda}$; the results are shown in Fig.~\ref{fig:ampClassifyJoint}(a). Clearly, an optimal range of nonlinearity strengths exists where $\cmax \to 1$. For the orange shaded region where $\cmax \geq 0.99$, we also plot $\nsmax$ along the right-hand axis (green curve). A nonlinearity strength exists for which $\nsmax$ is minimized, thereby reducing the measurement resources required for classification.

The binary nature of the current task allows us to extract further intuition about the observed QRC performance. As discussed earlier, the ability to linearly separate integrated output records processed by the QRC corresponding to the two amplifier states depends on the measured noisy distributions acquiring a sufficient displacement. The unconditional QRC state variable that is a measure of this displacement is simply the difference in QRC node amplitude for the two states, $\BD\equiv|\avg{\hat{b}_1^{(1)}}-\avg{\hat{b}_1^{(2)}}|$. We plot this quantity calculated using TEOMs in the top panel of Fig.~\ref{fig:ampClassifyJoint}(b) for the same parameter regime as (a), together with the yellow shaded region where $\cmax \geq 0.99$. The lower panel shows unconditional cumulants of the QRC state corresponding to the two amplifier states (blue and orange respectively). Outside the yellow shaded region, the QRC state is driven to distinct cumulants for the two amplifier states, but the magnitude of cumulants, and the resulting displacement of QRC node amplitudes $\BD$, is small. However, within the yellow shaded region, the distinct cumulants are much larger, leading to larger value of $\BD$. In contrast, for a linear QRC (dashed gray line), $\BD$ is always zero. 

We emphasize here that a nonzero value of this displacement is by itself insufficient for successful discrimination between the amplifier states. The displacement is contained within noisy measurement records, and classification amounts to being able to distinguish their filtered, ensemble-averaged distributions in measured phase space. These factors are all accounted for by our training on measured quadrature distributions, which yield all our results for $\cmax$ and $\nsmax$.

Operating regimes where $\BD$ becomes nonzero can once again be analyzed in the context of the classical phase diagram of the single-node QRC. To this end, we introduce an effective drive $\eta_{\rm eff}$ experienced by the QRC due to its coupling to the amplifier mode $\hat{a}_1$. Unlike the QND interaction employed in Sec.~\ref{sec:cav} which coupled only the cavity $\hat{X}_a$ quadrature to the QRC, the directional hopping interaction here ensures that mode $\hat{b}_1$ sees both quadratures of mode $\hat{a}_1$. The effective drive $\eta_{\rm eff}^{(\sigma)}$ seen by the QRC for each amplifier state then takes the form
\begin{align}
    \eta_{\rm eff}^{(\sigma)} = -\frac{\Gamma_c}{\sqrt{2}} \!\left( \avg{\hat{X}_{a_1}^{(\sigma)}} + i\avg{\hat{P}_{a_1}^{(\sigma)}} \right) \! = \frac{2\Gamma_c\kappa\cdot\eta^{(\sigma)}}{4(G^{(\sigma)}_{12})^2 + \kappa(2G_1^{(\sigma)}\!\!-\!\kappa)}
\end{align}
where the second equality follows from the steady-state solutions for $\avg{\hat{X}_{a_1}^{(\sigma)}}$, $\avg{\hat{P}_{a_1}^{(\sigma)}}$ (see Appendix~\ref{app:amp}). The current task is specifically chosen so that $\eta_{\rm eff}^{(1)}=\eta_{\rm eff}^{(2)}$; we hence set $\eta_{\rm eff} \equiv |\eta_{\rm eff}^{(1,2)}|$ to once again define an effective nonlinearity $\CNLE$ that defines the classical phase diagram,
\begin{align}
    \CNLE = \frac{\eta_{\rm eff}}{\bar{\gamma}+\Gamma_c}\sqrt{\frac{\bar{\Lambda}}{\bar{\gamma}+\Gamma_c}}
    \label{eq:cnleAmp}
\end{align}
similar to Eq.~(\ref{eq:cnleCoh}), but now with $\bar{\gamma} \to \bar{\gamma} + \Gamma_c$ to account for the modified decay rate of the QRC due to the different form of coupling $\Lc$. The resulting phase diagram is plotted in the inset of Fig.~\ref{fig:ampClassifyJoint}(a). The dashed vertical line marks the cross-section along which Fig.~\ref{fig:ampClassifyJoint}(a) is calculated, while the solid blue line coincides with the yellow shaded region where $\cmax \geq 0.99$. Clearly, this region places the QRC in the vicinity of the classical bistability, where cumulants are amplified, precisely as observed in Sec.~\ref{sec:singleNodeQRC} for a single-node QRC under coherent driving. 





\subsubsection{Dependence on detuning $\bar{\Delta}$}

Next we explore the classification performance as a function of detuning $\bar{\Delta}$ for fixed nonlinearity strength $\bar{\Lambda}/\bar{\gamma} = 0.0027$, which sets $\CNLE\simeq 0.354$ via Eq.~(\ref{eq:cnleAmp}). These results are shown in Fig.~\ref{fig:ampClassifyJoint}(c). Once more, we find that an optimal range of detuning values exists for which $\cmax$ approaches unity and $\nsmax$ is minimized. Analysis of $\BD$ as a function of detuning values in Fig.~\ref{fig:ampClassifyJoint}(d) shows how the QRC node amplitude displacement peaks around a detuning value $\bar{\Delta}/\bar{\gamma} \simeq -1.0$ for the specific $\CNLE$ value considered. This operating regime corresponds to $\cmax \geq 0.99$ (yellow shaded region), and once again places the QRC in the vicinity of the classical bistability (solid blue line, Fig.~\ref{fig:ampClassifyJoint}(c) inset).

\subsubsection{Dependence on damping $\bar{\gamma}$}

The QRC node damping rate appears as the natural scaling parameter for QRC hyperparameters, just as in the pointer state classification task. Hence if $\bar{\gamma}$ is varied, the remaining QRC hyperparameters $\{\CNLE,\bar{\Delta}/\bar{\gamma}\}$ can be suitably adjusted to ensure that the QRC operates in a nonlinear processing regime, enabling successful classification.

Beyond parameterizing such nonlinear processing regimes, $\bar{\gamma}$ also influences the QRC node response time and the amplitude of QRC output fields for fixed input signal strengths. Recall that our setup of the amplifier state classification task - using measurement records only beyond a time $t_0$ once transients have settled - effectively analyzes the QRC's processing of \textit{steady-state} signals from the two-mode measured quantum system. Hence considerations of the QRC responding promptly to a dynamically-evolving signal as in the pointer state classification task do not apply directly. Regardless, the QRC damping rate $\bar{\gamma}$ can influence the time $t_0$ if it is \textit{slower} than the amplifier response rate for the states under consideration; more precisely, in this case we would choose $t_0 \gg 1/\bar{\gamma}$. This in principle increases the total required measurement time. On the other hand, larger $\bar{\gamma}$ for fixed strength of incident drives will suppress the amplitude of QRC output, which is in general detrimental to classification. These considerations imply that $\bar{\gamma}$ should ideally be chosen to be on the order of the response rate of the amplifier (or the measured quantum system in general).

\subsubsection{Summary}

The results of this section identify a range of QRC hyperparameters $\{\CNLE,\bar{\Delta}/\bar{\gamma}\}$ where the QRC is able to successfully classify amplifier states with equal means, finding a strong connection to operating points in the vicinity of the classical bistability of the single coherently-driven QRC node, similar to the pointer state classification task. Importantly, the mechanism of classification is distinct: the input-dependent enhancement of second-order cumulants of the QRC state, whose nonlinearity leads to this difference manifesting via the QRC node amplitudes.

\subsection{Classical vs. quantum regimes of operation}
\label{subsec:ampquantum}

As with the task of pointer state classification, we now analyze QRC performance for amplifier state classification across decreasing input drives to the measurement chain. To this end, we consider different instances of this task, defined by the same two-mode amplifier interaction strengths $G_1^{(\sigma)}, G_{12}^{(\sigma)}$, but different values of drive input to the measurement chain, $(\eta^{(1)}/\bar{\gamma},\eta^{(2)}/\bar{\gamma})  \in \{(2.0,3.2),(5.0,8.0),(10.8,12.0)\}$. Via Eq.~(\ref{eq:cnleAmp}), these instances yield effective QRC drive strengths $\eta_{\rm eff}/\bar{\gamma} \in \{20.0,12.5,5.0\}$, for which we calculate the classification accuracy metrics $\cmax$ and $\nsmax$ as a function of nonlinearity $\bar{\Lambda}$. We have fixed $\bar{\Delta}/\bar{\gamma} = 0.0$ and $\bar{\gamma} = \kappa$ as before.

\begin{figure}[t]
    \centering
    \includegraphics[scale=1.0]{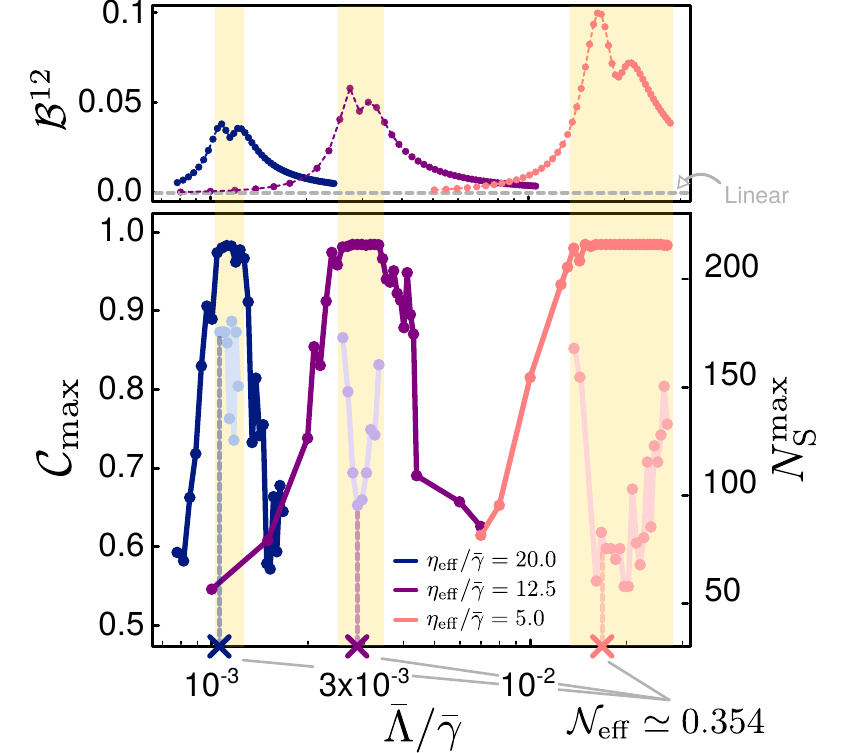}
    \caption{$\cmax$ (solid curves) and $N_{\rm S}^{\rm max}$ (lighter curves, right-hand axis) as a function of average nonlinearity hyperparameter $\bar{\Lambda}$ for instances of the amplifier state classification task with different effective drive amplitudes $\eta_{\rm eff}$ as indicated. Crosses indicate QRCs with fixed $\CNLE\simeq 0.354$ for the three considered task instances; for weaker $\eta_{\rm eff}/\bar{\gamma}$, corresponding QRCs have stronger nonlinearity and operate in more quantum regimes (see text for more details). Top panel plots the unconditional QRC node amplitude difference $\BD=|\avg{\hat{b}_1^{(1)}}-\avg{\hat{b}_1^{(2)}}|$, which increases for QRCs operating in more quantum regimes. The dashed gray line indicating the (vanishing) value of $\BD$ for linear QRCs.  }
    \label{fig:ampClassifyDrive}
\end{figure}

We plot $\cmax$ in the lower panel of Fig.~\ref{fig:ampClassifyDrive} for the corresponding values of $\eta_{\rm eff}/\bar{\gamma}$ (solid blue, purple, and peach respectively), with the yellow shaded region indicating nonlinearity strengths for which $\cmax \geq 0.99$. Similar to the pointer state classification task, we see that for lower $\eta_{\rm eff}$, the nonlinearity strength required for successful classification increases. This is captured by the form of effective nonlinearity $\CNLE$; colored crosses indicate QRCs with the same value of $\CNLE \simeq 0.354$ for the different instances of this classification task, which all succeed at classification but require stronger $\bar{\Lambda}$ for weaker $\eta_{\rm eff}$.  

Decreasing the input power therefore requires QRCs with stronger nonlinearity, providing a natural probe of the quantum regime of QRC operation. To analyze QRC performance across this transition, we also plot $\nsmax$ in the lower panel of Fig.~\ref{fig:ampClassifyDrive} (corresponding lighter colors, right-hand axis). Interestingly, we find that for decreasing $\bar{\Lambda}$ indicating the classical limit of QRC operation, the ampifier state classification task becomes more difficult, as the number of shots $\nsmax$ required to obtain $\cmax \geq 0.99$ increases. Conversely, for operation in more quantum regimes, $\nsmax$ decreases, so that the task has a reduced resource cost. 



To understand this observation further, we use the TEOMs to calculate the difference in the unconditional QRC node amplitude for the two states, $\BD = |\avg{\hat{b}_1^{(1)}}-\avg{\hat{b}_1^{(2)}}|$. The result is plotted in the top panel of Fig.~\ref{fig:ampClassifyDrive} for each considered instance of the classification task. We clearly see that $\BD$ increases as we move towards QRCs operating deeper in the quantum regime (blue to purple to peach). For conditional evolution, this increase manifests in a larger relative displacement of measured quadrature distributions, resulting in a reduction in ensemble-averaging required for successful classification. Conversely, extrapolating the results to the classical limit ($\bar{\Lambda}\to 0$ for fixed $\CNLE$), we see that $\BD \to 0$, so that the QRC is unable to distinguish the amplifier states in the classical limit.
 
 
We thus observe a qualitative difference in QRC performance across classical and quantum regimes when compared to the pointer state classification task. This is because unlike the latter, the information that distinguishes amplifier states is encoded in their second-order cumulants. In the classical limit governed primarily by first-order cumulants of the measurement chain, the influence of these second-order cumulants is suppressed; this directly reduces $\BD$, rendering the classification task more difficult (requiring larger $\nsmax$). Conversely, in more quantum regimes the influence of second-order cumulants on QRC node amplitudes is enhanced, reducing $\nsmax$. These results indicate that the nature of input encoding into QRC evolution can play an important role in extracting possible processing advantages in quantum operating regimes.
 
 
However, one aspect of probing the quantum regime is shared across both classification tasks: the absolute amplitude of measured QRC outputs is reduced for weaker input drives in this regime. For the amplifier state classification task, the \textit{relative} difference in QRC node amplitudes $\BD$ is still larger in this regime, which is why the required $\nsmax$ decreases in our simulations of the complete measurement chain. In a real cQED experiment, however, the reduced amplitude can lead to the desired signal being swamped by noisy (classical) HEMT amplifiers that are also part of the measurement chain. As discussed earlier, this limitation is overcome by including quantum-limited amplifiers as part of the measurement chain, in this case to amplify the QRC output at cryogenic temperatures. We discuss the inclusion of such an amplifier for the current task in Appendix~\ref{app:postamp}. We find that even when using phase-preserving quantum-limited amplifiers that add unavoidable amplified quantum noise to the measured quadrature distributions, the associated amplification of the difference in QRC node amplitudes $\BD$ ensures that the benefit of operating the QRC in more quantum regimes is preserved.




\section{Discussion and Outlook}

In this paper we present and analyze a framework for quantum reservoir computing that we believe is timely within the current landscape of quantum computation. Research efforts geared towards the realization of a universal quantum computer to outperform classical computational paradigms have accelerated in recent years~\cite{arute_quantum_2019, wu_strong_2021}; however, several important challenges remain in scaling nascent quantum computers to larger and larger numbers of quantum degrees of freedom, in particular maintaining coherence times, correcting for the inevitable errors that occur during computation, and managing growing hardware complexity~\cite{berke_transmon_2020}. Nevertheless, as part of these and related efforts, modern experiments have continued to evolve in scope and complexity, realizing extended multimode nonlinear quantum systems with an unprecedented level of control. This naturally raises the question of whether such near-term quantum devices can already be used for nontrivial computation. This paper lays the theoretical foundations to assess the potential computational capabilities of physical quantum systems as they are currently realized, with constrained input and output control and subject to specific measurement schemes. The machine-learning paradigm of reservoir computing using physical systems, with relaxed requirements over internal device optimization, is ideally suited to harness these capabilities.

To this end, our strategy is to develop a framework that (1) is built upon a quantum-mechanical description of a physical quantum systems as part of its complete quantum measurement chain, outputs from which are sampled via a prescribed measurement scheme, and (2) has a physically-motivated classical limit of operation. Such a framework allows us to determine performance metrics, such as fidelity and time-to-solution, of performing a computational task subject to operational constraints on inputs and outputs, and using only accessible information obtained via measurement, therefore accounting for experimental resource costs such as measurement time and repeated runs. Then, comparing these metrics across well-defined classical and quantum regimes of reservoir operation pinpoints advantages of quantum reservoir processing for a given computational task.

We believe such an approach can be applied broadly in seeking advantages or disadvantages to computation using quantum devices across very general computational tasks. For a concrete analysis, in this paper we realize this framework in the cQED architecture, and apply it to two quantum state classification tasks: states with distinct means (task I: pointer state classification), and states with equal means but distinct variances (task II: amplifier state classification). This requires developing an efficient computational approach to simulate the conditional dynamics of nonlinear quantum measurement chains incorporating Josephson-junctions, signal-generating quantum systems, with the entire chain evolving under continuous heterodyne measurement. We present such an approach based on a set of Stochastic Truncated Equations Of Motion (STEOMs) for measured observables and quantum cumulants. Different from earlier quantum reservoir computing approaches, this allows us to consider training of a linear output layer directly on measured outputs, which in addition to measurement noise include quantum fluctuations non-linearly transformed by the action of the quantum reservoir itself. 

The imposition of a trainable \textit{linear} output layer applied to obtained measurement records is an important choice: it allows any useful nonlinear processing that enables quantum state classification to be attributed to nonlinearity in the quantum measurement chain, and not to post-processing operations. In our realization, this nonlinearity is provided by the QRC, and is in fact critical: considering linear QRCs under the same operating conditions and output processing does not yield perfect classification. We note that the required nonlinearity could in principle be provided by other elements of a measurement chain, for example via alternate measurement schemes such as photodetection, or distinct input encoding schemes. This intuition is supported by work on classical reservoirs, which have demonstrated the power of computation with entirely linear reservoirs but employing non-linear readout layers~\cite{gauthier_next_2021, Bollt2021}. This flexibility speaks to the ability to implement physical reservoir computing across a range of quantum device platforms, tailored to specific available nonlinearities and measurement schemes. A comparison of the relative performance of different quantum reservoir computing implementations is an interesting avenue for future exploration. 



The ability to analyze QRC performance across classical-to-quantum transitions within this framework provides a systematic approach to identifying the role of reservoir processing in regimes where the operation is governed by quantum statistical features. Carrying out this analysis, we find that pointer states, distinguished by their first-order cumulants, can be successfully classified even by QRCs operating in the classical limit. In addition, from the point of view of time-to-solution ($\tmax$), it is more advantageous to operate deeper in the classical limit. This may be perceived as a natural outcome of higher signal-to-noise ratio in the measured outputs, but the analysis shows the impact of non-linearly transformed noise should be considered carefully when reaching conclusions about the information processing capacity of a physical systems in such limits. In contrast, the classification of amplifier states distinguished only by their second-order cumulants are performed in a more sample-efficient manner (lower $\nsmax$) deeper in the quantum regime of operation. From the point of view of power-efficiency, Task II provides an example where operating deeper in the quantum regime is advantageous but comes at the expense of the need for a QRC with a stronger non-linearity.

Finally, the computational efficiency of STEOMs permits us to train and test QRCs on sets of several stochastic measurement outputs across a wide range of parameter values beyond simply its bare nonlinearity. We identify regimes of operation where the QRC is successful at the chosen classification tasks, and find an interesting unifying feature: near classical bifurcation points, the QRC exhibits an enhanced nonlinear response to its quantum inputs, which proves useful for quantum state classification. Such observations can aid in the design and operation of physical quantum reservoir computers.

Looking ahead, the presented framework is ideally suited to theoretical studies of the possible processing advantage afforded by quantum reservoirs when compared to their classical counterparts, across a variety of general computational tasks. This provides the basis for future studies of quantum reservoir processing capabilities as a function of reservoir size and hardware complexity. More importantly, the role of resources special to quantum reservoirs can be addressed: entanglement amongst the reservoir nodes, as well as between the quantum signal-generating system and the reservoir degrees of freedom. These studies are essential to identify the input and output schemes, and nature of quantum reservoirs that will be able to best extract the benefits of reservoir processing in the quantum regime. This is the subject of future work.

In view of practical implementations, two distinct information processing paradigms are made accessible: Firstly, it becomes possible to model quantum reservoirs as standalone quantum processors of data encoded in classical inputs, or quantum inputs via the state of a quantum system that is part of the same measurement chain and coupled to the quantum reservoir. Secondly, our analysis naturally provides a basis for studying quantum reservoirs as nonlinear quantum measurement devices, embedded into existing measurement chains for quantum systems, effectively describing an efficiently-trainable nonlinear readout layer. Our results indicate that simple QRCs can carry out useful nonlinear processing at cryogenic temperatures in a low noise environment, allowing experimentalists to avoid nonlinear processing of obtained noisy measurement records at room temperature.

\begin{acknowledgements}
We would like to thank Dan Gauthier, Luke Govia, Michael Hatridge, Peter McMahon, Ioan Pop, Graham Rowlands and Guilhem Ribeill for useful discussions. This work is supported by AFOSR under Grant No. FA9550-20-1-0177 and the Army Research Office under Grant No. W911NF18-1-0144. Simulations in this paper were performed using the Princeton Research Computing resources at Princeton University, which is a consortium of groups led by the Princeton Institute for Computational Science and Engineering (PICSciE) and Office of Information Technology's Research Computing.
\end{acknowledgements}

\begin{center}
    \rule{30mm}{1pt}
\end{center}

\appendix

\section{Quantum cumulants and the generating function}
\label{app:cumulants}

The starting point for our discussion of cumulants is via the characteristic function $\chi(z,z^*)$, defined in terms of the density operator $\rhou$ for an arbitrary quantum state as
\begin{align}
    \chi(z,z^*) = {\rm tr}\left\{ \rhou e^{i z^* \hat{b}^{\dagger}} e^{i z \hat{b}}. \right\}
    \label{appeq:chi}
\end{align}
The characteristic function as defined in Eq.~(\ref{appeq:chi}) has the property that it defines all \textit{normal-ordered} operator averages with respect to the quantum state $\rhou$; simply evaluating the derivative of the characteristic function~\cite{carmichael_statistical_2002} at $z=z^* = 0$ yields
\begin{equation}
  \left. \frac{\partial^{p + q}}{\partial (iz^*)^p \partial (iz)^q} (\mathrm{tr} \{ \rho e^{i z^* \hat{b}^{\dagger}} e^{i z \hat{b}}\} )  \right|_{z = z^* = 0} = \langle \hat{b}^{\dagger p} \hat{b}^q \rangle.
\end{equation}
It is then possible to define a distribution function for calculating these normal-ordered moments,
\begin{align}
    P (\beta, \beta^{\dagger}) &=  \frac{1}{\pi^2}  \int d^2z~\chi (z, z^{\ast}) e^{- i z^{\ast} \beta^{\ast}} e^{- i z \beta} \label{eq:chi2P} 
\end{align}
for which we use the suggestive notation $P(\zeta,\zeta^*)$, as it is easily shown to be none other than the Glauber-Sudarshan $P$ distribution~\cite{glauber_coherent_1963,sudarshan_equivalence_1963,carmichael_statistical_2002}:
\begin{align}
    \rhou = \int d^2\beta~P(\zeta,\zeta^{\dagger})|\beta\rangle\langle\beta|
\end{align}



Having recalled the connection of the characteristic function to the quantum state defined by $\rhou$ and its normal-ordered moments, we now define a quantum generating function $G : \mathbb{C} \times \mathbb{C} \rightarrow
\mathbb{R}$,
\begin{equation}
  G (z, z^{\ast}) = \ln \mathrm{tr} \{ \chi(z,z^*) \} = \ln \mathrm{tr} \{\hat{\rho} e^{i z^{\ast} \hat{b}^{\dagger}} e^{i z \hat{b}}\}. \label{eq:GeneratingFunction}
\end{equation}
The \textit{quantum cumulants} are defined through Eq.~(\ref{eq:GeneratingFunction}) by the series expansion of the generating function $G (z, z^{\ast})$ which yields all quantum cumulants $C_{a^{\dagger p} a^q}$ for any $p,q \in \mathbb{N}^2$
\begin{equation}
  \ln \mathrm{tr} \{\hat{\rho} e^{i z^{\ast} \hat{b}^{\dagger}} e^{i z \hat{b}}\} = \sum_{n = 1}^{\infty}\!\!
  \sum_{\scriptsize{\begin{array}{c}
    p, q \in \mathbb{N} \\
    p\!+\!q \!=\! n
  \end{array}}} \frac{1}{p!q!} C_{b^{\dagger p} b^q} (i z^{\ast})^p (i z)^q,
\end{equation}
or equivalently, 
\begin{equation}
    C_{b^{\dagger p} b^q} \equiv \left. \frac{\partial^{p+q} G}{\partial (i z^{\ast})^p \partial (i z)^q} (z, z^{\ast})\right|_{z=z^{\ast}=0} 
\end{equation}

So far, cumulants may simply appear to be a mathematical construct. However, it can be shown that normal-ordered cumulants and normal-ordered moments can be mapped to each other one-to-one. To see this, we can think of $G (z,z^{\ast})$ as the composite of the logarithm function $\mathrm{log} (\chi) = \ln \chi$ and $\chi(z,z^*) = {\rm tr}\left\{ \rhou e^{i z^* \hat{b}^{\dagger}} e^{i z \hat{b}} \right\}$, and employ multi-variable version of Fa{\`a} di Bruno's formula for the $n$-th order partial derivative where $n=p+q$,
\begin{align}
    & \frac{\partial^{p+q} G}{\partial (i z^{\ast})^p \partial (i z)^q} (z, z^{\ast}) \nonumber \\
    =~& \sum_{\pi} \left( \mathrm{log}^{(| \pi |)} (\chi (z, z^{\ast})) \prod_{B \in \pi} \frac{\partial^{p' + q'} \chi}{\partial (i z^{\ast})^{p'} \partial (i z)^{q'}} (z, z^{\ast}) \right),
\end{align}
where $\pi$ defines the set of possible partitions of the ordered set $\{ iz^{\ast}, \cdots, iz^{\ast}, iz, \cdots iz \}$ with $iz^{\ast}$ appearing $p$ times and $iz$ appearing $q$ times. $B$ indicates elements in this set of partitions, and $p'$ and $q'$ respectively counts the number of $iz^{\ast}$ and $iz$ in each partition element $B$. Finally, $f^{(|\pi|)}(\cdot)$ denotes the $|\pi|$-order derivative of the function $f(\cdot)$, where $|\pi|$ denotes the cardinality of the set $\pi$. By noting that $\chi (0, 0) = \mathrm{tr} (\rhou) = 1$, and then using $\mathrm{log}^{(| \pi |)} (\chi (0, 0)) = (\ln \chi)^{(| \pi |)} |_{\chi = 1} = (- 1)^{| \pi | - 1} (| \pi | - 1) !$ we obtain
\begin{align}
    C_{b^{\dagger p} b^q} & = \left. \frac{\partial^{p+q} G}{\partial (i z^{\ast})^p \partial (i z)^q} (z, z^{\ast})\right|_{z=z^{\ast}=0} \nonumber \\
    & = \sum_{\pi} \left( (- 1)^{| \pi | - 1} (| \pi | - 1) ! \prod_{B \in \pi} \langle \hat{b}^{\dagger p'} \hat{b}^{q'} \rangle \! \right). \label{eq:moment2cumulant}
\end{align}
As such, cumulants can be thought of as a reparameterization of the normal-ordered moments that typically appear in observables measured in an experiment. We will see in the next two appendices that this reparameterization leads to a much more efficient description of certain quantum states, which can then be leveraged to construct an efficient computational framework.

For our purposes, we will find it more useful to express quantum moments in terms of quantum cumulants. This inverse transformation can be performed by thinking of $\chi(z,z^*) = {\rm tr}\left\{ \rhou e^{i z^* \hat{b}^{\dagger}} e^{i z \hat{b}} \right\}$ as the composite of the exponential function $\mathrm{exp} (G) = e^G$ and $G (z, z^{\ast}) = \ln \mathrm{tr} \{\hat{\rho} e^{i z^{\ast} \hat{b}^{\dagger}} e^{i z \hat{b}}\}$. Again, Fa{\`a} di Bruno's formula gives
\begin{align}
    & \frac{\partial^{p+q} \chi}{\partial (i z^{\ast})^p \partial (i z)^q} (z, z^*) \\
    =~& \sum_{\pi} \!\! \left( \mathrm{exp}^{(| \pi |)} (G (z, z^*)) \prod_{B \in \pi} \frac{\partial^{p' + q'} G}{\partial (i z^{\ast})^{p'} \partial (i z)^{q'}} (z, z^*) \! \right)
\end{align}
where $\pi$ defines the set of possible partitions of the ordered set $\{ iz^{\ast}, \cdots, iz^{\ast}, iz, \cdots iz \}$ with $iz^{\ast}$ appearing $p$ times and $iz$ appearing $q$ times, $B$ indicates elements in this set of partitions, and $p'$ and $q'$ respectively counts the number of $iz^{\ast}$ and $iz$ in each partition element $B$. Again, noting that $G (0, 0) = \ln \mathrm{tr} (\rhou) = 0$, and then using $\mathrm{exp}^{(| \pi |)} (G(0, 0)) = (e^G)^{(| \pi |)} |_{G = 0} = 1$ we get for $n$-th order partial derivative where $n=p+q$:
\begin{align}
    \langle \hat{b}^{\dagger p} \hat{b}^q \rangle & = \left. \frac{\partial^{p+q} \chi}{\partial (i z^{\ast})^p \partial (i z)^q} (z, z
    ^{\ast}) \right|_{z=z^{\ast}=0} \nonumber \\
    & = \left. \sum_{\pi} \prod_{B \in \pi} \frac{\partial^{p' + q'} G}{\partial (i z^{\ast})^{p'} \partial (i z)^{q'}} (z, z^{\ast}) \right|_{z=z^{\ast}=0} \nonumber \\
    & = \sum_{\pi} \prod_{B \in \pi} C_{b^{\dagger p'} b^{q'}}. \label{eq:cumulant2moment}
\end{align}

Finally, it is straightforward to generalize the definition of quantum cumulants to the $N$-node bosonic mode case. For a complex-valued vector $\boldsymbol{z} = (z_1, z_2, \cdots, z_N)$, we can define the generating function $G: \mathbb{C}^{N} \times \mathbb{C}^{N} \rightarrow \mathbb{R}$ as: 
\begin{align}
     G (\boldsymbol{z}, \boldsymbol{z}^{\ast}) = \ln \mathrm{tr} \left\{ \rhou e^{iz^{\ast}_1 \hat{b}_1^{\dagger}} \cdots e^{iz^{\ast}_N \hat{b}_N^{\dagger}} e^{iz_1 \hat{b}_1} \cdots e^{iz_N \hat{b}_N} \right\}. \label{eq:GenNmodecumulants}
\end{align}
The series expansion of $G(\boldsymbol{z}, \boldsymbol{z}^{\ast})$ yields all quantum cumulants
\begin{align}
    & \left. \frac{\partial^n G}{\partial (iz_1^{\ast})^{p_1} \!\cdots\! \partial (iz_N^{\ast})^{p_N} \partial (iz_1)^{q_1} \!\cdots\! \partial (iz_N)^{q_N}} (\boldsymbol{z}, \boldsymbol{z}^{\ast}) \right| _{\boldsymbol{z}=\boldsymbol{z}^{\ast}=\boldsymbol{0}} \nonumber \\
    & = C_{b_1^{\dagger p_1} \cdots b_N^{\dagger p_N} b_1^{q_1} \cdots b_N^{q_N}}. \label{eq:Nmodecumulants}
\end{align}
where $p_1+\cdots+p_N+q_1+\cdots+q_N = n$. The transformation between moments and cumulants in multi-node case can be obtained by simply replacing all single-node moments and cumulants in Eq.(\ref{eq:moment2cumulant}),~(\ref{eq:cumulant2moment}) with multi-node moments and cumulants corresponding to normal-ordered operator $ \hat{o}_1 \hat{o}_2 \cdots \hat{o}_n = \hat{b}_1^{\dagger p_1} \cdots \hat{b}_N^{\dagger p_N} \hat{b}_1^{q_1} \cdots \hat{b}_N^{q_N}$: 
\begin{align}
    C_{o_1 o_2 \cdots o_n} & = \sum_{\pi} \left( (- 1)^{| \pi | - 1} (| \pi | - 1) ! \prod_{B \in \pi} \langle \hat{o}_{i}: i \in B \rangle \right), \label{eq:moment2cumulantN} \\
    \avg{\hat{o}_1 \hat{o}_2 \cdots \hat{o}_n} & = \sum_{\pi}\prod_{B\in \pi} C_{o_i : i \in B}, \label{eq:cumulant2momentN}
\end{align}
where $\pi$ defines the set of possible partitions of operators in the $n$-order moment, and $B$ indicates elements in this set of partitions.

\section{Cumulants for some simple quantum states}
\label{app:cumulantsQS}

Eq.(\ref{eq:moment2cumulant}),~(\ref{eq:cumulant2moment}) and their multi-node case generalization Eq.(\ref{eq:moment2cumulantN},\ref{eq:cumulant2momentN}) show that quantum cumulants and quantum moments have one-to-one correspondence through logarithm and exponential. On the other hand, for two complex-valued vectors $\boldsymbol{z} = (z_1, z_2, \cdots, z_N)$ and $\boldsymbol{\beta} = (\beta_1, \beta_2, \cdots, \beta_N)$, the $P$-representation of quantum state and quantum characteristic function are mutually related by Fourier transformation \cite{carmichael_statistical_2002}:
\begin{align}
    P (\boldsymbol{\beta}, \boldsymbol{\beta}^{\ast}) & = \frac{1}{\pi^2}  \int \chi (\boldsymbol{z}, \boldsymbol{z}^{\ast}) e^{- i \boldsymbol{z} \cdot \boldsymbol{\beta}^{\ast}} e^{- i \boldsymbol{z}^{\ast} \boldsymbol{\beta}}~d^2 \boldsymbol{z}, \label{eq:chi2P} \\
    \chi (\boldsymbol{z}, \boldsymbol{z}^{\ast}) & = \int P (\boldsymbol{\beta}, \boldsymbol{\beta}^{\ast}) e^{ i \boldsymbol{z} \cdot \boldsymbol{\beta}^{\ast}} e^{ i \boldsymbol{z}^{\ast} \boldsymbol{\beta}}~d^2 \boldsymbol{\beta} . \label{eq:P2chi}
\end{align}
where the inner product is conventionally $\boldsymbol{z} \cdot \boldsymbol{\beta}^{\ast} = z^{\ast}_1 \beta_1^{\ast} + \cdots + z^{\ast}_N \beta_N^{\ast}$ and $\boldsymbol{z}^{\ast} \cdot \boldsymbol{\beta} = z_1 \beta_1 + \cdots + z_N \beta_N$. These result ensures the one-to-one correspondence between density matrix and quantum cumulants.

Importantly, the truncation of cumulants to certain discrete orders naturally characterizes some special types of quantum states. In this section, we will show that states with only nonzero first order cumulants correspond to coherent states, while states with only nonzero first and second cumulants orders correspond to Gaussian states.

We first consider a product of coherent states $\ket{\bm{\beta}_0} \equiv \ket{\beta_1, \beta_2, \cdots, \beta_N}$. In Eq.~(\ref{eq:GenNmodecumulants}), we then set $\hat{\rho} = \ket{\boldsymbol{\beta}_0} \bra{\boldsymbol{\beta}_0}$, following which the RHS gives
\begin{align}
    & \ln \mathrm{tr} \left\{ \ket{\boldsymbol{\beta}_0} \bra{\boldsymbol{\beta}_0} e^{iz^{\ast}_1 \hat{b}_1^{\dagger}} \cdots e^{iz^{\ast}_N \hat{b}_N^{\dagger}} e^{iz_1 \hat{b}_1} \cdots e^{iz_N \hat{b}_N} \right\} \nonumber\\
    =~& \ln \left( \bra{\boldsymbol{\beta}_0} e^{iz^{\ast}_1 \hat{b}_1^{\dagger}} \cdots e^{iz^{\ast}_N \hat{b}_N^{\dagger}} e^{iz_1 \hat{b}_1} \cdots e^{iz_N \hat{b}_N} \ket{\boldsymbol{\beta}_0} \right) \nonumber\\
    =~& \ln e^{iz^{\ast}_1 \beta_1^{\ast} + \cdots + iz^{\ast}_N \beta_N^{\ast} + iz_1 \beta_1 + \cdots + iz_N \beta_N} \nonumber\\
    =~& iz^{\ast}_1 \beta_1^{\ast} + \cdots + iz^{\ast}_N \beta_N^{\ast} + iz_1 \beta_1 + \cdots + iz_N \beta_N,
\end{align}
From Eq.~(\ref{eq:Nmodecumulants}), we obtain non-trivial derivative contributions only for first-order derivatives. Thus for all $k \in \{1,2,\cdots,N\}$, we have $C_{b_k} = \beta_k$ and $C_{b_k^{\dagger}} = \beta_k^{\ast}$, so that 
\begin{subequations}
\begin{align}
    \left(C_{b_1}, C_{b_2}, \cdots, C_{b_N} \right) & = \boldsymbol{\beta}_0, \\ 
    \left(C_{b^{\dagger}_1}, C_{b^{\dagger}_2}, \cdots, C_{b^{\dagger}_N} \right) & = \boldsymbol{\beta}^{\ast}_0
\end{align}
\end{subequations}

We can also show that physical states with vanishing cumulants of order greater than one are necessarily coherent states. We begin with the normal-ordered characteristic function for such states, which is given by $\chi (\boldsymbol{z}, \boldsymbol{z}^{\ast}) = e^{G (i \boldsymbol{z}^{\ast}, i \boldsymbol{z})} = e^{i \boldsymbol{z} \cdot \boldsymbol{\beta}^{\ast}_0 + i \boldsymbol{z}^{\ast} \cdot
\boldsymbol{\beta}_0}$. The corresponding $P$-representation is simply the Fourier transform of the characteristic function, and is given by
\begin{align}
    P (\boldsymbol{\beta}, \boldsymbol{\beta}^{\ast}) & = \frac{1}{\pi^2} \int \chi(\boldsymbol{z}, \boldsymbol{z}^{\ast}) e^{- i \boldsymbol{z} \cdot \boldsymbol{\beta}^{\ast}} e^{- i \boldsymbol{z}^{\ast} \cdot \boldsymbol{\beta}}~d^2\boldsymbol{z} \nonumber\\ 
    & = \frac{1}{\pi^2}  \int e^{i \boldsymbol{z} \cdot \boldsymbol{\beta}^{\ast}_0 + i \boldsymbol{z}^{\ast} \cdot \boldsymbol{\beta}_0} e^{- i \boldsymbol{z} \cdot \boldsymbol{\beta}^{\ast}} e^{- i \boldsymbol{z}^{\ast} \cdot \boldsymbol{\beta}}~d^2 \boldsymbol{z} \nonumber\\
    & = \delta (\boldsymbol{\beta} - \boldsymbol{\beta}_0),
\end{align}
which describes a product of coherent states. As a result, the states with all cumulants or order greater than one vanishing are exactly the collection of all coherent states.

Next, we consider quantum states whose quasi-probability distribution is Gaussian. Recall that the $P$, $Q$, and Wigner phase-space representations are related by the Weierstrass transformation, which always maps one Gaussian distribution to another Gaussian distribution \cite{carmichael_statistical_2002}: 
\begin{align}
    W (\boldsymbol{\beta}, \boldsymbol{\beta}^{\ast}) & = \frac{2}{\pi}  \int d^2\boldsymbol{\widetilde{\beta}}~e^{- 2 | \boldsymbol{\beta} - \boldsymbol{\widetilde{\beta}} |^2} P (\boldsymbol{\widetilde{\beta}}, \boldsymbol{\widetilde{\beta}}^{\ast}) , \\
    Q (\boldsymbol{\beta}, \boldsymbol{\beta}^{\ast}) & = \frac{2}{\pi}  \int d^2\boldsymbol{\widetilde{\beta}}~e^{- 2 | \boldsymbol{\beta} - \boldsymbol{\widetilde{\beta}} |^2} W (\boldsymbol{\widetilde{\beta}}, \boldsymbol{\widetilde{\beta}}^{\ast}). 
\end{align}
The relation between the $P$-representation and characteristic function in Eqs.~(\ref{eq:chi2P}),~(\ref{eq:P2chi}) respectively is via Fourier transform, which also maps one Gaussian distribution to another Gaussian distribution. 

Via this series of transformations, we therefore see that a quantum state with a Gaussian Wigner distribution will also have a Gaussian characteristic function, and hence must be entirely defined by cumulants of up to second-order (by definition of a Gaussian distribution).

\section{Working with cumulants - TEOMs for single coherently-driven Kerr oscillator}
\label{app:mToC}

Our approach requires the expression of moments of a multimode quantum system in terms of their associated cumulants. Moments and cumulants are related via their generating functions, which allows one to write a general prescription for writing an arbitrary normal-ordered joint system moment of order $n$ in terms of a series of cumulants of order $1$ up to $n$. This takes the specific form given by Eq.~(\ref{eq:cumulant2momentN}),
\begin{align}
    \avg{\hat{o}_1\hat{o}_2\ldots \hat{o}_n} = \sum_{\pi}\prod_{B\in \pi} C_{o_i : i \in B},
    \label{appeq:mtoC}
\end{align}
where $\pi$ defines the set of possible partitions of operators in the $n$-order moment, and $B$ indicates elements in this set of partitions. The rather opaque form can be simplified by considering a specific case. We will use the above prescription to obtain the Eqs.~(\ref{eq:b})-(\ref{eq:CbdbFull}) in the main text for the single-node QRC. This minimal system already includes the key nonlinear element of the QRC, and therefore the most complex terms that need to be analyzed will arise already in these equations. We start by writing down the equation for $\avg{\hat{b}} = {\rm tr}\{\mathcal{L}\rhou\hat{b}\}$,
\begin{align}
    \avg{\dot{\hat{b}}} = \left(i\Delta - \frac{\widetilde{\gamma}}{2} \right)\avg{\hat{b}} + i\Lambda \avg{\hat{b}^{\dagger}\hat{b}\hat{b}} - i\eta.
    \label{appeq:b}
\end{align}
To rewrite the above equation in terms of cumulants, we must express the the third-order moment $\avg{\hat{b}^{\dagger}\hat{b}\hat{b}}$ in terms of cumulants. The possible partitions of the third-order moment are given by: $\pi \in \{(\hat{b}^{\dagger}\hat{b}\hat{b}),(\hat{b}^{\dagger}\hat{b},\hat{b})\times 2,(\hat{b}\hat{b},\hat{b}^{\dagger}),(\hat{b}^{\dagger},\hat{b},\hat{b}) \}$. Then, using Eq.~(\ref{appeq:mtoC}), we obtain
\begin{align}
    \avg{\hat{b}^{\dagger}\hat{b}\hat{b}} = C_{b^{\dagger}bb} + 2C_{b^{\dagger}b}\avg{\hat{b}} + C_{bb}\avg{\hat{b}^{\dagger}} + \avg{\hat{b}^{\dagger}}\avg{\hat{b}}^2.
\end{align}
Using the above, we can now rewrite Eq.~(\ref{appeq:b}) in terms of cumulants as
\begin{align}
    \avg{\dot{\hat{b}}} &= \left(i\Delta - \frac{\widetilde{\gamma}}{2} \right)\avg{\hat{b}} + i\Lambda \avg{\hat{b}^{\dagger}}\avg{\hat{b}}^2 - i\eta \nonumber \\
    &~~~~+i\Lambda\left(C_{bb}\avg{\hat{b}^{\dagger}} + 2C_{b^{\dagger}b}\avg{\hat{b}} + C_{b^{\dagger}bb} \right)
\end{align}
which is as given in Eq.~(\ref{eq:b}). 

To similarly obtain dynamical equations for cumulants $C_{b^{\dagger}b}, C_{bb}$, we write equations of motion for the second-order moments $\avg{\hat{b}^{\dagger}\hat{b}}$ and $\avg{\hat{b}\hat{b}}$ respectively. Starting with the former, we find:
\begin{align}
    \frac{d}{dt} \avg{\hat{b}^{\dagger}\hat{b}} &= -\widetilde{\gamma}\avg{\hat{b}^{\dagger}\hat{b}} +i\eta\avg{\hat{b}} - i\eta\avg{\hat{b}^{\dagger}}  \nonumber \\
    &= -\widetilde{\gamma}C_{b^{\dagger}b} - \widetilde{\gamma}\avg{\hat{b}^{\dagger}}\avg{\hat{b}} + i\eta\avg{\hat{b}} - i\eta\avg{\hat{b}^{\dagger}}
\end{align}
where in the second line we rewrite the second-order moment in terms of cumulants. From this we can easily write $dC_{b^{\dagger}b} = d\avg{\hat{b}^{\dagger}\hat{b}} - (d\avg{\hat{b}^{\dagger}})\avg{\hat{b}} - (d\avg{\hat{b}})\avg{\hat{b}^{\dagger}}$, which yields the dynamical equation for $C_{b^{\dagger}b}$:
\begin{align}
    \dot{C}_{b^{\dagger}b} &= -\widetilde{\gamma}C_{b^{\dagger}b} -i\Lambda\left(C_{bb}\avg{\hat{b}^{\dagger}}^2 - C_{bb}^*\avg{\hat{b}}^2 \right) \nonumber \\ &~~~~- i\Lambda \left( C_{b^{\dagger}bb}\avg{\hat{b}^{\dagger}} - C_{b^{\dagger}bb}^*\avg{\hat{b}}   \right)
\end{align}
which is as written in Eq.~(\ref{eq:CbdbFull}) of the main text.

Finally, we write the equation of motion for $\avg{\hat{b}\hat{b}}$,
\begin{align}
    \frac{d}{dt} \avg{\hat{b}\hat{b}} &= \left(2i\Delta-\widetilde{\gamma}\right)\avg{\hat{b}\hat{b}} -2i\eta\avg{\hat{b}} + i\Lambda \avg{\hat{b}\hat{b}} + i2\Lambda\avg{\hat{b}^{\dagger}\hat{b}\hat{b}\hat{b}} \nonumber \\
    &= \left(2i\Delta-\widetilde{\gamma}+i\Lambda\right)(C_{bb}+\avg{\hat{b}}\avg{\hat{b}}) -2i\eta\avg{\hat{b}} + i2\Lambda\avg{\hat{b}^{\dagger}\hat{b}\hat{b}\hat{b}}
\end{align}
where we once again rewrite second-order moments in terms of cumulants in the second line. We now need to express the remaining fourth-order moment $\avg{\hat{b}^{\dagger}\hat{b}\hat{b}\hat{b}}$ in terms of cumulants. The possible partitions of this fourth-order moment are given by $\pi \in \{(\hat{b}^{\dagger}\hat{b}\hat{b}\hat{b}),(\hat{b}^{\dagger}\hat{b}\hat{b},\hat{b})\times 3,(\hat{b}\hat{b}\hat{b},\hat{b}^{\dagger}),(\hat{b}^{\dagger}\hat{b},\hat{b}\hat{b})\times 3,(\hat{b}^{\dagger}\hat{b},\hat{b},\hat{b})\times 3,(\hat{b}\hat{b},\hat{b}^{\dagger},\hat{b})\times 3,(\hat{b}^{\dagger},\hat{b},\hat{b},\hat{b}) \}$, so that
\begin{align}
    \avg{\hat{b}^{\dagger}\hat{b}\hat{b}\hat{b}} &= C_{{b}^{\dagger}{b}{b}{b}} +3\avg{\hat{b}}C_{b^{\dagger}bb} + \avg{\hat{b}^{\dagger}}C_{bbb} \nonumber \\
    &+ 3C_{b^{\dagger}b}C_{bb} + 3C_{b^{\dagger}b}\avg{\hat{b}}^2 + 3C_{bb}|\avg{\hat{b}}|^2 + \avg{\hat{b}^{\dagger}}\avg{\hat{b}}\avg{\hat{b}}\avg{\hat{b}}
\end{align}

From this we can finally write $dC_{bb} = d\avg{\hat{b}\hat{b}} - 2(d\avg{\hat{b}})\avg{\hat{b}}$, which yields the dynamical equation for $C_{bb}$:
\begin{align}
    \dot{C}_{bb} &= \left(2i\Delta-\widetilde{\gamma}+i\Lambda\right)(C_{bb}+\avg{\hat{b}}^2) -2i\eta\avg{\hat{b}} \nonumber \\
    &~~~~+ i2\Lambda\left( C_{{b}^{\dagger}{b}{b}{b}} +3\avg{\hat{b}}C_{b^{\dagger}bb} + \avg{\hat{b}^{\dagger}}C_{bbb} \right) \nonumber \\
    &~~~~+ i2\Lambda\left( 3C_{b^{\dagger}b}C_{bb} + 3C_{b^{\dagger}b}\avg{\hat{b}}^2 + 3C_{bb}|\avg{\hat{b}}|^2 + \avg{\hat{b}^{\dagger}}\avg{\hat{b}}^3 \right) \nonumber \\
    &~~~~-2\left(i\Delta - \frac{\widetilde{\gamma}}{2} \right)\avg{\hat{b}}^2 - i2\Lambda \avg{\hat{b}^{\dagger}}\avg{\hat{b}}^3 + 2i\eta\avg{\hat{b}} \nonumber \\
    &~~~~-i2\Lambda\left(C_{bb}|\avg{\hat{b}}|^2 + 2C_{b^{\dagger}b}\avg{\hat{b}}^2 + C_{b^{\dagger}bb}\avg{\hat{b}} \right)
\end{align}
which finally simplifies to:
\begin{align}
    &\dot{C}_{bb} = \nonumber \\
    &\left(2i\Delta-\widetilde{\gamma}+i\Lambda\right)C_{bb} +i\Lambda\avg{\hat{b}}^2 +i 4\Lambda|\avg{\hat{b}}|^2C_{bb} + i 6\Lambda C_{b^{\dagger}b}C_{bb} \nonumber \\
    &+i2\Lambda C_{b^{\dagger}b}\avg{\hat{b}}^2+i2\Lambda C_{b^{\dagger}bbb} + i4\Lambda C_{b^{\dagger}bb}\avg{\hat{b}} +i 2\Lambda C_{bbb}\avg{\hat{b}^{\dagger}}
\end{align}
which can be re-arranged to the form written in Eq.~(\ref{eq:CbbFull}) of the main text.



Our computer algebra approach automates the above process of calculating contributions to equations of motion and expressing arbitrary moments in terms of cumulants, thus allowing the systematic truncation necessary to arrive at TEOMs (and STEOMs) introduced in the main text.

\section{Calculating STEOMs - stochastic measurement contributions to truncated cumulant dynamics}
\label{app:stochcumulants}

To illustrate the salient features of the calculation, a single homodyne measurement superoperator suffices:
\begin{align}
    \mathcal{S}_{k,\rm meas}(dW)\rhou = \sqrt{\frac{\gamma_k}{2}} \left( \bk{k} \rhou + \rhou \bkd{k} - \avg{\bk{k}+\bkd{k}} \right) dW_{k}^{X}(t)
\end{align}
For convenience, we will then write a general stochastic master equation in the form:
\begin{align}
    d\rhoc = \mathcal{L}\rhoc~dt + \mathcal{S}_{k,\rm meas}(dW)\rhoc
\end{align}
where $\mathcal{L}$ defines all contributions to the master equation governing deterministic evolution. We can thus write for the differential of conditional expectation of an arbitrary operator $\hat{o}$:
\begin{align}
    d\avgc{\hat{o}} = {\rm tr} \{\mathcal{L}\rhoc\hat{o}\}~dt + \sqrt{\frac{\gamma_k}{2}} \left( \Cc{o b_k} + \Cc{b_k^{\dagger}o} \right)dW_{k}^{X}(t)
    \label{appeq:do}
\end{align}
Analogously, we can write for arbitrary second-order moments:
\begin{align}
    &d\avgc{\hat{o}_1\hat{o}_2} = {\rm tr} \{\mathcal{L}\rhoc\hat{o}_1\hat{o}_2\}~dt + \sqrt{\frac{\gamma_k}{2}}\times \nonumber \\
    &\left(\! \avgc{\hat{o}_1\hat{o}_2\hat{b}_k} \!+\! \avgc{\hat{b}_k^{\dagger}\hat{o}_1\hat{o}_2} \!-\! \avgc{\hat{o}_1\hat{o}_2}\avgc{\hat{b}_k} \!-\! \avgc{\hat{b}_k^{\dagger}}\avgc{\hat{o}_1\hat{o}_2} \! \right)\! dW_{k}^{X}(t)
    \label{appeq:do1o2}
\end{align}

We will now use the above results obtain the equation of motion for an arbitrary \textit{normal-ordered} cumulant $\Cc{o_1o_2} = \avgc{\hat{o}_1\hat{o}_2} - \avgc{\hat{o}_1}\avgc{\hat{o}_2}$. The important step arises in writing the differential of this cumulant using Ito's lemma:
\begin{align}
    d\Cc{o_1o_2} &= d\avgc{\hat{o}_1\hat{o}_2} - (d\avgc{\hat{o}_1})\avgc{\hat{o}_2} - \avgc{\hat{o}_1}(d\avgc{\hat{o}_2}) \nonumber \\
    & - (d\avgc{\hat{o}_1})(d\avgc{\hat{o}_2})
    \label{appeq:dco1o2}
\end{align}
where the term on the second line arises from the modified chain rule in Ito calculus. We can now proceed to obtaining the individual contributions to the above equation.

We start with the first term in Eq.~(\ref{appeq:dco1o2}), which was calculated in Eq.~(\ref{appeq:do1o2}). To proceed, it will prove convenient to write moments higher than first-order in terms of their corresponding cumulants. We begin with the second-order moments:
\begin{align}
    &d\avgc{\hat{o}_1\hat{o}_2} = {\rm tr} \{\mathcal{L}\rhoc\hat{o}_1\hat{o}_2\}~dt + \nonumber \\
    &\sqrt{\frac{\gamma_k}{2}}\Big( \avgc{\hat{o}_1\hat{o}_2\hat{b}_k} + \avgc{\hat{b}_k^{\dagger}\hat{o}_1\hat{o}_2} - \Cc{{o}_1{o}_2}\avgc{\hat{b}_k} - \avgc{\hat{b}_k^{\dagger}}\Cc{{o}_1{o}_2} \nonumber \\
    &- \avgc{\hat{o}_2}\avgc{\hat{o}_1}\avgc{\hat{b}_k} - \avgc{\hat{o}_2}\avgc{\hat{o}_1}\avgc{\hat{b}_k^{\dagger}}  \Big)dW_{k}^{X}(t)
\end{align}
Next, we write down the expressions relating third-order moments to their corresponding cumulants using Eq.~(\ref{appeq:mtoC}):
\begin{subequations}
\begin{align}
    \Cc{o_1o_2b_k} &= \avgc{\hat{o}_1\hat{o}_2\hat{b}_k} - \Cc{{o}_1{o}_2}\avgc{\hat{b}_k} - \Cc{{o}_1{b}_k}\avgc{\hat{o}_2} - \Cc{{o}_2{b}_k}\avgc{\hat{o}_1} \nonumber \\
    & - \avgc{\hat{o}_2}\avgc{\hat{o}_1}\avgc{\hat{b}_k} \\
    \Cc{b_k^{\dagger}o_1o_2} &= \avgc{\hat{b}_k^{\dagger}\hat{o}_1\hat{o}_2} - \Cc{{o}_1{o}_2}\avgc{\hat{b}^{\dagger}_k} - \Cc{{b}_k^{\dagger}{o}_1}\avgc{\hat{o}_2} - \Cc{{b}_k^{\dagger}{o}_2}\avgc{\hat{o}_1} \nonumber \\
    & - \avgc{\hat{o}_2}\avgc{\hat{o}_1}\avgc{\hat{b}^{\dagger}_k}
\end{align}
\end{subequations}
employing which, Eq.~(\ref{appeq:dco1o2}) becomes:
\begin{align}
    &d\avgc{\hat{o}_1\hat{o}_2} = {\rm tr} \{\mathcal{L}\rhoc\hat{o}_1\hat{o}_2\}~dt + \nonumber \\
    &\sqrt{\frac{\gamma_k}{2}}\Big( \Cc{o_1o_2b_k} + \Cc{{o}_1{b}_k}\avgc{\hat{o}_2} + \Cc{{o}_2{b}_k}\avgc{\hat{o}_1}  \Big)dW_{k}^{X}(t) \nonumber \\
    +&\sqrt{\frac{\gamma_k}{2}}\Big( \Cc{b_k^{\dagger}o_1o_2} + \Cc{{b}_k^{\dagger}{o}_1}\avgc{\hat{o}_2} + \Cc{{b}_k^{\dagger}{o}_2}\avgc{\hat{o}_1}   \Big)dW_{k}^{X}(t)
    \label{appeq:do1o2C}
\end{align}
where terms involving only first-order moments cancel. Note that the above includes terms of up to $O(dt)$, as $dW_k^X$ is formally $O(dt^{1/2})$.

Next, we can calculate the second and third terms in Eq.~(\ref{appeq:dco1o2}), using Eq.~(\ref{appeq:do}). These take the simple forms:
\begin{subequations}
\begin{align}
    (d\avgc{\hat{o}_1})\avgc{\hat{o}_2} &= \avgc{\hat{o}_2}{\rm tr} \{\mathcal{L}\rhoc\hat{o}_1\}~dt \nonumber \\
    &+ \sqrt{\frac{\gamma_k}{2}} \left( \Cc{o_1 b_k}\avgc{\hat{o}_2} + \Cc{b_k^{\dagger}o_1}\avgc{\hat{o}_2} \right)dW_{k}^{X}(t) \label{appeq:do1xo2} \\
    (d\avgc{\hat{o}_2})\avgc{\hat{o}_1} &= \avgc{\hat{o}_1}{\rm tr} \{\mathcal{L}\rhoc\hat{o}_2\}~dt \nonumber \\
    &+ \sqrt{\frac{\gamma_k}{2}} \left( \Cc{o_2 b_k}\avgc{\hat{o}_1} + \Cc{b_k^{\dagger}o_2}\avgc{\hat{o}_1} \right)dW_{k}^{X}(t) \label{appeq:do2xo1}
\end{align}
\end{subequations}
Both equations above once again contain terms up to $O(dt)$. 

Finally, we can write down the term arising from Ito's lemma. The requiring differential is that given by Eq.~(\ref{appeq:do}). However, we need only retain the lowest $O(dt)$ term here, which is given by:
\begin{align}
    &(d\avgc{\hat{o}_1})(d\avgc{\hat{o}_2}) =\nonumber \\
    &\frac{\gamma_k}{2}\left( \Cc{o_1 b_k} + \Cc{b_k^{\dagger}o_1} \right)\left( \Cc{o_2 b_k} + \Cc{b_k^{\dagger}o_2} \right)~dt + O(dt^{3/2})
    \label{appeq:do1do2}
\end{align}
where we have used $(dW_k^X)^2 = dt$.

Finally, we can write down Eq.~(\ref{appeq:dco1o2}) by combining Eqs.~(\ref{appeq:do1o2C}),~(\ref{appeq:do1xo2}),~(\ref{appeq:do2xo1}), and (\ref{appeq:do1do2}). This finally yields:
\begin{align}
    &d\Cc{o_1o_2} = {\rm tr} \{\mathcal{L}\rhoc\hat{o}_1\hat{o}_2 \!-\! \avgc{\hat{o}_2}\mathcal{L}\rhoc\hat{o}_1  \!-\! \avgc{\hat{o}_1}\mathcal{L}\rhoc\hat{o}_2 \}dt \nonumber \\
    &~~~~~~~~~~~~- \frac{\gamma_k}{2}\left( \Cc{o_1 b_k} + \Cc{b_k^{\dagger}o_1} \right)\left( \Cc{o_2 b_k} + \Cc{b_k^{\dagger}o_2} \right)dt \nonumber \\
    &~~~~~~~~~~~~+\sqrt{\frac{\gamma_k}{2}}\left( \Cc{o_1o_2b_k} + \Cc{b_k^{\dagger}o_1o_2} \right)dW_{k}^{X}(t)
\end{align}
The first term on the right hand side describes deterministic dynamics governed by $\mathcal{L}$. The second term is due to the measurement, but does not depend on the stochastic Wiener increment $dW_k^X(t)$. The last term is a stochastic contribution; however note that it is related only to third-order cumulants. Within the ans\"atz used in this paper, this explicitly stochastic term vanishes.

\section{Complex-$P$ representation of single coherently-driven QRC node}
\label{app:complexP}

The complex-$P$ representation associated phase-space variables $(\beta,\beta^{\dagger})$ with operators $(\hat{b},\hat{b}^{\dagger})$. For a single node of our QRC model, defined as a single coherently-driven Kerr oscillator coupled to a zero temperature bath via Eq.~(\ref{eq:singleQRCME}), which we reproduce explicitly,
\begin{align}
     \Ltot\hat{\rho} = &-i\left[-\Delta\hat{b}^{\dagger}\hat{b} - \frac{\Lambda}{2}\hat{b}^{\dagger}\hat{b}^{\dagger}\hat{b}\hat{b} + \eta(e^{-i\varphi_{\eta}}\hat{b} + e^{i\varphi_{\eta}}\hat{b}^{\dagger}),\hat{\rho} \right] \nonumber \\
     &+ (\gamma+\Gamma) \mathcal{D}[\hat{b}]\hat{\rho}.
\end{align}
We note first the the drive phase $\varphi_{\eta}$ can simply be absorbed into the definition of the operators $\hat{b} \to \hat{b}e^{i\varphi_{\eta}},\hat{b}^{\dagger} \to \hat{b}^{\dagger}e^{-i\varphi_{\eta}}$, while leaving their commutator unchanged. The drive amplitude can thus be chosen to be completely real. The steady-state complex-$P$ distribution $P_{\rm ss}(\beta,\beta^{\dagger})$ for the resulting system, in the phase space of variables $\beta,\beta^{\dagger}$ associated with operators $\hat{b},\hat{b}^{\dagger}$ respectively, can be found exactly by the method of potentials~\cite{drummond_quantum_1980, gardiner_stochastic_2009}, and takes the form
\begin{align}
    &P_{\rm ss}(\beta,\beta^{\dagger}) = \nonumber \\
    &(\beta)^{(c-2)}(\beta^{\dagger})^{\left(c^*-2\right)}\exp\left\{\frac{2\eta}{\Lambda}\!\left(\frac{1}{\beta}+\frac{1}{\beta^{\dagger}} \right) + 2\beta^{\dagger}\beta \right\}
\end{align}
where 
\begin{align}
    c = \frac{-i\Delta + \frac{\gamma+\Gamma}{2}}{-i\Lambda}.
\end{align}
Knowledge of the exact steady-state complex $P$-distribution allows one to calculate arbitrary moments of the quantum steady-state of the driven nonlinear mode by integrating over complex phase space. We find for arbitrary normal-ordered steady-state moments~\cite{drummond_quantum_1980}
\begin{align}
    &\avg{(\hat{b}^{\dagger})^j(\hat{b})^i} = \nonumber \\ 
    & e^{-i\varphi_{\eta}(i-j)}\left|\frac{2\eta}{\Lambda}\right|^2\!\!\!\! \frac{\Gamma(c)\Gamma(c^*)}{\Gamma(c+i)\Gamma(c^*+j)}\frac{h(c+i,c^*+j,8|\eta/\Lambda|^2)}{h(c,c^*,8|\eta/\Lambda|^2)},
    \label{eq:prepmoments}
\end{align}
where we have reintroduced the drive phase by simply undoing the earlier transformation on operators $\hat{b},\hat{b}^{\dagger}$. Here the function $h(x,y,z)$ is given by the hypergeometric series
\begin{align}
    h(x,y,z) = \sum_{n=0}^{\infty} \frac{z^n}{n!}\frac{\Gamma(x)\Gamma(y)}{\Gamma(x+n)\Gamma(y+n)}
\end{align}
and $\Gamma(x)$ is the Gamma function.

Eq.~(\ref{eq:prepmoments}) is used to calculate steady-state first-order moments and cumulants plotted in Fig.~\ref{fig:singleQRCCumulants} in Sec.~\ref{sec:singleNodeQRC} of the main text.

\section{Supplementary benchmarking simulations}
\label{app:verify}

In Fig.~\ref{fig:singleQRCSim} of the main text, we provided results benchmarking the performance of STEOMs against full SME simulations for a single coherently-driven QRC node undergoing heterodyne measurement. In this appendix section, we present results for a more complex measurement chain that includes the key features common to the classification tasks considered in this paper: inclusion of a measured quantum system and its coupling to the QRC.

\begin{figure}[t]
    \centering
    \includegraphics[scale=1.0]{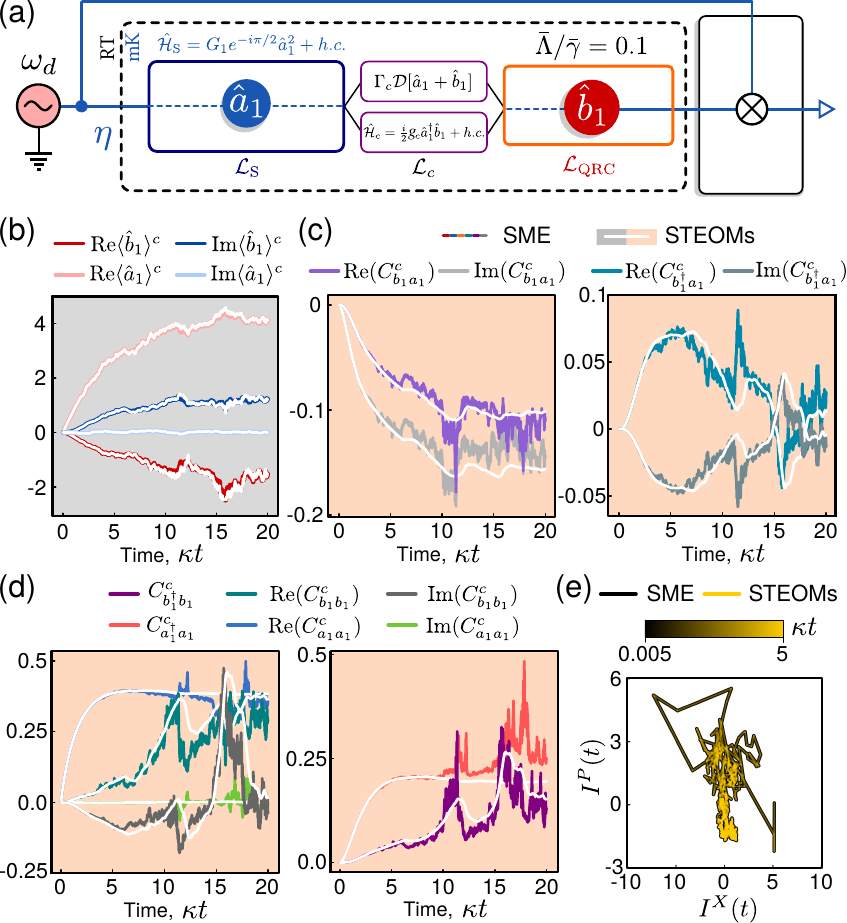}
    \caption{Benchmarking STEOMs against full SME simulations for the two-mode measurement chain depicted in (a). The single-mode measured system is an amplifier with parameters given by $\kappa_1 = \kappa/2$, $G_{1}/\kappa = 0.3$, and $\Delta_1/\kappa = 0.0$. The amplifier is coupled non-reciprocally (by setting $\varg_c = \Gamma_c = \kappa/2$) to a single-node QRC undergoing continuous measurement. The QRC parameters are given by $\bar{\gamma} = \kappa$, $\bar{\Lambda}/\bar{\gamma} = 0.1$, $\bar{\Delta}/\bar{\gamma} = -1.0$, such that $\CNLE = 0.385$. Simulation results for a single measurement trajectory showing (b) first-order cumulants, (c) second-order cross-cumulants, (d) second-order self-cumulants, and (e) measured quadratures $\{I^X(t),I^P(t)\}$.}
    \label{fig:appverification}
\end{figure}

The specific measurement chain we consider is a simplified version of the full measurement chain employed for the task of amplifier state classification in Fig.~\ref{fig:classifyAmpStates}(a) of the main text, and is depicted in Fig.~\ref{fig:appverification}(a). More precisely, it is defined by Eq.~(\ref{eq:sme}), with the system dynamics governed by Eq.~(\ref{eq:lsysamp}) but with $G_{12} = 0.0$, thereby reducing the amplifier system to a single mode $\hat{a}_1$. The rest of the measurement chain is unchanged: the mode $\hat{a}_1$ is coupled to the single QRC node via a directional hopping interaction given by Eq.~(\ref{eq:couplingCirc}), and the QRC node undergoes continuous measurement as defined by Eq.~(\ref{eq:Lqrc}) with $K=1$.


For fair benchmarking comparisons, we choose parameters where the QRC enables classification of the amplifier states; from results in the main text, we choose $\bar{\Delta}/\bar{\gamma} = -1.0$ and $\CNLE = 0.385$.  Unfortunately, increasing the number of modes immediately makes the SME simulations much more computationally taxing. To lower the modal occupation numbers, we keep $\CNLE$ fixed while increasing the nonlinearity strength to $\bar{\Lambda}/\bar{\gamma} = 0.1$, and simultaneously decreasing the drive strength to $\eta/\bar{\gamma} = 0.894$. We note that this nonlinearity strength is about five times stronger than the largest nonlinearity used in the main text. All other parameters of the measurement chain are summarized in the caption of Fig.~\ref{fig:appverification}.

The complete measurement chain of $N_{\rm T}=2$ quantum modes is determined by $2N_{\rm T}^2+3N_{\rm T}=14$ independent degrees of freedom: $\{\avgc{\hat{b}_1},\avgc{\hat{a}_1},\Cc{b_1b_1},\Cc{a_1a_1},\Cc{b_1a_1},\Cc{b_1^{\dagger}a_1} \}$ and their complex conjugates, as well as the real-valued cumulants $\{\Cc{b_1^{\dagger}b_1},\Cc{a_1^{\dagger}a_1}\}$.  We plot these quantities in Fig.~\ref{fig:appverification}(b)-(e) for a single measurement trajectory obtained using the STEOMs (white) and full SME simulations (color). We find that even though the nonlinearity is much stronger than those considered in the main text, there is in general good agreement between the methods. We also emphasize that for the same time step, SME simulations with a Hilbert space cutoff of $40$ photons per quantum mode take about 37 hours to complete~\cite{johansson_qutip_2013}; the STEOMs require 4 minutes.

\section{Classical phase diagram for coherently-driven QRCs}
\label{app:phaseDiagram}

In this appendix section we present the calculations required to obtain the classical phase diagram of the general two-node coherently-driven QRC analyzed in the main text. Our starting point is with the TEOMs describing the \textit{unconditional} dynamics of the two-node QRC within the classical limit (i.e. also neglecting second-order cumulants):
\begin{eqnarray}
  \frac{d}{d t} \langle \hat{b}_1 \rangle & = & \left( +i \Delta_1 - \frac{\gamma_1}{2} \right)\langle \hat{b}_1 \rangle  - i \varg_{12} \langle \hat{b}_2 \rangle - i_{} \eta_1 \nonumber\\
  &  & + i \Lambda_1 \langle \hat{b}_1^{\dagger} \rangle \langle \hat{b}_1 \rangle \langle \hat{b}_1 \rangle, \label{appeq:cb1} \\
  \frac{d}{d t} \langle \hat{b}_2 \rangle & = & \left( +i \Delta_2 - \frac{\gamma_2}{2} \right)\langle \hat{b}_2 \rangle - i \varg_{12} \langle \hat{b}_1 \rangle - i_{} \eta_2, \nonumber\\
  &  & + i \Lambda_2 \langle \hat{b}_2^{\dagger} \rangle \langle \hat{b}_2 \rangle \langle \hat{b}_2 \rangle. \label{appeq:cb2} 
\end{eqnarray}
We now introduce scaled variables that highlight the dependence of the classical dynamics of the two-node QRC on the nonlinearity strength and coherent drives. For convenience we set $\gamma_1 = \gamma_2 = \bar{\gamma}$, and once again perform the scaling introduced in Sec.~\ref{sec:singleNodeQRC}, introducing dimensionless time and analogous energy scales:
\begin{subequations}
\begin{align}
  t' &=  \bar{\gamma} t, \\
  ({\Delta}'_{k},{\Lambda}'_{k},\eta_k',{\varg}'_{12}) &= ({\Delta}_{k},{\Lambda}_{k},\eta_k,{\varg}_{12})/\bar{\gamma}
\end{align}
\end{subequations}
If we now scale $\avg{\hat{b}_{k}}' \to \sqrt{{\Lambda}'_{k}}\avg{\hat{b}_{k}}$, Eqs.~(\ref{appeq:cb1}),~(\ref{appeq:cb2}) transform to:
\begin{eqnarray}
  \frac{d}{d t} \avg{\hat{b}_1}' & = & \left( +i \Delta'_1 - \frac{1}{2} \right)\avg{\hat{b}_1}'  - i \varg'_{12}\sqrt{\frac{\Lambda'_1}{\Lambda'_2}} \avg{\hat{b}_2}' - i_{} \sqrt{\Lambda'_1}\eta'_1 \nonumber\\
  &  & + i  \avg{\hat{b}_1^{\dagger}}'\avg{\hat{b}_1}' \avg{\hat{b}_1}', \label{appeq:cb1S}  \\
  \frac{d}{d t} \avg{\hat{b}_2}'  & = & \left( +i \Delta'_2 - \frac{1}{2} \right)\avg{\hat{b}_2}' - i \varg'_{12}\sqrt{\frac{\Lambda'_2}{\Lambda'_1}} \avg{\hat{b}_1}' - i_{} \sqrt{\Lambda'_2}\eta'_2 \nonumber\\
  &  & + i  \avg{\hat{b}_2^{\dagger}}'\avg{\hat{b}_2}' \avg{\hat{b}_2}'. \label{appeq:cb2S}  
\end{eqnarray}
It is immediately clear that the classical dynamics described by the above equations do \textit{not} depend on the magnitude of the nonlinearity strengths $\Lambda'_k$ \textit{alone}; instead, these appear as a scaling factor to the drive amplitudes, and via their \textit{relative} strength ${\Lambda'_1}/{\Lambda'_2}$. This indicates that \textit{provided} ${\Lambda'_1}/{\Lambda'_2}$ is held fixed, decreasing $\Lambda'_k$ while simultaneously increasing $\eta'_k$ so that $\sqrt{\Lambda'_k}\eta'_k$ remain constant will leave the classical dynamics unchanged, apart from a scaling of node amplitudes $\avg{\hat{b}_k}'$. Analogously to the single-node QRC case analyzed in Sec.~\ref{sec:singleNodeQRC} of the main text, this transformation will reduce the influence of cumulants (not shown here), smoothly transitioning the two-node QRC to an effectively classical regime of operation.

We now specialize to the case of the two-node QRC analyzed in the main text, considering specific parameters , $\Delta'_1 = \Delta'_2 =\bar{\Delta}'$, $\Lambda'_1 = \Lambda'_2 = \bar{\Lambda}'$, and real coupling $\varg'_{12} = \bar{\varg}'$. Additionally, we consider only the $\hat{b}_1$ to be driven with a real-valued drive $(\eta'_1, \eta'_2) = (i\eta'_{\rm eff}, 0)$, $\eta'_{\rm eff} \in \mathbb{R}$, consistent with the coupling scheme employed in the main text. Then, the scaled TEOMs in Eqs.~(\ref{appeq:cb1S}),~(\ref{appeq:cb2S}) and their conjugates simplify to:
\begin{align}
  \frac{d}{d t'} \langle \hat{b}_1 \rangle' & =  \left(+i\bar{\Delta}'- \frac{1}{2}\right) \langle \hat{b}_1 \rangle' - i \bar{\varg}' \langle \hat{b}_2 \rangle' + \CNLE \nonumber\\
  &   + i \langle \hat{b}_1^{\dagger} \rangle' \langle \hat{b}_1 \rangle' \langle \hat{b}_1 \rangle', \label{eq:dynamics-b1} \\
  \frac{d}{d t'} \langle \hat{b}_1^{\dagger} \rangle^{\prime} & =  \left(-i\bar{\Delta}'- \frac{1}{2}\right) \langle \hat{b}_1^{\dagger} \rangle^{\prime} + i \bar{\varg}'  \langle \hat{b}_2^{\dagger} \rangle^{\prime} + \CNLE
  \nonumber\\
  &   - i \langle \hat{b}_1^{\dagger} \rangle^{\prime} \langle \hat{b}_1^{\dagger} \rangle^{\prime } \langle \hat{b}_1 \rangle', \label{eq:dynamics-b1D}\\
  \frac{d}{d t'} \langle \hat{b}_2 \rangle' & = \left(+i\bar{\Delta}'- \frac{1}{2}\right) \langle \hat{b}_2 \rangle' - i
  \bar{\varg}' \langle \hat{b}_1 \rangle' \nonumber \\
  &   + i \langle \hat{b}_2^{\dagger} \rangle^{\prime} \langle \hat{b}_2 \rangle' \langle \hat{b}_2 \rangle', \label{eq:dynamics-b2}\\
  \frac{d}{d t'} \langle \hat{b}_2^{\dagger} \rangle^{\prime} & =  \left(-i\bar{\Delta}'- \frac{1}{2}\right) \langle \hat{b}_2^{\dagger} \rangle' + i \bar{\varg}' \langle \hat{b}_1^{\dagger} \rangle^{\prime} \nonumber\\
  &   - i \langle \hat{b}_2^{\dagger} \rangle^{\prime} \langle \hat{b}_2^{\dagger} \rangle^{\prime} \langle \hat{b}_2 \rangle' . \label{eq:dynamics-b2D}
\end{align}
where $\CNLE$ takes the form given in Eq.~(\ref{eq:cnleCoh}) of the main text. We see once more that the classical dynamics of the two-node QRC depends on the drive and nonlinearity only via the effective nonlinear cooperativity $\CNLE$.


We now wish to analyze the stability of the two-node QRC and construct its classical phase diagram. To this end, we first obtain the steady state classical fixed points $(\langle \hat{b}_1 \rangle_{\mathrm{ss}}', \langle \hat{b}_2 \rangle_{\mathrm{ss}}')$ of the two-node QRC by setting $\frac{d}{d t'} \langle \hat{b}_k \rangle_{\mathrm{ss}}' = 0$ for $k=1,2$: 
\begin{align}
  &\! 0  =  \left( i\bar{\Delta}' - \frac{1}{2} \right)\! \langle \hat{b}_1 \rangle_{\rm ss}'  - i \bar{\varg}' \langle \hat{b}_2 \rangle_{\rm ss}'   + i |\langle \hat{b}_1 \rangle_{\rm ss}^{\prime}|^2 \langle \hat{b}_1 \rangle_{\rm ss}' + \CNLE , \\
  &\! 0  =  \left( i\bar{\Delta}' - \frac{1}{2} \right)\! \langle \hat{b}_2 \rangle_{\rm ss}'  - i
  \bar{\varg}' \langle \hat{b}_1 \rangle_{\rm ss}'  + i |\langle \hat{b}_2 \rangle_{\rm ss}^{\prime}|^2 \langle \hat{b}_2 \rangle_{\rm ss}',
\end{align}
which describes a system of algebraic equations defining the steady-state amplitudes $\langle \hat{b}_k \rangle_{\mathrm{ss}}'$ defining the classical fixed points.

\begin{widetext}
To analyze the stability of these fixed points of the classical QRC dynamics, we evaluate the eigenvalues of Jacobian of the dynamical system. More explicitly, we treat the right hand side of Eq.~(\ref{eq:dynamics-b1}-\ref{eq:dynamics-b2D}) as a map $\mathbb{C}^4 \to \mathbb{C}^4$ with four variables $\langle \hat{b}_1 \rangle'_{\rm ss}, \langle \hat{b}_1^{\dagger} \rangle'_{\rm ss}, \langle \hat{b}_2 \rangle'_{\rm ss}, \langle \hat{b}_2^{\dagger} \rangle'_{\rm ss}$. The first-order derivative or Jacobian of this map gives
\begin{equation}
  J [\avg{\hat{b}_1}'_{\rm ss}, \avg{\hat{b}_2}'_{\rm ss}] 
  = \left(\begin{array}{cccc}
    - \frac{1}{2} + i \bar{\Delta}' + 2 i | \langle \hat{b}_1 \rangle'_{\rm ss} |^2 & i \langle \hat{b}_1 \rangle^{\prime 2}_{\rm ss} & - i \bar{\varg} & 0\\
    - i \langle \hat{b}_1^{\dagger} \rangle^{\prime 2}_{\rm ss} & - \frac{1}{2} - i \bar{\Delta}' - 2 i | \langle \hat{b}_1 \rangle'_{\rm ss} |^2 & 0 & i \bar{\varg} \\
    - i \bar{\varg} & 0 & - \frac{1}{2} + i \bar{\Delta}' + 2 i | \langle \hat{b}_2 \rangle'_{\rm ss} |^2 & i \langle \hat{b}_2 \rangle^{\prime 2}_{\rm ss} \\
    0 & i \bar{\varg} & - i \langle \hat{b}_2^{\dagger} \rangle^{\prime 2}_{\rm ss} & - \frac{1}{2} - i \bar{\Delta}' - 2 i | \langle \hat{b}_2 \rangle'_{\rm ss} |^2
  \end{array}\right).
\end{equation}
\end{widetext}
Note that the Jacobian does not explicitly depend on the effective nonlinearity $\CNLE$; this dependence enters implicitly via the steady state amplitudes $\langle \hat{b}_1 \rangle_{\mathrm{ss}}', \langle \hat{b}_2 \rangle_{\mathrm{ss}}'$. 

The matrix $J [\langle \hat{b}_1 \rangle_{\mathrm{ss}}', \langle \hat{b}_2 \rangle_{\mathrm{ss}}']$ has four complex eigenvalues $\lambda_{s}[\langle \hat{b}_1 \rangle_{\mathrm{ss}}', \langle \hat{b}_2 \rangle_{\mathrm{ss}}']$ where $s=1,2,3,4$ for each fixed point $\langle \hat{b}_1 \rangle_{\mathrm{ss}}', \langle \hat{b}_2 \rangle_{\mathrm{ss}}'$. A fixed point is stable if and only if the largest real parts of all four eigenvalues is smaller than zero (indicating decaying fluctuations),
\begin{equation}
    \max_{s=1,2,3,4} {\mathrm{Re}(\lambda_{s}[\langle \hat{b}_1 \rangle_{\mathrm{ss}}', \langle \hat{b}_2 \rangle_{\mathrm{ss}}'])} < 0. 
\end{equation}
The classical phase diagram is thus simply constructed by calculating the total number of fixed points of the system for given QRC parameters and $\CNLE$, and then counting the number of stable fixed points. The classical bistable region corresponds to regions where two fixed points are stable. Note that the single-node QRC phase diagram can be constructed using the above approach by setting $\bar{\varg} = 0$. 

\section{Pointer state classification as a function of QRC damping hyperparameter $\bar{\gamma}$}
\label{app:damping}

In this appendix section, we explore the role of the QRC nodes damping hyperparameter $\bar{\gamma}$ on the task of classification of cavity states studied in Sec.~\ref{sec:cav}. We have already seen that the classification performance is strongly dependent on where the QRC is being operated with respect to its classical phase diagram, determined by parameters $\{\CNLE, \bar{\varg}/\bar{\gamma},\bar{\Delta}/\bar{\gamma}\}$. The simplest role of $\bar{\gamma}$ is to simply modify these parameters and lead to operation of the QRC in a different part of the classical phase diagram, thereby affecting its performance in the same way as explored in Sec.~\ref{sec:cav}.

However, we also know that $\bar{\gamma}$ controls the relaxation rate of QRC nodes and thus should play some role in the response of the QRC to time-dependent signals such as those relevant for the pointer state classification task. To be able to isolate this effect from that due to different operating points on the classical phase diagram, we must choose parameters that fix the QRC operating regime. This requires holding $\{\CNLE, \bar{\varg}/\bar{\gamma},\bar{\Delta}/\bar{\gamma}\}$ all fixed while $\bar{\gamma}$ is varied. The coupling and detuning hyperparameters can be set independently of any other components in the measurement chain; we therefore set them according to
\begin{align}
    \frac{\bar{\varg}}{\bar{\gamma}} = 1.0, \frac{\bar{\Delta}}{\bar{\gamma}} = 0.0,
\end{align}
while $\bar{\gamma}$ is varied (relative to $\kappa$). However, recall that the effective nonlinearity $\CNLE$, defined in Eq.~(\ref{eq:cnleCoh}),
\begin{align}
    \CNLE = \frac{\eta_{\rm eff}}{\bar{\gamma}}\sqrt{ \frac{\bar{\Lambda}}{\bar{\gamma}} }
\end{align}
depends on the cavity drive and detuning, and the amplification strength, via $\eta_{\rm eff}$, in addition to QRC nonlinearity and damping hyperparameters. We must hold $\CNLE$ fixed \textit{without} modifying cavity or amplifier parameters (thus leaving $\eta_{\rm eff}$ unchanged), since changing either effectively changes the classification task required of the QRC, thus not allowing for a sensible comparison between QRC performance for different $\bar{\gamma}$ values.  Then, when changing $\bar{\gamma}$, we simultaneously modify $\bar{\Lambda}$ such that $\CNLE$ is held fixed. This is achieved by setting
\begin{align}
    \frac{\bar{\Lambda}}{\bar{\gamma}} = 0.005\cdot\left(\frac{\bar{\gamma}}{\kappa} \right)^{\! 2} \implies \CNLE = \frac{\eta_{\rm eff}}{\kappa}\sqrt{0.005} \simeq 1.27
\end{align}
so that $\CNLE$ is now constant regardless of the value of $\bar{\gamma}$; in calculating $\eta_{\rm eff}$, we have used cavity parameters from Sec.~\ref{sec:cav}. In effect, for a fixed drive $\eta_{\rm eff}$, changing the damping rate modifies the QRC internal field amplitude, so that the QRC node nonlinearity must be increased to compensate and still achieve a strongly nonlinear response.  

Of course the reduction in QRC field for increased damping rate will lead to a smaller QRC output field. This is true even though in our formulation, the damping hyperparameter $\bar{\gamma}$ also controls the coupling to the output channel. As a result, the main consequence of our choice of scaling QRC hyperparameters is that we will remain on a fixed position of the \textit{steady-state} classical phase diagram, but with increasing damping rate, the effective output field from the QRC will be reduced.


\begin{figure}[t]
    \centering
    \includegraphics[scale=1.0]{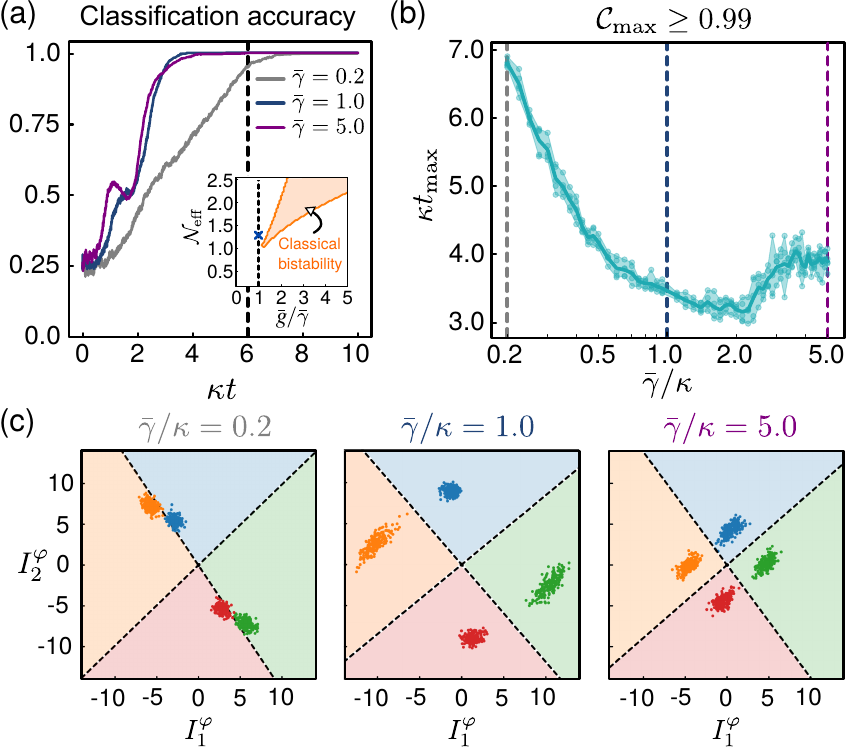}
    \caption{(a) Classification accuracy as a function of time for $\kappa t \in [0,10]$ and specific damping hyperparameter values $\bar{\gamma} \in [0.2,1.0,5.0]\kappa$. Inset shows the classical phase diagram for the 2-node QRC, with the effective operating regime marked by the blue cross. (b) Time to classification $\kappa \tmax$ as a function of damping hyperparameter $\bar{\gamma}$. (c) QRC outputs in measured phase space $\{I_1^{\varphi},I_2^{\varphi}\}$ at time $\kappa t = 6.0$ for the specific damping hyperparameter values in (a). }
    \label{fig:ampDamping}
\end{figure}


To analyze how these various considerations come in to play in determining QRC performance, we obtain the maximum classification accuracy $\cmax$ and the time to reach this maximum $\kappa \tmax$ for the pointer state classification task as before. By virtue of fixing $\CNLE$ so that we are in a nonlinear operating regime, we find that for all $\bar{\gamma}$ values considered, we reach $\cmax \geq 0.99$ within the measurement time of $\kappa t \in [0,10]$. Classification accuracy as a function of time at representative values of $\bar{\gamma} \in [0.2,1.0,5.0]\kappa$ is plotted in Fig.~\ref{fig:ampDamping}(a), indicating the approach to successful classification. From here, however, it is immediately clear that the performance of QRCs as a function of $\bar{\gamma}$ differ in the time taken to achieve perfect classification. 

The extracted value of $\kappa\tmax$ is plotted in Fig.~\ref{fig:ampDamping}(a) as a function of $\bar{\gamma}/\kappa$. We clearly observe a non-monotonic dependence of $\kappa \tmax$ on the damping rate. For $\bar{\gamma} < \kappa$, $\tmax$ increases; this is expected, since the QRC is responding slower than the rate of evolution of the cavity signal. For larger $\bar{\gamma}$, the QRC response time is reduced, leading to a decrease in $\tmax$. We clearly see that a minimum is reached around $\bar{\gamma} \simeq 2\kappa$. For larger $\bar{\gamma}$, we see that $\tmax$ begins to increase once more, even though the QRC response time is still fast relative to the quantum system evolution timescale. 

To understand this, we consider the measured phase space plot of integrated heterodyne records at time $\kappa t= 6.0$ plotted in Fig.~\ref{fig:ampDamping}(c) for the specific values of $\bar{\gamma} \in [0.2,1.0,5.0]\kappa$ from left to right. The plotted phase space area is the same across all panels. We clearly see that for all regimes of damping rate $\bar{\gamma}$ relative to $\kappa$, the QRC is operating in a nonlinear regime enabling classification of the four cavity signals, as arranged by fixing $\CNLE$. For $\bar{\gamma}/\kappa = 0.2$, the slow QRC response time means that this separation occurs much more slowly in time. For $\bar{\gamma}/\kappa = 5.0$, the states are well-separated into distinct areas of phase space; however we see clearly that the means of the output distributions are smaller than their values at lower QRC damping rates. This is because for fixed input signal strength $\eta$, the output field from the QRC is reduced in amplitude. As a result, the signal-to-noise ratio of measured quadratures is reduced for the same integration time. This manifests in a requirement of increased time $\tmax$ to successfully classify the cavity states.

\section{Quantum states for amplifier state classification task: additional details}
\label{app:amp}

In this section, we will analyze the unconditional dynamics of the two-mode amplifier analyzed in Sec.~\ref{sec:amp} to justify parameter choices that general Gaussian states of the amplifier with equal first-order moments but distinct second-order cumulants.

We begin by writing down the \textit{unconditional} equations of motion for the amplifier mode expectation values; these are given by
\begin{align}
    \frac{d}{dt}\avg{\hat{a}_1} &= {\rm tr}\{\Lsys\rhou\hat{a}_1 \} + {\rm tr}\{\Lc\rhou\hat{a}_1 \} \nonumber \\
    &= \left(i\Delta_{a1}-\frac{\kappa_1+\Gamma_c}{2}\right)\avg{\hat{a}_1} + G_1 \avg{\hat{a}_1^{\dagger}} - iG_{12} \avg{\hat{a}_2^{\dagger}}  + \eta \nonumber \\ 
    &~~~~+ \frac{\varg_c-\Gamma_c}{2}\avg{\hat{b}_1},  \\
    \frac{d}{dt}\avg{\hat{a}_2} &= {\rm tr}\{\Lsys\rhou\hat{a}_2 \} + {\rm tr}\{\Lc\rhou\hat{a}_2 \} \nonumber \\
    &=  \left(i\Delta_{a2}-\frac{\kappa_2}{2}\right)\avg{\hat{a}_2} - iG_{12} \avg{\hat{a}_1^{\dagger}}.
\end{align}
Requiring a directional interaction between the two-mode amplifier and the QRC, we impose $\varg_c = \Gamma_c$. Furthermore, we now set $\kappa_1 + \Gamma_c = \kappa_2 \equiv \kappa$, so that the effective damping rate of both modes $\hat{a}_1$ and $\hat{a}_2$ is equal. Finally, for simplicity we also choose $\kappa_1 = \Gamma_c = \kappa/2$ and $\Delta_{a1} = \Delta_{a2} = 0$. . Then, the simplified system above can be conveniently written in the quadrature basis as
\begin{align}
\frac{d}{dt}
    \begin{pmatrix}
    \avg{\hat{X}_{a_1}} \\
    \avg{\hat{P}_{a_1}} \\
    \avg{\hat{P}_{a_2}} \\
    \avg{\hat{X}_{a_2}} 
    \end{pmatrix}
    =
    \mathbf{M}
    \begin{pmatrix}
    \avg{\hat{X}_{a_1}} \\
    \avg{\hat{P}_{a_1}} \\
    \avg{\hat{X}_{a_2}} \\
    \avg{\hat{P}_{a_2}} 
    \end{pmatrix}
    + 
    \begin{pmatrix}
    \sqrt{2}\eta \\
    0 \\
    0 \\
    0
    \end{pmatrix},
    \label{appeq:amp}
\end{align}
where the dynamical matrix $\mathbf{M}$ takes the form:
\begin{align}
    \mathbf{M} = \begin{pmatrix}
    G_1 -\frac{\kappa}{2} & 0 & 0 & -G_{12} \\
    0 & -G_1-\frac{\kappa}{2} & -G_{12} & 0 \\
    0 & -G_{12} & -\frac{\kappa}{2} & 0 \\
    -G_{12} & 0 & 0 & -\frac{\kappa}{2}
    \end{pmatrix}.
\end{align}
The form of $\mathbf{M}$ already clarifies that the interaction $\propto G_1$ amplifies the $\hat{X}_{a_1}$ quadrature and squeezes the $\hat{P}_{a1}$ quadrature. We are interested in the steady-state value of the quadrature expectation value $\avg{\hat{X}_{a_1}}$ that couples to the QRC. This is easily found by setting the left hand side of Eq.~(\ref{appeq:amp}) to zero and solving the resulting system
\begin{align}
\begin{pmatrix}
    \avg{\hat{X}_{a_1}} \\
    \avg{\hat{P}_{a_1}} \\
    \avg{\hat{X}_{a_2}} \\
    \avg{\hat{P}_{a_2}} 
\end{pmatrix} = 
-\mathbf{M}^{-1}
\begin{pmatrix}
    \sqrt{2}\eta \\
    0 \\
    0 \\
    0
    \end{pmatrix}.
    \label{appeq:amp2}
\end{align}
This yields the steady-state value of $\avg{\hat{X}^{(\sigma)}_{a_1}}$ for parameters corresponding to the $j$th state generated by the two-mode amplifier
\begin{align}
    \avg{\hat{X}^{(\sigma)}_{a_1}} = -\frac{2\kappa}{4(G^{(\sigma)}_{12})^2 + \kappa(2G_1^{(\sigma)}-\kappa)}\cdot\sqrt{2}\eta^{(\sigma)}.
\end{align}
It is easy to check that for the chosen parameter values corresponding to Fig.~\ref{fig:classifyAmpStates}, namely $(\eta^{(\sigma)},G_1^{(\sigma)},G_1^{(\sigma)}) \in  \{(5.0,0.3,0.0), (8.0,0.0,0.3)\}\kappa$, we have $\avg{\hat{X}^{(1)}_{a_1}} = \avg{\hat{X}^{(2)}_{a_1}}$. By extracting the steady-state solution for $\avg{\hat{P}_{a_1}}$ from Eq.~(\ref{appeq:amp2}), we similarly find $\avg{\hat{P}^{(1)}_{a_1}} = \avg{\hat{P}^{(2)}_{a_1}} = 0$.

\section{Training and testing details}
\label{app:training}

In this section, we provide details on the linear output layer for the processing of measurement records obtained from the QRC measurement chain, as applied to a $C$-state classification task. 

\textit{QRC outputs.}$-$ We begin by defining the QRC output vector $\mathbf{x}(t)$ spanning the measured phase space, constructed using measured quadratures,
\begin{align}
    \mathbf{x}(t) = 
    \begin{pmatrix}
    I^X_1(t) \\
    I^P_1(t) \\
    \vdots \\
    I_K^X(t) \\
    I_K^P(t) \\
    \end{pmatrix}
    =
    \begin{pmatrix}
    \avg{\frac{1}{t}\int_0^t d\tau~J^X_1(\tau)}_{N_{\rm S}} \\
    \avg{\frac{1}{t}\int_0^t d\tau~J^P_1(\tau)}_{N_{\rm S}} \\
    \vdots \\
    \avg{\frac{1}{t}\int_0^t d\tau~J^X_K(\tau)}_{N_{\rm S}} \\
    \avg{\frac{1}{t}\int_0^t d\tau~J^P_K(\tau)}_{N_{\rm S}} \\
    \end{pmatrix}.
    \label{appeq:x}
\end{align}
For a $K$-node QRC under heterodyne measurement, $\mathbf{x}(t)$ as defined above is thus a $2K$-dimensional vector at each $t$. Single-shot readout corresponds to setting $N_{\rm S}=1$.


Note that individual runs of the measurement chain for each of the $C$ classes comprising the classification tasks yield distinct stochastic trajectories of measured quadratures as defined in Eq.~(\ref{appeq:x}). Thus, we can index QRC output vectors $\mathbf{x}^{(\sigma)}_{(q)}(t)$ by the corresponding class label $\sigma=1,2,\ldots,C$, and with the index $q$ labelling the stochastic trajectory.

Once the measured QRC output is obtained, an output layer acts on these outputs to ideally return the corresponding state label $\sigma$; the result of the computation can therefore formally be written as
\begin{align}
    \sigma = f_N\{\WO\mathbf{x}^{(\sigma)}_{(q)}(t) + \mathbf{b}\}
    \label{appeq:trainW}
\end{align}
where $\WO$ is the matrix of trainable, time-independent output weights, $\mathbf{b}$ is a vector of trainable biases.


In Eq.~(\ref{appeq:trainW}), $f_N\{\cdot\}$ is a normalizing function that maps the vector of measured QRC outputs to a discrete, scalar state label $\sigma \in [1,\ldots,C]$. This mapping is carried out via two operations. First, the QRC outputs $\mathbf{x}^{(\sigma)}_{(q)}(t)$ are mapped to an intermediate $C$-dimensional target vector $\mathbf{y}^{(\sigma)}_{(q)}(t)$ employing a `one-hot' encoding (conventional for classification tasks): The $i$th element of this target vector $\mathbf{y}^{(\sigma)}_{(q)}(t)$ is given by:
\begin{align}
    [\mathbf{y}^{(\sigma)}_{(q)}(t)]_i = 
    \begin{cases}
    1~\forall~t~\text{if}~i = \sigma, \\
    0~\forall~t~\text{otherwise.}
    \end{cases}
\end{align}
Note that elements of target vectors with one-hot encoding sum to unity. The output layer mapping QRC outputs to target vectors therefore also includes a normalization function,
\begin{align}
    \mathbf{y}^{(\sigma)}_{(q)}(t) = S[\WO\mathbf{x}^{(\sigma)}_{(q)}(t) + \mathbf{b}] \equiv f_S[\mathbf{x}^{(\sigma)}_{(q)}(t)]
    \label{appeq:yfn}
\end{align}
where $S[\cdot]$ is the normalized exponential (or `softmax') function, defined by its action on an input vector $\mathbf{v}$:
\begin{align}
    S[\mathbf{v}]_i = \frac{e^{v_i}}{\sum_j e^{v_j}}
    \label{appeq:softmax}
\end{align}
This function therefore has the property that $\sum_i S[\mathbf{v}]_i = 1$; in other words, it normalizes entries of a vector $\vec{v}$ it acts upon such that the elements of the resulting vector add up to unity.

The reason to define target vectors with one-hot encoding is the property that the corresponding class label $\sigma$ is simply given by
\begin{align}
    \sigma &= {\rm arg~max}~\{ \mathbf{y}^{(\sigma)}_{(q)}(t) \} 
    \label{appeq:sigmafn}
\end{align}
where ${\rm arg~max}\{\mathbf{y}\}$ provides the index of the element of $\mathbf{y}$ with largest magnitude. Therefore, combining Eqs.~(\ref{appeq:sigmafn}),~(\ref{appeq:yfn}) completes the definition of a trainable output layer mapping vector QRC outputs to discrete state labels,
\begin{align}
    \sigma = {\rm arg~max}~\{ S[\WO\mathbf{x}^{(\sigma)}_{(q)}(t) + \mathbf{b}]\} 
    \label{appeq:outputFull}
\end{align}
Comparing with Eq.~(\ref{appeq:trainW}), we see that $f_N\{\cdot\} \equiv {\rm arg~max}~\{ S[\cdot]\}$, and is therefore a \textit{nonlinear} normalizing operation on measured QRC outputs. However, note that it is fixed (not trained), and hence does not increase training complexity. More importantly, by analyzing linear QRCs followed by the output layer defined by Eq.~(\ref{appeq:outputFull}), we show in the main text that the nonlinearity of $f_N\{\cdot\}$ is insufficient to successfully perform the classification tasks considered in this paper.


\textit{Training.}$-$ The goal of training can now be defined: to learn trainable weights $\WO$ and biases $\mathbf{b}$ to optimally map measured QRC outputs $\mathbf{x}^{(\sigma)}_{(q)}(t)$ to target vectors $\mathbf{y}^{(\sigma)}_{(q)}(t)$; the fidelity of the mapping is determined by minimizing a suitable cost function. We now discuss how measured QRC outputs are assembled in a form that aids this optimization on a digital computer.

To this end, we first note that the time label in the above definitions is actually discrete; the measurement window $t\in [t_0,t_f]$ is sampled at $M_T$ temporal grid points (determined by the temporal resolution of measurement apparatus in experiments, and by the temporal step size of integration of STEOMs in simulations). Any segment of this measurement window, defined by times $t \in [t_s,t_e]$ with a total of $m_T$ grid points, can be used for training; for example, for the pointer state classification task in the main text, the entire measurement window was used (so $t_s=t_0, t_e=t_f$, and $m_T = M_T$), while for amplifier state measurement, only the final integrated time point was used ($t_s=t_e=t_f$, and $m_T = 1$). The specific choice can be task-dependent.

Thus we can define a matrix $\mathbf{X}_{(q)}$ for the $q$th run of the measurement chain for state $\sigma$:
\begin{align}
    &\mathbf{X}_{(q)} = \nonumber \\
    &
    \begin{pmatrix}
        \mathbf{x}_{(q)}^{(\sigma=1)}(t_s) \cdots  \mathbf{x}_{(q)}^{(\sigma=1)}(t_e) & \cdots & \mathbf{x}_{(q)}^{(\sigma=C)}(t_s)  \cdots  \mathbf{x}_{(q)}^{(\sigma=C)}(t_e)
    \end{pmatrix}
\end{align}
which is evidently a $2K$-by-$(C\times m_T)$ matrix. The corresponding matrix of output vectors takes the analogous form:
\begin{align}
    &\mathbf{Y}_{(q)} = \nonumber \\
    &
    \begin{pmatrix}
        \mathbf{y}_{(q)}^{(\sigma=1)}(t_s) \cdots  \mathbf{y}_{(q)}^{(\sigma=1)}(t_e) & \cdots & \mathbf{y}_{(q)}^{(\sigma=C)}(t_s)  \cdots  \mathbf{y}_{(q)}^{(\sigma=C)}(t_e)
    \end{pmatrix}
\end{align}
and is a $C$-by-$(C\times m_T)$ matrix.

A training set is constructed by obtaining a total of $\QTR$ measured QRC outputs, indexed as $q=1,\ldots,\QTR$. For measured quadratures constructed by averaging over $N_{\rm S}$ individual shots from the measurement chain, this requires $\QTR \times N_{\rm S}$ runs of the measurement chain per state $\sigma$. For each $q$ we obtain a matrix of QRC outputs $\mathbf{X}_{(q)}$ and corresponding targets $\mathbf{Y}_{(q)}$. These can be compiled into a composite QRC output matrix $\mathbf{X}$ by concatenation of the individual matrices $\mathbf{X}_{(q)}$:
\begin{align}
    \mathbf{X}  = 
    \begin{pmatrix}
        \mathbf{X}_{(q=1)} &\cdots & \mathbf{X}_{(q=Q_T)}
    \end{pmatrix}
\end{align}
which is clearly a $2K$-by-$(\QTR \times C\times m_T)$ matrix. The corresponding target matrix $\mathbf{Y}$ is similarly defined as:
\begin{align}
    \mathbf{Y} = \begin{pmatrix}
        \mathbf{Y}_{(q=1)} &\cdots & \mathbf{Y}_{(q=\QTR)}
    \end{pmatrix}
\end{align}
which is a $C$-by-$(\QTR\times C\times m_T)$ matrix. 

The optimal matrix of weights $\WO^{\rm opt}$ and biases $\mathbf{b}^{\rm opt}$ are finally determined by least squares optimization:
\begin{align}
    \{\WO^{\rm opt},\mathbf{b}^{\rm opt}\} = {\rm min}_{\WO,\mathbf{b}}\{||\mathbf{Y}-f_S[\mathbf{X}]||^2 \}
    \label{appeq:wopt}
\end{align}
where $f_S[\mathbf{X}]$ is to be interpreted as operating column-wise on $\mathbf{X}$, in accordance with Eq.~(\ref{appeq:yfn}).

A final note before discussing the results of this training procedure relates to the particular form of the measured subspace for the pointer state classification task analyzed in Sec.~\ref{sec:cav}. The full measured phase space is projected onto a set of optimal quadratures determined by phases $\{\varphi_k\}$. This projection is achieved by simply rewriting the matrix of trainable weights $\WO$ as:
\begin{align}
    \WO = \widetilde{\mathbf{W}}_{\rm O}\mathcal{M}^{\varphi}
    \label{appeq:wop}
\end{align}
where $\mathcal{M}^{\varphi}$ is the $K$-by-$2K$ matrix defined in terms of $K$ phase angles introduced in Eq.~(\ref{eq:Mphi}) of the main text, and $\widetilde{\mathbf{W}}_{\rm O}$ is now a $C$-by-$K$ matrix of weights. With this reparameterization of $\WO$, the rest of the training procedure remains unchanged, and the optimal matrix $\WO^{\rm opt}$ is still given by Eq.~(\ref{appeq:wopt}). Formally, the only change is that the number of independent trainable weights in $\WO$ is reduced from $C\times 2K$ to $C\times K$ weights $\widetilde{\mathbf{W}}_{\rm O}$ and $K$ projection phases $\{\varphi_k\}$, and that the projection phases must be trained with bounds $[0,2\pi]$. 

\textit{Linear decision boundaries.}$-$ Once the weight matrix (and biases) are trained via Eq.~(\ref{appeq:wopt}), they define linear decision boundaries that separate different classes and allow classification, as we will now show explicitly. We begin with the following observation: \textit{on} the decision boundary that separates an arbitrary class $i$ from class $j$, QRC outputs corresponding to $\sigma=i$ could be equally classified as belonging to $\sigma=j$, and vice versa. Mathematically, this implies $\mathbf{y}^{(\sigma=i)}(t) = \mathbf{y}^{(\sigma=j)}(t)$ on the decision boundary; from Eq.~(\ref{appeq:yfn}), this condition can be written as:
\begin{align}
    S[\WO\mathbf{x}(t) + \mathbf{b}]_i = S[\WO\mathbf{x}(t) + \mathbf{b}]_j
\end{align}
or, using the definition of the softmax function in Eq.~(\ref{appeq:softmax}),
\begin{align}
    \exp\{  [\WO\mathbf{x}(t) + \mathbf{b}]_i \} = \exp \{ [\WO\mathbf{x}(t) + \mathbf{b}]_j \}
\end{align}
where the normalization factor is common to both sides and cancels. This immediately requires the arguments of the exponentials to be equal, a requirement that can be rewritten in the form:
\begin{align}
    (b_i - {b}_j) + \sum_{k=1}^{2K} ( (\WO)_{ik}-(\WO)_{jk}){x}_k = 0
    \label{appeq:ldb}
\end{align}
which is simply the definition of a hyperplane in the measured phase space spanned by $\{x_1,\ldots,x_k,\ldots,x_{2K}\}$. This hyperplane defines the linear decision boundary that separates arbitrary classes $i$ and $j$.

We now specialize the above general prescription to the case of pointer state classification considered in the main text. First, we find that the bias terms vanish due to the symmetry of the task being considered. The minor change is the introduction of trainable phase angles $\{\varphi_1,\varphi_2\}$, as mentioned earlier. To proceed, we simply rewrite $\WO$ in Eq.~(\ref{appeq:ldb}), in accordance with Eq.~(\ref{appeq:wop});
\begin{align}
    \sum_{l=1}^{K}\sum_{k=1}^{2K} ( (\widetilde{\mathbf{W}}_{\rm O})_{il}\mathcal{M}^{\varphi}_{lk}-(\widetilde{\mathbf{W}}_{\rm O})_{jl}\mathcal{M}^{\varphi}_{lk}){x}_k &= 0 \nonumber \\
    \implies \sum_{l=1}^{K} ( (\widetilde{\mathbf{W}}_{\rm O})_{il}-(\widetilde{\mathbf{W}}_{\rm O})_{jl}){I}_l^{\varphi} &= 0
\end{align}
In going from the first to the second line the sum over $k$ is performed, and simply yields $\sum_k \mathcal{M}^{\varphi}_{lk}x_k = I_l^{\varphi}$ as defined by Eq.~(\ref{eq:Mphi}) of the main text. Recalling that $K=2$, the linear decision boundaries separating classes $i$ and $j$ are now lines in the measured phase space $\{I_1^{\varphi},I_2^{\varphi}\}$, given by:
\begin{align}
    I_2^{\varphi} = -\frac{(\widetilde{\mathbf{W}}_{\rm O})_{i1}-(\widetilde{\mathbf{W}}_{\rm O})_{j1}}{(\widetilde{\mathbf{W}}_{\rm O})_{i2}-(\widetilde{\mathbf{W}}_{\rm O})_{j2}} \cdot I_1^{\varphi}
\end{align}



\textit{Testing.}$-$ Once $\WO^{\rm opt}$, $\mathbf{b}^{\rm opt}$ have been learned, the now-trained output layer can be used to test classification performance. The class label $\sigma^{p(\sigma)}_{(q)}$ predicted from a measured QRC output trajectory $\{\mathbf{x}^{(\sigma)}_{(q)}\}$ that is part of the test set ($q=1,\ldots,\QTE$) is obtained by passing the QRC trajectory through the trained output layer defined in Eq.~(\ref{appeq:outputFull}),
\begin{align}
    \sigma^{p(\sigma)}_{(q)} = {\rm arg~max}~\{ \mathbf{y}^{(\sigma)}_{(q)} \} = {\rm arg~max}~\{ S[\WO^{\rm opt}\mathbf{x}^{(\sigma)}_{(q)} + \mathbf{b}^{\rm opt}]\}.
\end{align}
Classification is correct if $\sigma^{p(\sigma)}_{(q)} = \sigma$. We have suppressed time labels on measured QRC trajectories and predicted class labels for brevity.

By predicting class labels using the entire test set of size $\QTE$, the mean classification accuracy can be computed as
\begin{align}
    {\rm Classification~accuracy} = \frac{1}{C}\sum_{\sigma=1}^C\left[ \frac{1}{\QTE}\sum_{q=1}^{\QTE} \delta_{\sigma^{p(\sigma)}_{(q)},\sigma} \right],
\end{align}
where $\delta_{i,j}$ is the Kronecker delta function, here used simply to count instances where $\sigma^{p}_{(q)} = \sigma$ indicating correct classification. 


\begin{figure}[t]
    \centering
    \includegraphics[scale=1.0]{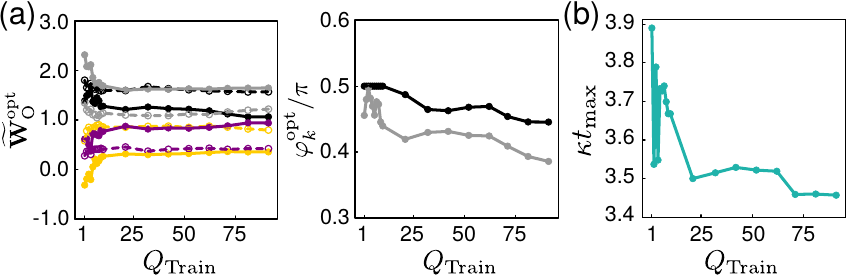}
    \caption{(a) Convergence of trained optimal weights $\WO^{\rm opt}$ for the pointer state classification task as a function of \textit{training} set size $\QTR$, parameterized as defined in Eq.~(\ref{appeq:wop}) in terms of elements of $\widetilde{\mathbf{W}}_{\rm O}^{\rm opt}$ (left panel) and phases $\{\varphi_k^{\rm opt}\}$ (right panel). (b) Convergence of $\kappa \tmax$ using optimal trained weights as a function of training set size $\QTR$.}
    \label{fig:weightsConv}
\end{figure}



\textit{Convergence with training set size.}$-$ Having discussed the formal training protocol we analyze briefly how training results vary with training set size, applied to the pointer state classification task. The left panel of Fig.~\ref{fig:weightsConv}(a) shows elements of the trained $C$-by-$K$ matrix $\widetilde{\mathbf{W}}^{\rm opt}_{\rm O}$ (solid lines are first column elements, dashed lines are second column elements) as a function of training set size $\QTR$. The right panel shows trained projection phases $\{\varphi_k^{\rm opt}\}$. In Fig.~\ref{fig:weightsConv}(b) we plot the metric $\tmax$ that characterizes classification performance on the \textit{training} set (for the QRC used to obtain this training set, $\cmax \geq 0.99$). We see that the obtained optimal weights in particular change rapidly for short training set sizes $\QTR$, with a corresponding irregular change in $\tmax$. However with increasing training set size, both the the trained weights and, more importantly, $\tmax$ converge. We find that the optimization landscape for projection phases is relatively shallow, so that small changes in $\{\varphi_k^{\rm opt}\}$ do not significantly change $\tmax$. We use this approach of testing classification performance using the training data set to determine $\QTR$ required for training in our classification tasks.

\section{Training on finitely-sampled measured distributions vs. training using expectation values}
\label{app:trainingTypes}

In this appendix section, we analyze the difference in training using expectation values and using noisy data sets of consisting of measured quadrature records. For simplicity, we consider the amplifier state classification task of Sec.~\ref{sec:amp}, for $\bar{\Lambda}/\bar{\gamma} = 0.02$. All other QRC and amplifier hyperparameters are as mentioned in Sec.~\ref{subsec:ampChain}.

\begin{figure}[t]
    \centering
    \includegraphics[scale=1.0]{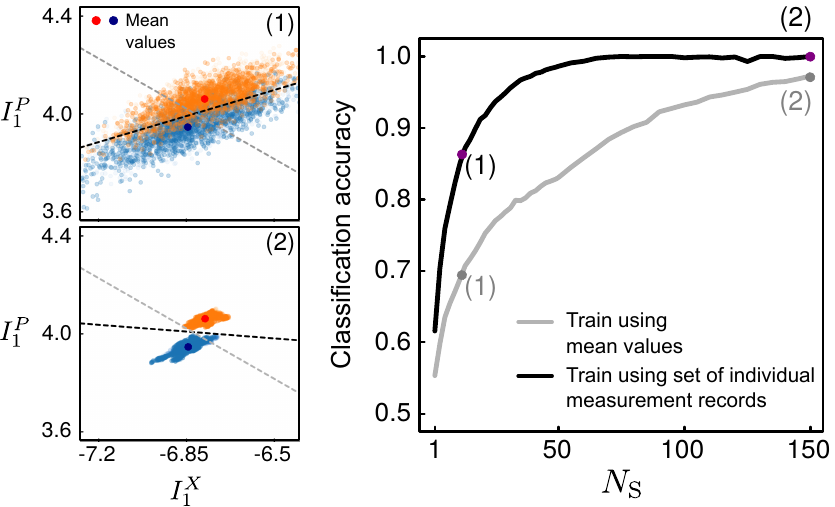}
    \caption{Classification accuracy as a function of number of shots $N_{\rm S}$. Linear decision boundaries are obtained by training using either $\QTR$ measured quadrature records (black), or using only indicated mean values (gray). }
    \label{fig:trainingTypes}
\end{figure}

We begin by recalling the approach to training in our quantum reservoir computing framework. Consider the measured quadrature distributions shown in measured phase space in (1), the left panel of Fig.~\ref{fig:trainingTypes}. Each colored point is one of $1500$ measured quadrature records per amplifier state (in orange and blue respectively), each constructed using $N_{\rm S}=10$ individual measurement records. The linear decision boundary learned based on the measured \textit{distributions} is shown in dashed black. For (2), $N_{\rm S} = 150$, the measured quadrature distributions have a reduced variance due to increased ensemble averaging. The decision boundary trained on these measured distributions is different from (1), being tailored to the experimental resources used in output processing. We can test the classification accuracy of trained QRCs as in Sec.~\ref{sec:amp}, using $\QTE=3000$ measured quadrature records per state, each averaged over $N_{\rm S}$ shots. This result is plotted in the right panel of Fig.~\ref{fig:trainingTypes} in solid black, and shows the same qualitative form familiar from Fig.~\ref{fig:classifyAmpStates}.


We now consider how training on expectation values might be carried out. In practice, expectation values must also be obtained from measurement records. Here, we consider expectation values of the measured quadratures as being approximately obtained once a large number of measurement records are averaged over; here we consider $N_{\rm S} = 3000$. In measured phase space, these expectation values are marked by the red and blue dots respectively in the left panel of Fig.~\ref{fig:trainingTypes}. The linear decision boundary in this case is simply the perpendicular bisector of the line joining these two mean values, and is shown by the dashed gray line. 

The classification accuracy plotted in solid gray in the right panel of Fig.~\ref{fig:trainingTypes} corresponds to testing using this linear decision boundary learned using the two expectation values; we see that it underperforms relative to the case of linear decision boundaries trained using measured quadrature distributions. Note that the linear decision boundary learned using expectation values has no information about the non-isotropic (i.e. squeezed) distribution of measured quadratures; how this is detrimental is obvious from the distributions and decision boundaries in (1) and (2). Note that gray curve asymptotically approaches the black curve as $N_{\rm S} \to \infty$ limit. In this limit, the measured distributions approach their expectation values, and the difference in the two training protocols is suppressed. 



\section{Post-Amplification}
\label{app:postamp}

\begin{figure}[t]
    \centering
    \includegraphics[scale=1.0]{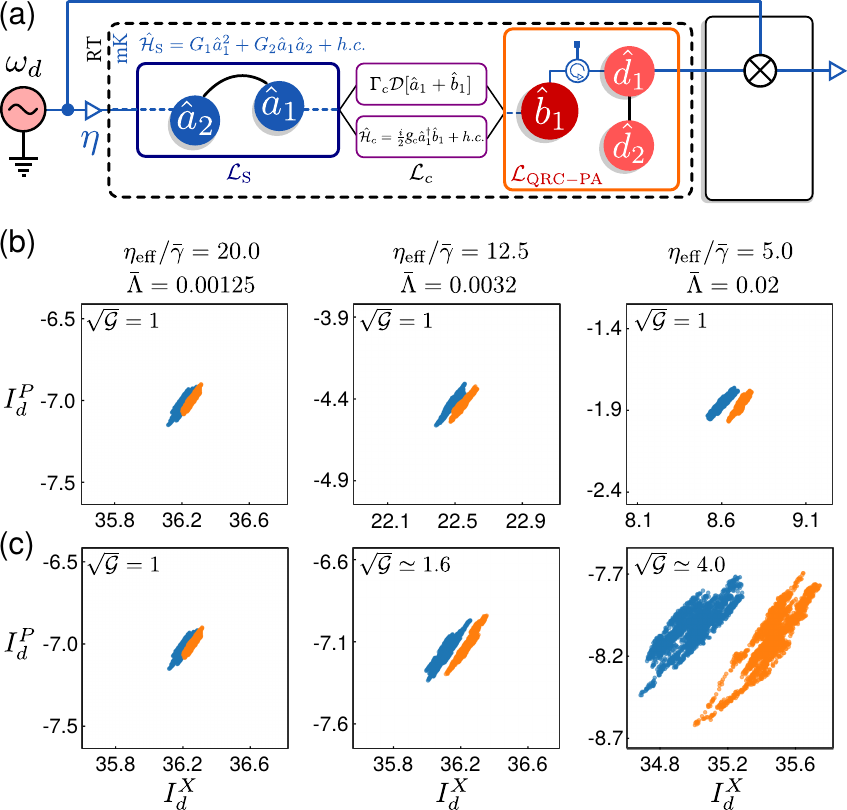}
    \caption{(a) Schematic using a non-degenerate post-amplifier (PA) to amplify output from the single nonlinear QRC node processing inputs from a measured quantum system. (b) Measured phase space outputs for unit gain $\sqrt{\mathcal{G}}=1$ for different instances of the amplifier state classification task, carried out using three QRCs with $\CNLE \simeq 0.385$. Equal areas of measured phase space are shown; note, however, the different regions, as the amplitudes of measured heterodyne records are different due to the different $\eta_{\rm eff}$ strengths. (c) Measured phase space outputs with PA gain $\sqrt{\mathcal{G}}$ (as indicated) chosen to yield approximately equal output amplitudes. Again, equal areas of phase space are shown, now also in approximately the same region of phase space, as the PA appropriately amplifies the QRC output for the different $\eta_{\rm eff}$ strengths. }
    \label{fig:postAmp}
\end{figure}

In this appendix section we consider the amplifier state classification task discussed in Sec.~\ref{sec:amp} of the main text, but now including a post-amplifier (PA), directionally-coupled to the single-node QRC, to amplify the QRC output. The corresponding schematic of the complete measurement chain is shown in Fig.~\ref{fig:postAmp}(a). 

We model the PA as a non-degenerate (phase-preserving) quantum-limited amplifier consisting of two modes $\hat{d}_1$, $\hat{d}_2$. The inclusion of the PA modifies the Liouvillian describing the QRC. The SME describing the measurement chain is now given by
\begin{align}
d\rhoc =  \Lsys\rhoc~dt + \Lc\rhoc~dt + \mathcal{L}_{\rm QRC-PA}\rhoc~dt +\widetilde{\mathcal{S}}_{\rm meas}(dW)\rhoc.
\end{align}
$\Lsys$ and $\Lc$ are as defined by Eq.~(\ref{eq:lsysamp}) and Eq.~(\ref{eq:couplingCirc}) of the main text respectively. The modified Liouvillian $\mathcal{L}_{\rm QRC-PA}$ describes the evolution of the single-node QRC coupled to the PA, and is given by
\begin{align}
    \mathcal{L}_{\rm QRC-PA}\rhou = \widetilde{\mathcal{L}}_{\rm QRC}\rhou + \widetilde{\mathcal{L}}_{c}\rhou + \mathcal{L}_{\rm PA}\rhou.
\end{align}
Here, $\widetilde{\mathcal{L}}_{\rm QRC}$ describes the uncoupled evolution of single-node QRC as before, but has a slightly different form than that in Eq.~(\ref{eq:Lqrc}) considered in the main text, since the single-node QRC is incorporated differently in the measurement chain. More precisely,
\begin{align}
    \widetilde{\mathcal{L}}_{\rm QRC}\rhou = -i\Big[ & \omega_1 \bkd{1}\bk{1} -  \frac{\Lambda_1}{2}\bkd{1}\bkd{1}\bk{1}\bk{1} , \rhou \Big].
\end{align}
Next, $\mathcal{L}_{\rm PA}$ describes the evolution of the PA, defined as a non-degenerate (phase-preserving) parametric amplifier with interaction strength $G_{\rm PA}$,
\begin{align}
    \mathcal{L}_{\rm PA}\rhou = -i\Big[ & \omega_{d1} \hat{d}^{\dagger}_1\hat{d}_1 + \omega_{d2} \hat{d}^{\dagger}_2\hat{d}_2  + (G_{\rm PA}\hat{d}_1\hat{d}_2 + h.c.), \rhou \Big]  \nonumber \\
 ~~+ &\gamma_{d1} \mathcal{D}[\hat{d}_1]\rhou + \gamma_{d2} \mathcal{D}[\hat{d}_2]\rhou,
 \label{eq:lqrcpa}
\end{align}
with frequencies $\omega_{d1},\omega_{d2}$ and damping rates $\gamma_{d1},\gamma_{d2}$ for the two PA modes. Finally, the coupling Liouvillian $\widetilde{\mathcal{L}}_{c}$ between the single nonlinear node of the QRC and mode $\hat{d}_1$ of the PA takes the form
\begin{align}
    \widetilde{\mathcal{L}}_{c}\rhou = 
    -i\Big[&\frac{i}{2}\widetilde{\varg}_{c}\hat{b}_1^{\dagger}\hat{d}_1 + h.c., \rhou \Big] + \widetilde{\Gamma}_{c}\mathcal{D}[\hat{b}_1 + \hat{d}_1]\rhou,
\end{align}
which is exactly the same form as the coupling between the two-mode amplifier system and the QRC, and represents a circulator.

Finally, we now consider heterodyne measurements of a single node of the PA, in this case $\hat{d}_1$, so that the stochastic measurement superoperator $\widetilde{\mathcal{S}}_{\rm meas}(dW)$ takes the form
\begin{align}
&\widetilde{\mathcal{S}}_{\rm meas}(dW)\rhoc = \nonumber \\
& \sqrt{\frac{\gamma_{d1}}{2}} \left( \hat{d}_1 \rhoc + \rhoc \hat{d}_1^{\dagger} - \avgc{\hat{d}_1+\hat{d}_1^{\dagger}} \right) dW_{d}^{X}(t) + \nonumber \\
& \sqrt{\frac{\gamma_{d1}}{2}} \left( -i\hat{d}_1 \rhoc + i\rhoc \hat{d}_1^{\dagger} - \avgc{-i\hat{d}_1+i\hat{d}_1^{\dagger}} \right) dW_{d}^{P}(t)
\end{align}
and the obtained measurement currents $J_d^{X,P}(t)$ are given by:
\begin{align}
    J_{d}^{X}(t)~dt &= \sqrt{\frac{\gamma_{d1}}{2}}\avgc{\hat{d}_1+\hat{d}_1^{\dagger}}~dt + dW_d^{X}(t), \\
    J_{d}^{P}(t)~dt &= \sqrt{\frac{\gamma_{d1}}{2}}\avgc{-i\hat{d}_1+i\hat{d}_1^{\dagger}}~dt + dW_d^{P}(t).
\end{align}


Having defined the SME for the measurement chain depicted in Fig.~\ref{fig:postAmp}(a), we now discuss the choice of parameters to operate the measurement chain in the regime of interest. We first write down the unconditional equations of motion for first-order moments of the single nonlinear node $\hat{b}_1$ and the amplifier mode $\hat{d}_1$ in the frame rotating with the drive:
\begin{align}
    \frac{d}{dt}\avg{\hat{b}_1} &= {\rm tr}\{\mathcal{L}_{\rm QRC}\rhou\hat{b}_1 \} + {\rm tr}\{\Lc\rhou\hat{b}_1 \} \nonumber \\
    &= {\rm tr}\Big\{ -i\Big[  \omega_1 \bkd{1}\bk{1} -  \frac{\Lambda_1}{2}\bkd{1}\bkd{1}\bk{1}\bk{1} , \rhou \Big] \hat{b}_1 \Big\} \nonumber \\
    &~~~~-\frac{\Gamma_c+\widetilde{\Gamma}_c}{2}\avg{\hat{b}_1} + \frac{\widetilde{\varg}_c-\widetilde{\Gamma}_c}{2}\avg{\hat{d}_1} \label{appeq:bPA}  \\
    \frac{d}{dt}\avg{\hat{d}_1} &= {\rm tr}\{\mathcal{L}_{\rm PA}\rhou\hat{d}_1 \} + {\rm tr}\{\Lc\rhou\hat{d}_1 \} \nonumber \\
    &= \left(i\Delta_{d1}-\frac{\gamma_{d1}+\widetilde{\Gamma}_c}{2}\right)\avg{\hat{d}_1} - iG_{\rm PA} \avg{\hat{d}_2^{\dagger}}  \nonumber \\ 
    &~~~~+ \frac{\widetilde{\varg}_c+\widetilde{\Gamma}_c}{2}\avg{\hat{b}_1}  \\
    \frac{d}{dt}\avg{\hat{d}_2} &= {\rm tr}\{\mathcal{L}_{\rm PA}\rhou\hat{d}_2 \} + {\rm tr}\{\Lc\rhou\hat{d}_2 \} \nonumber \\
    &=  \left(i\Delta_{d2}-\frac{\gamma_{d2}}{2}\right)\avg{\hat{d}_2} - iG_{12} \avg{\hat{d}_1^{\dagger}}
\end{align}
where we have introduced the detunings $\Delta_{d1,d2} = \omega_d - \omega_{d1,d2}$, which will be taken to vanish for convenience. For a directional coupling such that the QRC node field drives the PA, but signals from the PA do not impact the unconditional dynamics of the QRC node, we once more choose
\begin{align}
    \widetilde{\varg}_{c} = \widetilde{\Gamma}_{c}.
\end{align}
As is clear from Eq.~(\ref{appeq:bPA}), the decay rate of the QRC node is now given by $\Gamma_c + \widetilde{\Gamma}_{c}$. Recall that the decay rate of the QRC considered in Sec.~\ref{sec:amp} was given by $\Gamma_c + \gamma_1 = 3\kappa/2$, with $\Gamma_c = \kappa/2$ and $\gamma_1 = \kappa$. We will require the single QRC node in this measurement chain to have the same total decay rate, namely $\Gamma_c + \widetilde{\Gamma}_{c} = 3\kappa/2$; keeping the same value of $\Gamma_c = \kappa/2$, we therefore must have $\widetilde{\Gamma}_c = \kappa$. For convenience, we also enforce both PA modes to have the same total decay rate, so that $\gamma_{d1}+\widetilde{\Gamma}_{c} = \gamma_{d2} \equiv \gamma_d$. We do so by requiring $\gamma_{d1} = \kappa/2, \gamma_{d2} = 3\kappa/2$.

With these parameter choices, we can proceed to analyze the dynamics of the PA nodes when driven by signals from the single nonlinear QRC node. In particular, we can calculate the gain by which the post-amplifier enhances the QRC signal. This can be extracted simply from the equations of motion of the first-order moments of the post-amplifier, written for convenience in the quadrature basis,
\begin{align}
\frac{d}{dt}
    \begin{pmatrix}
    \avg{\hat{X}_{d_1}} \\
    \avg{\hat{P}_{d_1}} \\
    \avg{\hat{P}_{d_2}} \\
    \avg{\hat{X}_{d_2}} 
    \end{pmatrix}
    =
    \widetilde{\mathbf{M}}
    \begin{pmatrix}
    \avg{\hat{X}_{d_1}} \\
    \avg{\hat{P}_{d_1}} \\
    \avg{\hat{X}_{d_2}} \\
    \avg{\hat{P}_{d_2}} 
    \end{pmatrix}
    + \Gamma'_c
    \begin{pmatrix}
    \avg{\hat{X}_1} \\
    \avg{\hat{P}_1} \\
    0 \\
    0
    \end{pmatrix}
    \label{appeq:amp}
\end{align}
where the dynamical matrix $\widetilde{\mathbf{M}}$ takes the form:
\begin{align}
    \widetilde{\mathbf{M}} = \begin{pmatrix}
     -\frac{\gamma_{d}}{2} & 0 & 0 & -G_{\rm PA} \\
    0 & -\frac{\gamma_{d}}{2} & -G_{\rm PA} & 0 \\
    0 & -G_{\rm PA} & -\frac{\gamma_{d}}{2} & 0 \\
    -G_{\rm PA} & 0 & 0 & -\frac{\gamma_{d}}{2}
    \end{pmatrix}
\end{align}
Solving for the steady-state first-order moments of the PA mode $\hat{d}_1$, we immediately find
\begin{align}
    \avg{\hat{X}_{d_1}} =  -\frac{2\gamma_d}{4G_{\rm PA}^2 -\gamma_d^2} \cdot\widetilde{\Gamma}_c\avg{\hat{X}_1}.
\end{align}
To define the transmission gain $\mathcal{G}$ of the PA, we can now simply write down the ratio of its output field amplitude, defined via the ensemble-averaged measurement record $\avg{J_d^X}$, relative to the input to be amplified, namely the single QRC node field:
\begin{align}
    \frac{\avg{J_d^X}}{\sqrt{\widetilde{\Gamma}_c}\avg{\hat{X}_1}} = \frac{\sqrt{\gamma_{d1}}\avg{\hat{X}_{d_1}}}{\sqrt{\widetilde{\Gamma}_c}\avg{\hat{X}_1}} = \left[ -\frac{2\gamma_d\sqrt{\widetilde{\Gamma}_c\gamma_{d1}}}{4G_{\rm PA}^2 -\gamma_d^2} \right] \equiv \sqrt{\mathcal{G}}
    \label{appeq:PAG}
\end{align}
For $G_{\rm PA} < \gamma_d/2$ (the instability threshold for the PA), the factor in square brackets is positive-definite. The same gain applies to the heterodyne measurement of the conjugate quadrature $\avg{\hat{J}_{d}^P}$ due to the non-degenerate (i.e. phase-preserving) nature of the PA chosen here. 

The required PA gain determines the interaction strength $G_{\rm PA}$, thus specifying the remaining operating parameters of the measurement chain for a given task. We will apply this measurement chain to the task of classifying amplifier states discussed in Sec.~\ref{sec:amp} of the main text, in particular when using a single-node QRC to classify states with different values of $\eta_{\rm eff}$, as analyzed in Sec.~\ref{subsec:ampquantum}. More precisely, we consider three instances of the task as before, defined by effective drive strengths $\eta_{\rm eff}/\bar{\gamma} \in \{5.0,12.5,20.0\}$. For each instance, the nonlinearity of the single-node QRC in the measurement chain is respectively chosen as $\bar{\Lambda}/\bar{\gamma} \in \{0.00125,0.0032,0.02\}$, to render the effective dimensionless nonlinearity parameter $\CNLE$ equal for QRCs carrying out the corresponding task; for the choice of parameters here, $\CNLE \simeq 0.385$. 

We first consider the case where the PA is operated to provide unit gain. This requires the factor in Eq.~(\ref{appeq:PAG}) to have a magnitude of one, which imposes the PA interaction strength:
\begin{align}
    G_{\rm PA}^{\rm \mathcal{G}=1} = \sqrt{  \frac{1}{4}\gamma_d^2 -\frac{1}{2}\gamma_d\sqrt{\widetilde{\Gamma}_c\gamma_{d1}}  } = \frac{\kappa}{4}\sqrt{9-6\sqrt{2}} \simeq 0.1794\kappa.
\end{align}
The measured QRC-PA output is constructed as before using measured quadratures obtained from the PA, defined analogously to Sec.~\ref{sec:amp},
\begin{align}
\mathbf{x}(t) = 
\begin{pmatrix}
 I_d^{X}(t) \\
 I_d^{P}(t) 
\end{pmatrix},~
    I_d^{X,P}(t) = \Big< \frac{1}{t}\int_{t_0}^t d\tau~J_d^{X,P}(\tau) \Big>_{N_{\rm S}},
\end{align}
and are plotted in Fig.~\ref{fig:postAmp}(b) for the three instances of the classification task from left to right using QRCs with $\CNLE simeq 0.385$, and for $N_{\rm S} = 200$. Due to the directionality of the coupling between the QRC and the PA, for unit gain operation the output of the measurement chain including the PA is quite similar to that without the PA, as considered in Sec.~\ref{sec:amp}. Hence the salient features of classification performance for different $\eta_{\rm eff}$ are the same as those observed in Fig.~\ref{fig:ampClassifyDrive}. In particular, output from the measured chain with the QRC possessing the strongest nonlinearity (rightmost panel) has the largest separation of measured distributions. However, as noted earlier, this output also has the lowest amplitudes of measured quadratures, as evident from the phase space axes, due to the weakest input drive strength. 

The observed difference can be overcome by choosing the PA interaction strength $G_{\rm PA}$ such that for all three instances of the classification task, the amplitude of measured outputs is approximately equal. For concreteness, we do so by requiring the measured output amplitudes for the weaker effective drives $\eta_{\rm eff}/\bar{\gamma} \in \{5.0,12.5\}$, to match the measured output amplitudes for the strongest effective drive, $\eta_{\rm eff}/\bar{\gamma} = 20.0$. This requires amplitude gains $\sqrt{\mathcal{G}}$ determined (approximately) by the ratio of the corresponding effective drives, $\sqrt{\mathcal{G}} \in \{20.0/5.0=4.0,20.0/12.5=1.6,1.0\}$ respectively for the three instances considered. 

Applying this PA gain, the measured quadrature distributions are shown in Fig.~\ref{fig:postAmp}(c). The measured output distributions are now in very similar regions of phase space, indicating similar magnitudes of measured outputs. The lack of an exact equivalence is due to the change in PA linewidth for increasing gain due to anti-damping, which modifies the PA response time. As a result, for larger gain, the original value of waiting time $t_0$ introduced in Eq.~(\ref{eq:IkXPNS}) and carried through here is no longer sufficient for the measurement chain output to settle to steady-state. Consequently, the observed gain is a a slight underestimate of the steady-state gain $\sqrt{\mathcal{G}}$, for large gains. 

Each panel in Figs.~\ref{fig:postAmp}(b),~(c) show the same total area in phase space, albeit centered around different points. We can thus compare the variance of distributions directly. When comparing each measured distribution to its counterpart at unit PA gain (the panel directly above), we clearly see that phase-preserving amplification for $\mathcal{G}>1$ adds amplified quantum fluctuations to both measured quadratures, leading to an increase in the variance of measured distributions. This is particularly clear for the measurement chain with weakest input drive, which requires the largest gain amplification. However, the displacement of distributions that enables classification is also enhanced. As a result, the measurement chain containing the QRC with strongest nonlinearity would still require fewer experimental resources for classification.

\bibliography{refs}
\end{document}